\documentclass[%
 reprint,
 amsmath,amssymb,
 aps,
]{revtex4-1}
\usepackage{graphicx}
\usepackage{dcolumn}
\usepackage{bm}
\usepackage{ragged2e}
\usepackage{chngcntr}
\usepackage{epsfig}
\usepackage{amssymb}
\usepackage{lineno}
\usepackage{blindtext}
\usepackage{mathtools}

\begin{document}


\title{Nonlinear dynamics of acoustic bubbles excited by their pressure dependent subharmonic resonance frequency: oversaturation and enhancement of the subharmonic signal}

\author{AJ Sojahrood$^{1,2}$}
\email{amin.jafarisojahrood@ryerson.ca}
\author{R.E. Earl$^{3}$}%

\author{Q. Li$^{4}$}%
\author{T.M. Porter$^{4}$}%
\author{M. C. Kolios$^{1,2}$}%
\author{R. Karshafian$^{1,2}$}%
\affiliation{$^{1}$ Department of Physics, Ryerson University, Toronto, Canada\\
	$^{2}$ Institute for Biomedical Engineering, Science and Technology (IBEST) a partnership between Ryerson University and St. Mike's Hospital, Toronto, Ontario, Canada
	\\
}%
\affiliation{$^{3}$ Department of Mechanical Engineering, McGill University, Montreal, Canada
}%
\affiliation{$^{4}$ Department of Biomedical Engineering, Boston University, Boston, MA, USA
}%

\date{\today}

\begin{abstract}
The acoustic bubble is an example of a highly nonlinear system which is the building block of several applications and phenomena ranging from underwater acoustics to sonochemistry and medicine. Nonlinear behavior of bubbles, and most importantly $\frac{1}{2}$ order subharmonics (SH), are used to increase the contrast to tissue ratio (CTR) in diagnostic ultrasound (US) and to monitor bubble mediated therapeutic US.  It is shown experimentally and numerically that when bubbles are sonicated with their SH resonance frequency ($f_{sh}=2f_r$ where $f_r$ is the linear resonance frequency), SHs are generated at the lowest excitation pressure. SHs then increase rapidly with pressure increase and reach an upper limit of the achievable SH signal strength. Numerous studies have investigated the pressure threshold of SH oscillations; however, conditions to enhance the saturation level of SHs has not been investigated. In this paper nonlinear dynamics of bubbles excited by frequencies in the range of $f_r<f<2f_r$ is studied for different sizes of bubbles (400nm-8 $\mu m$). We show that the SH resonance frequency is pressure dependent and decreases as pressure increases. When a bubble is sonicated with its pressure dependent SH resonance frequency, oscillations undergo a saddle node bifurcation from a P1 or P2 regime to a P2 oscillation regime with higher amplitude. The saddle node bifurcation is concomitant with over saturation of the SH and UH amplitude and eventual enhancement of the upper limit of SH and UH strength (e.g. $\approx$ 7 dB in UH amplitude).   This can increase the CTR and signal to noise ratio in applications. Here, we show that the highest non-destructive SH amplitude occurs when $f\approxeq1.5-1.8f_r$
\end{abstract}
\maketitle
\section{Introduction}
A bubble is a nonlinear oscillator that can exhibits complex and chaotic dynamics [1-8]. Bubbles are the building block of several applications and phenomena; they have applications in sonochemistry [9-15], ultrasonic cleaning [16,17] , sonoluminscence [14,15] and medical ultrasound [18-23]. Pioneering works of [1-8] have shown the nonlinear and chaotic properties of forced bubble oscillations which are followed by recent extensive studies on the nonlinear behavior of bubbles in water [23-27], coated bubbles [24,26] , bubbles in highly viscous media [28-33]; and bubbles sonicated with asymmetrical driving acoustic forces [34-37].  Complexity of the bubble dynamics makes it very difficult to effectively implement bubbles in applications; however, within this complexity there exists an opportunity for beneficial bubble behavior in applications.\\ 
Period doubling (PD) is an example of a beneficial nonlinear behavior. In the bubble oscillator PD results in generation of $\frac{1}{2}$ order SHs and $\frac{3}{2}$ order UHs.  SH oscillations of bubbles are highly desirable due to unique properties that makes them very useful in several applications. Ultrasound contrast agents (coated bubbles) are clinically used on a daily basis to image microvascular blood flow and quantify blood perfusion (e.g. in the liver, kidney and the myocardium) [38-40]. Due the absence of SHs and UHs in tissue’s response to diagnostic ultrasound [20,41-43]; SH and UH emissions by bubbles allow the detection of blood flow with exceptional contrast enhancement [20,41-43]. Furthermore, SH emissions have lower frequencies and are attenuated less by the tissue.\\  
In therapeutic ultrasound SH emissions are used for monitoring therapeutic applications of ultrasound and as an indicator for stable cavitation [44-45].  SHs and UHs are employed to measure the efficacy of blood brain barrier opening [46,47].  SHs are proposed for the non-invasive measurement of the pressure inside vessels [48-50], to image the microvasculature [41,51-52] and can be utilized in bubble sizing [53], among other applications.\\ 
Esche [54] was the first to characterize the SH bubble behavior through experimental observations in 1952. Pioneering theoretical works of Eller [55] and Prosperetti [56-59] showed through a weakly non-linear analysis of the Rayleigh-Plesset model [60] that the subharmonic behavior of bubbles can only exist if the driving pressure amplitude exceeds a threshold pressure; it has been  predicted theoretically that the pressure threshold is minimum for sonications with twice the linear resonance frequency ($f_r$) of the bubbles.  This frequency is referred to as the linear $\frac{1}{2}$ order SH resonance frequency ($f_{sh}$=2$f_r$). Recently, several experimental [61-63], numerical [64,65]  and theoretical [59,62] studies have investigated the pressure threshold of SH generation in bubble oscillations.  These works have shown that SH oscillations in bubbles have three stages:  initiation, rapid growth and saturation.  Numerical Investigation of the SH threshold in uncoated and coated bubbles in [64,65];  have shown that for small bubbles (less than 500nm),  increased damping weakens the bubble response at $f_{sh}$. This leads to a shift of the minimum SH pressure threshold from $f_{sh}$ towards $f_r$.\\ 
We have recently studied the bifurcation structure of the bubbles excited with their $f_r$ and $f_{sh}$ [66]. We have  shown that for uncoated bubbles in water, non-destructive ($\frac{R}{R_0}<$2 [67]) stable SH oscillations are less likely to develop if the bubble is sonicated with $f_r$.  This is because when $f=f_r$, SHs oscillations only developed for $\frac{R}{R_0}>$2.  When the bubble is sonicated with $f_{sh}$, PD occurs through a bow-tie shape bifurcation and at very gentle oscillation regimes ($\frac{R}{R_0}<1.05$). This suggest that bubbles are more likely to sustain stable SHs at $f_{sh}$ [66]. The generation of PD in bifurcation diagrams was concomitant with the initiation of SHs which rapidly grow with increased pressure and get saturated. In other words, there is an upper limit for the achievable SHs and UHs strength and acoustic pressure increase will not necessarily result in SHs increase. To the contrary, a pressure increase can result in chaotic oscillations and/or bubble destruction which will lead to subsequent decrease in SHs and UHs  strength [50,66]. 
Despite several studies investigating the SH threshold of the bubbles [56-59,61-65], conditions for the enhancement of the upper limit of non-destructive SHs and UHs oscillations are under-investigated.  Furthermore, the bifurcation structure of the bubble oscillator in the regime of $\frac{1}{2}$ order SHs when $f_r<f<\approxeq f_{sh}$ is not studied in detail. Due to the importance of the SH and UH oscillations of bubbles, comprehensive knowledge of the resonant period 2 (P2) oscillations and conditions to enhance non-destructive P2 oscillations can aid in understanding and optimizing bubble applications.\\
In this study, which follows upon  our previous work [26] where pressure dependent resonance ($PDf_r$) was used to increase the non-destructive scattered pressure ($P_{sc}$) by bubbles, we investigate the pressure dependence of SH resonance ($Pdf_{sh}$ ). Through numerically simulating the resonance curves of bubbles at different pressures, linear ($f_{sh}$) and pressure dependent ($Pdf_{sh}$ ) SH resonance frequencies of bubble diameters ranging from 400nm up to 8 microns are calculated ( 8 microns can be considered the upper limit of size used in biomedical applications [19]). The bifurcation structure of the bubble oscillations as a function of pressure is studied when $f=Pdf_{sh}$. The corresponding SH and UH strength of the $P_{sc}$ are calculated and studied in conjunction with the bifurcation diagrams. We show that $Pdf_{sh}$  decreases as pressure increases. Sonication of the bubble with $Pdf_{sh}$  results in a saddle node bifurcation from a period 1 (P1) or a P2 oscillation regime to a P2 oscillation of higher amplitude. This is concomitant by the oversaturation of the SHs and UHs strength of up to $\approxeq$4 and $\approxeq$7 dB. Additionally for each bubble size there is an optimum frequency between $1.5f_r-1.8f_r$which results in the maximum SH scattering cross section.
\section{Methods}
Similar to [66], we have chosen the uncoated bubble as the bubble oscillator of interest. Effect of coating is neglected as addition of the encapsulation introduces more complexity to the bubble dynamics. In order to have a detailed understanding on the dynamics of the bubble its preferable to separate the nonlinear effects of coating from that of the bubble itself. This makes it easier to understand the dynamics of the system and understand the basis of the bubble behavior. Furthermore, in future studies where coating is involved, information of the system behavior in the absence of coating will allow for a more straightforward understanding of the complex effects of coating on the system behavior. 
\subsection{The Bubble model}
The radial oscillations of the bubbles are numerically simulated by solving the Keller-Miksis equation [68]:
\justifying
\begin{equation}
\rho[(1-\frac{\dot{R}}{c})R\ddot{R}+\frac{3}{2}\dot{R}^2(1-\frac{\dot{R}}{3c})]=(1+\frac{\dot{R}}{c})(G)+\frac{R}{c}\frac{d}{dt}(G)
\end{equation}
where $G=P_g-\frac{4\mu_L\dot{R}}{R}-\frac{2\sigma}{R}-P_0-P_A sin(2 \pi f t)$. $P_g$ is the gas pressure in the bubble and is given by $P_g=(P_0+\frac{2\sigma}{R})*(\frac{R}{R_0})^{3\gamma}$\\
In this equation, R is radius at time t, $R_0$ is the initial bubble radius, $\dot{R}$ is the wall velocity of the bubble and $\ddot{R}$ is the wall acceleration	$\rho{}$ is the liquid density (998 $\frac{kg}{m^3}$), c is the sound speed (1481 m/s), $P_g$ is the gas pressure, $\sigma{}$ is the surface tension (0.0725 $\frac{N}{m}$), $\mu{}$ is the liquid viscosity (0.001 Pa.s), $P_A=$ and \textit{f} are the amplitude and frequency of the applied acoustic pressure. The values in the parentheses are for water at 293$^0$K. In this paper the gas inside the bubble is Air ($\gamma$=1.4) and water is the host media.\\
\begin{figure*}
	\begin{center}
		\scalebox{0.63}{\includegraphics{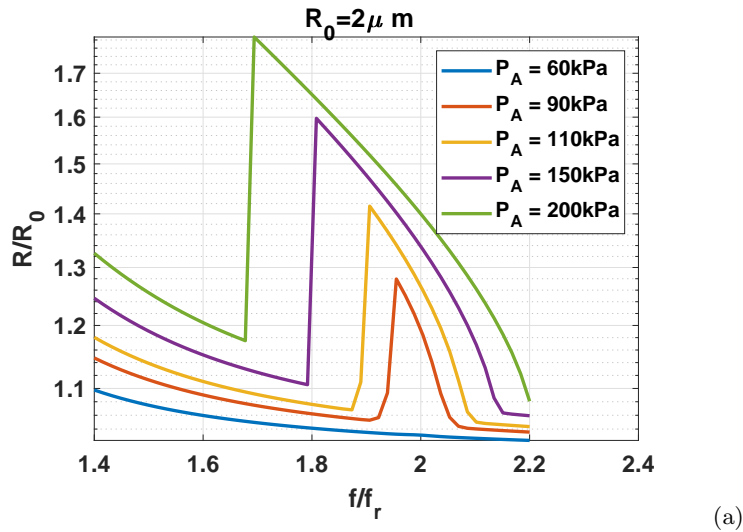}} 
		 (a) 
			\caption{SH resonance frequency of a bubble with $R_0=2 \mu m$ at different pressure amplitudes.}
	\end{center}
\end{figure*}
\subsection{Resonance curves}
It is shown that above a pressure threshold SH oscillations are generated and the threshold is minimum at $f_{sh}$ ($f=2f_r$). The radial oscillations of free uncoated air bubbles ($R_0$=200nm-4 $\mu m$) were numerically simulated by solving equation 1, for  $f_r<f<2.2f_r$ and for a range of pressure amplitudes. The resonance curves were only calculated for pressure amplitude and frequency combinations that result in non-destructive bubble oscillations $\frac{R}{R_0}<2$ [26,67]. The maximum bubble radius was calculated in the last 20 cycles of a 200 cycle driving pulse to ensure that there was no transient behavior (similar analysis to [66]) and then plotted as a function of frequency in each graph. At each pressure,$f_{sh}$ and $Pdf_{sh}$ were selected from the graphs and used to generate the bifurcation diagrams in the next step.
\begin{figure*}
	\begin{center}
		\scalebox{0.43}{\includegraphics{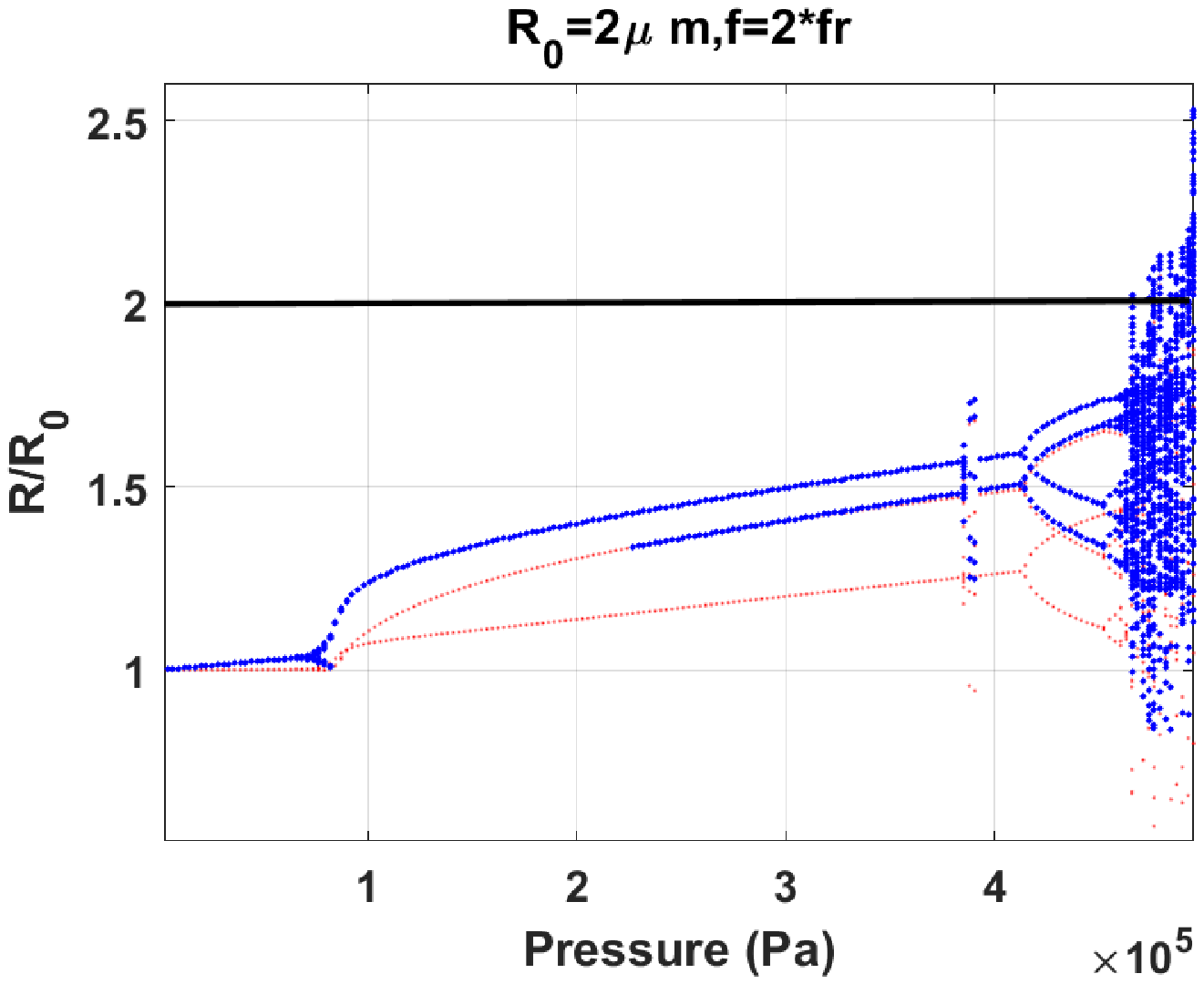}} \scalebox{0.43}{\includegraphics{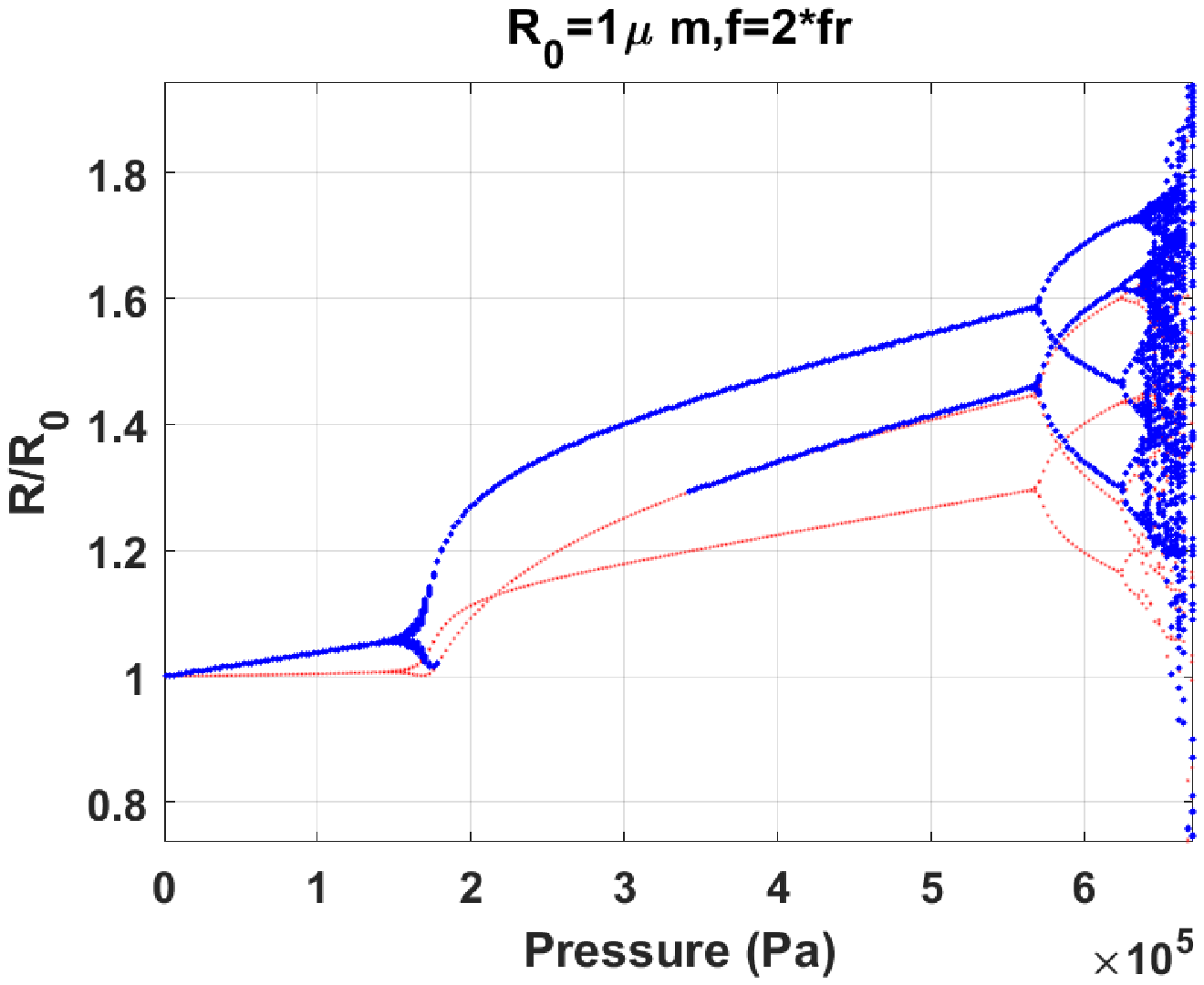}}\\
		\hspace{0.5cm} (a) \hspace{6cm} (b)\\
		\scalebox{0.43}{\includegraphics{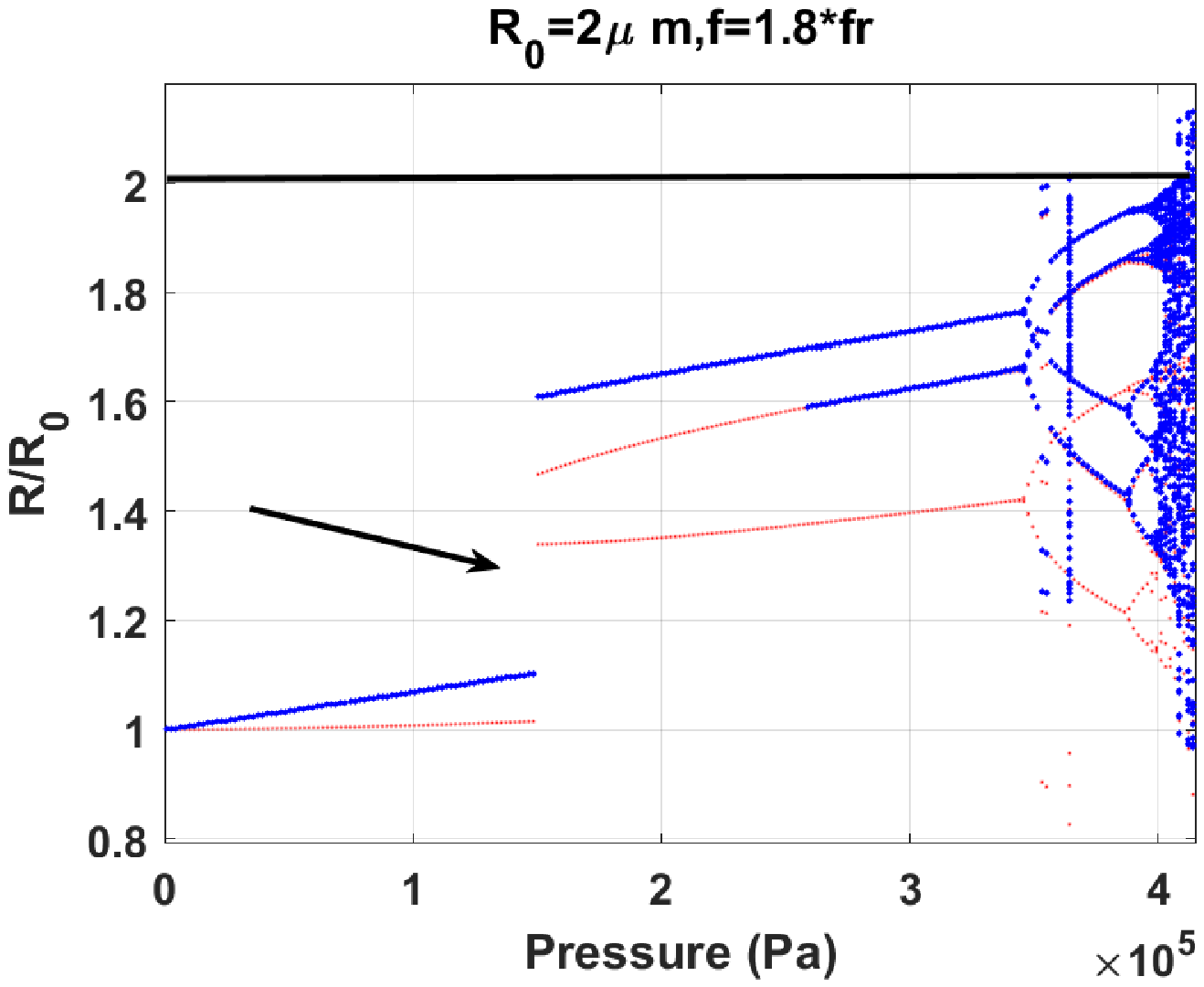}} \scalebox{0.43}{\includegraphics{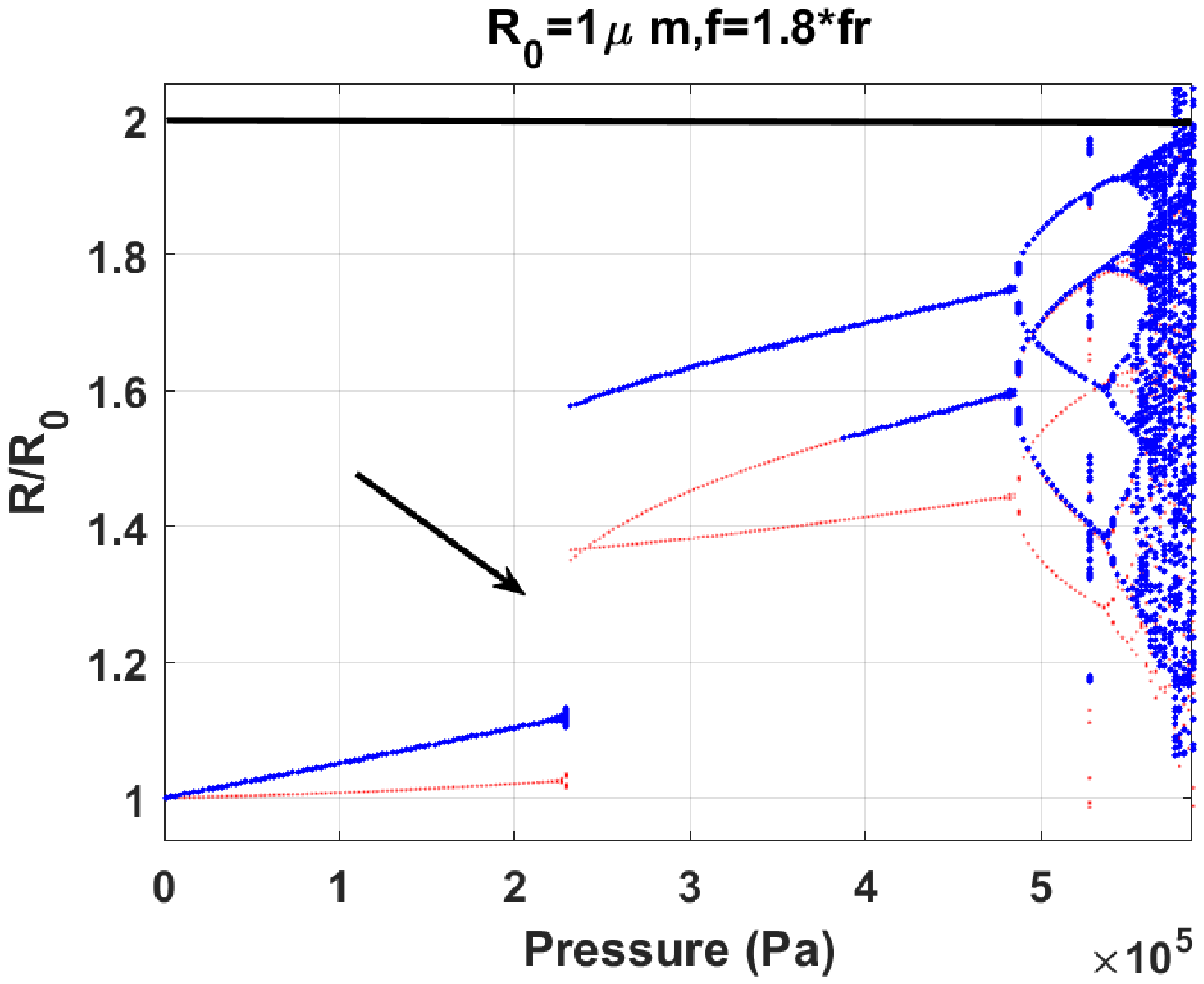}}\\
		\hspace{0.5cm} (c) \hspace{6cm} (d)\\
		\scalebox{0.43}{\includegraphics{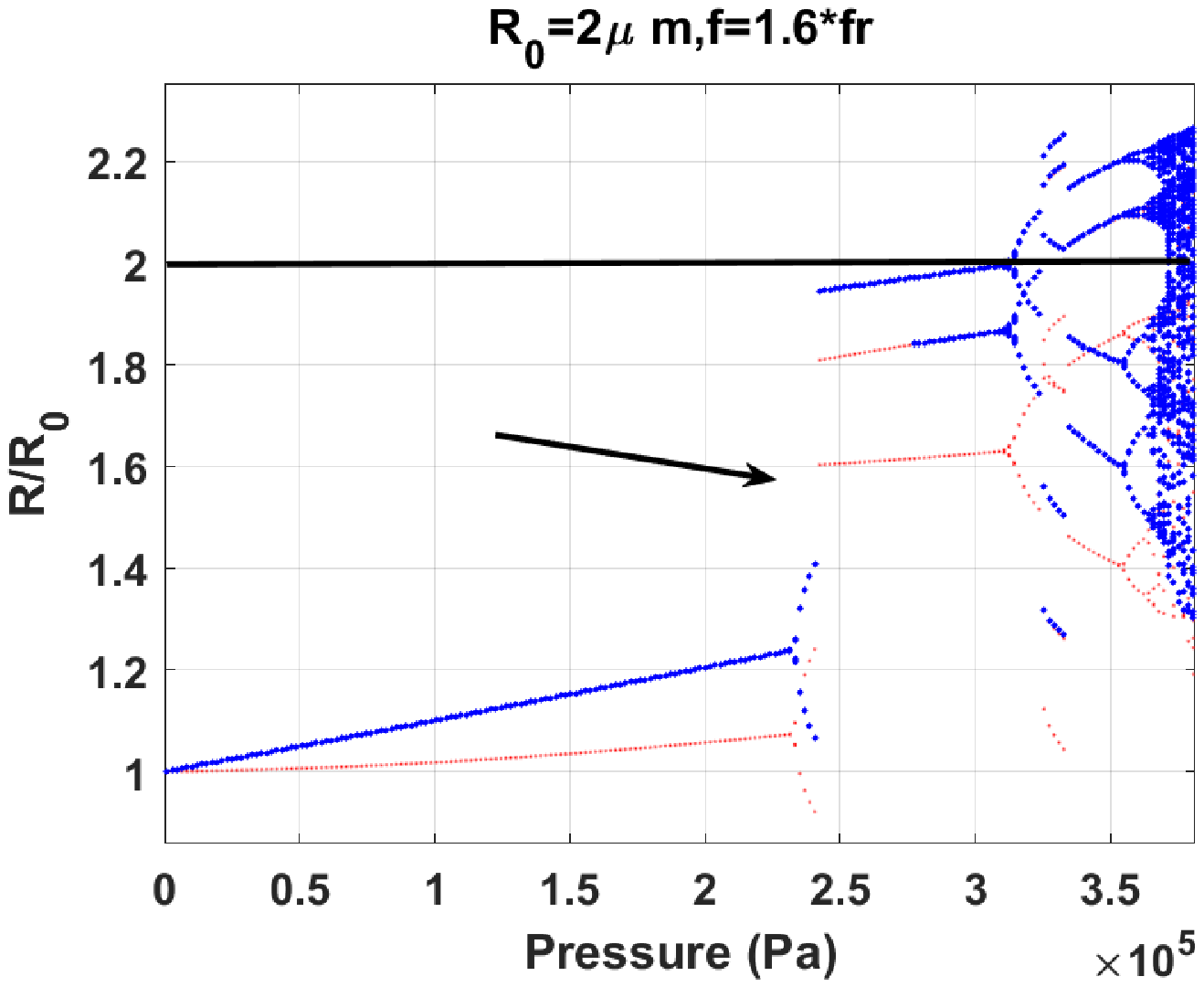}} \scalebox{0.43}{\includegraphics{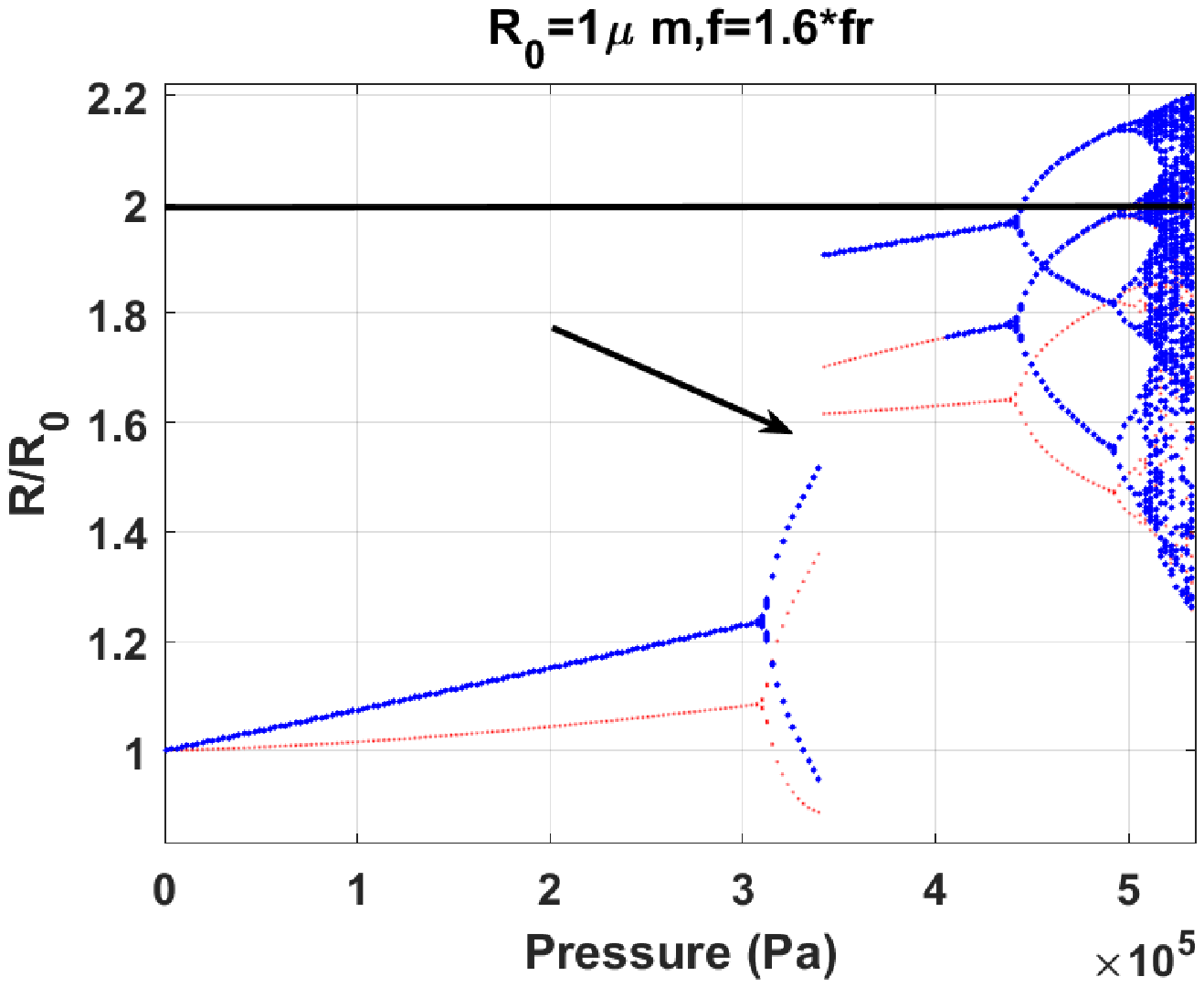}}\\
		\hspace{0.5cm} (e) \hspace{6cm} (f)\\
		\scalebox{0.43}{\includegraphics{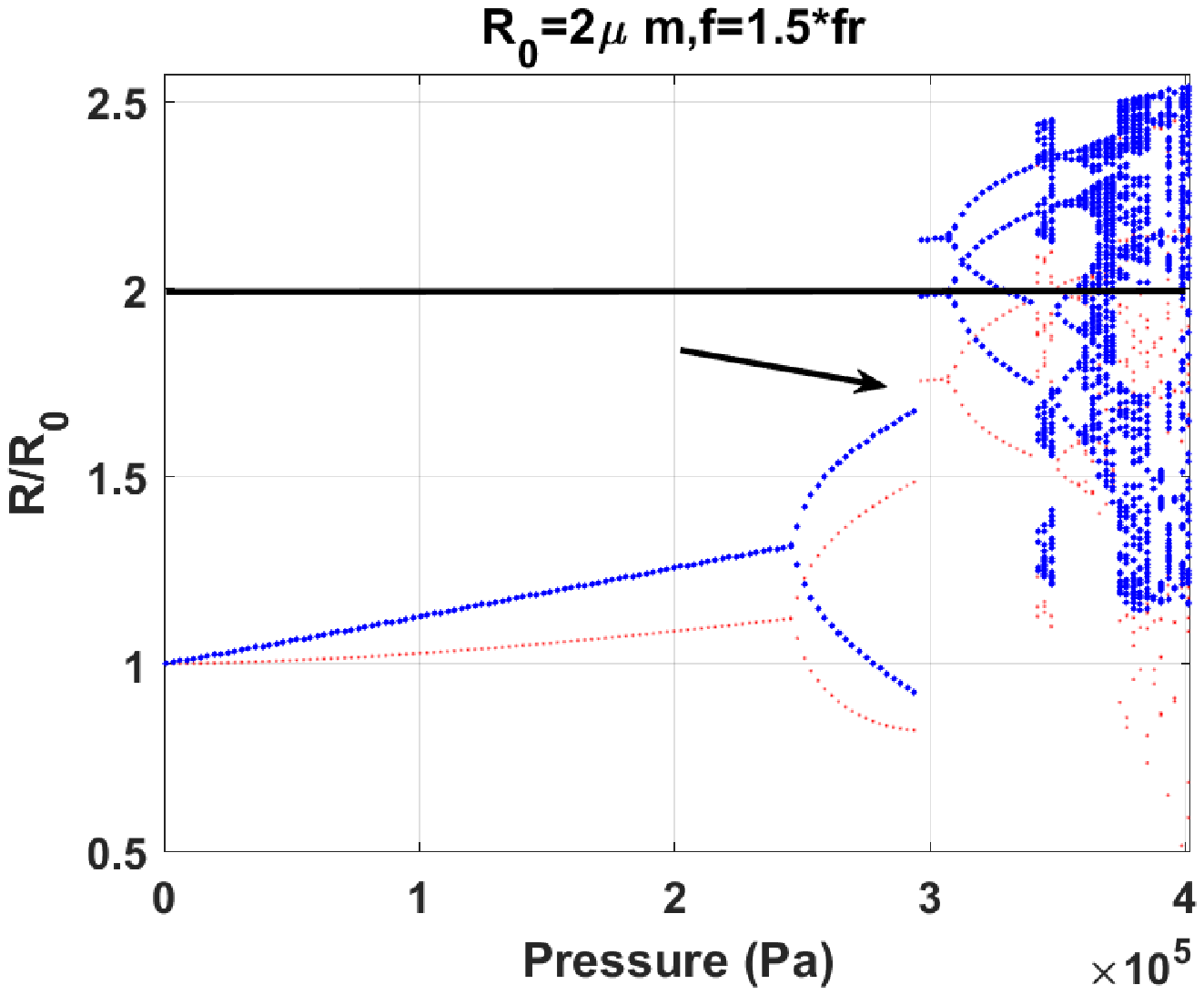}} \scalebox{0.43}{\includegraphics{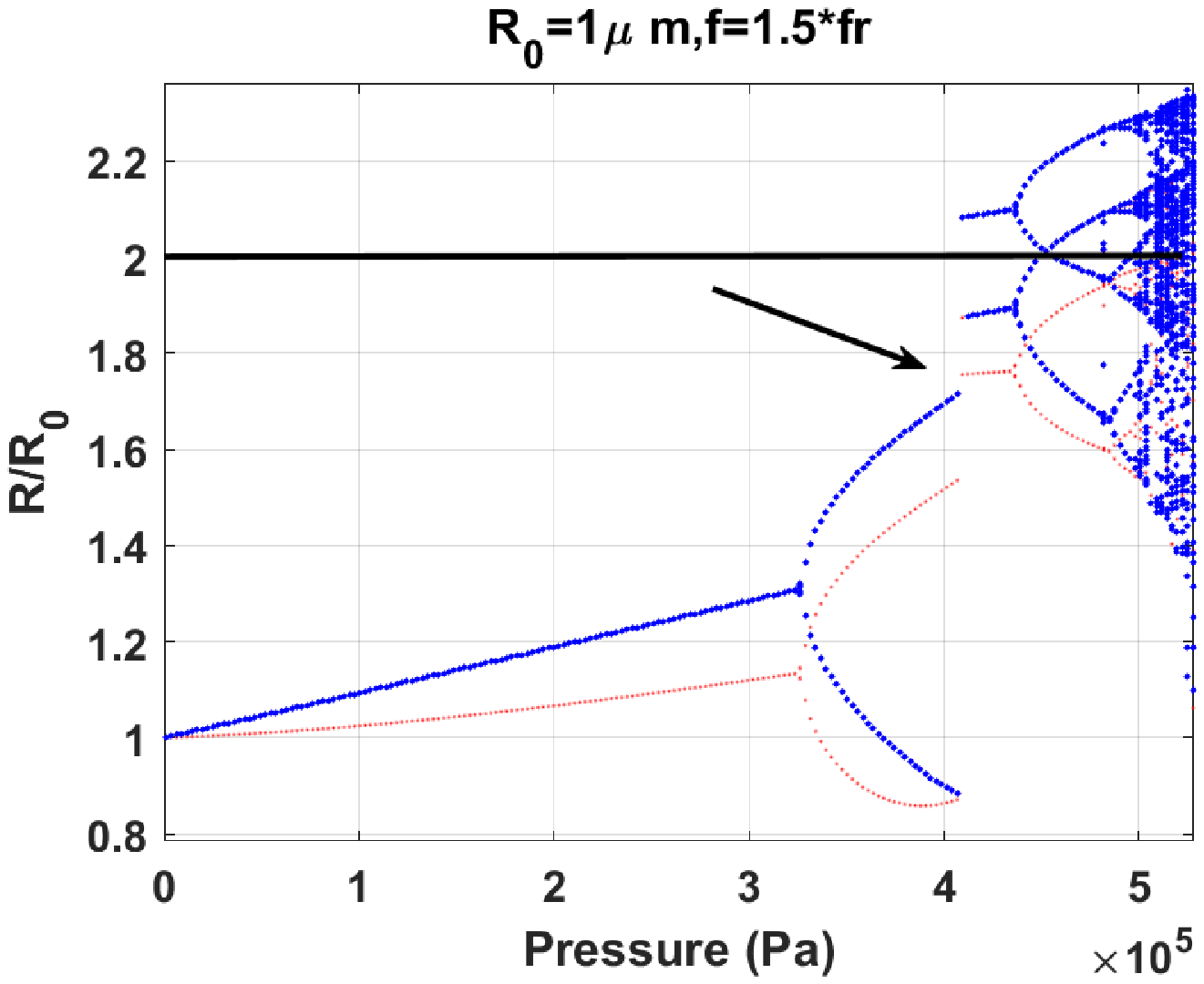}}\\
		\hspace{0.5cm} (e) \hspace{6cm} (f)\\
		\caption{Bifurcation structure (blue: method of peaks, red: conventional method) of the micron-size bubbles as a function of pressure when sonicated with $f_{sh}$ and $Pdf_{sh}$. Left column is for $R_0=2 \mu m$ and Right column is for $R_0=1 \mu m$ (arrow shows the pressure responsible for SN bifurcation)}
	\end{center}
\end{figure*}
\subsection{Bifurcation diagrams}
Bifurcation diagrams are valuable tools to analyze the dynamics of nonlinear systems where the qualitative and quantitative changes of the dynamics of the system can be investigated effectively over a wide range of the control parameters. In this paper, we employ a more comprehensive bifurcation analysis method introduced in [74,75].\\
\subsubsection{Conventional bifurcation analysis}
When dealing with systems responding to a driving force, to generate the points in the bifurcation diagrams vs. the control parameter, one option is to sample the R(t) curves using a specific point in each driving period. The approach can be summarized in:
\begin{equation}
P \equiv (R(\Theta))\{(R(t),  \dot{R}(t) ):\Theta= \frac{n}{f} \}\hspace{0.1cm} where \hspace{0.1cm} n=400,401...440
\end{equation}
Where $P$ denotes the points in the bifurcation diagram, $R$ and $\dot{R}$
are the time dependentradius and wall velocity of the bubble at a given
set of control parameters of ($R_{0}$, $P_{0}$, $P_{A}$, $c$, $k$, $\mu$,
$\sigma$, $f$) and $\Theta$ is given by $\frac{n}{f}$.  Points on the bifurcation diagram are constructed by plotting the solution of $R(t)$ at time points that are multiples of the driving acoustic period. The results are plotted for $n=400-440$ to ensure a steady state solution has been reached for all bubbles. Due to smaller viscous effects, bigger bubbles require longer cycles to reach steady state.\\
\subsubsection{Method of peaks}
As a more general method, bifurcation points can be constructed by setting one of the phase space coordinates to zero:      
\begin{equation}
Q \equiv max(R)\{(R, \dot{R} ):\dot{R}= 0\}
\end{equation}
In this method, the steady state solution of the radial oscillations for each control parameter is considered. The maxima of the radial peaks ($\dot{R}=0$) are identified (determined within 400-440 cycles of the stable oscillations) and are plotted versus the given control parameter in the bifurcation diagrams. 
The bifurcation diagrams of the normalized bubble oscillations ($\frac{R}{R_0}$) are calculated using both methods a) and b). When the two results are plotted alongside each other, it is easier to uncover more important details about the SuH and UH oscillations, as well as the SH and chaotic oscillations.\\
\subsection{Backscattered pressure}
Oscillations of a bubble generates a scattered pressure ($P_{sc}$) which can be calculated by [71]:
\begin{equation}
P_{sc}=\rho\frac{R}{d}(R\ddot{R}+2\dot{R}^2)
\end{equation}
where $d$ is the distance from the center of the bubble (and for simplicity is considered as 1m in this paper) [26]. 
Equation 1 is solved using the 4th order Runge-Kutta technique using the ode45 function in Matlab (this function also has a 5th order estimation); the control parameters of interest are $R_0$, $f$ and $P_A$.  The resulting radial bubble oscillations are visualized using resonance curves and  bifurcations diagrams. Bifurcation diagrams of the normalized bubble oscillations $\frac{R}{R_0}$ are presented as a function of the driving pressure in conjunction with the SH and UH amplitude of the $P_{sc}$. The scattered pressure ($P_{sc}$) is calculated alongside the bifurcation structure only for pressures that result in non-destructive oscillations ($\frac{R}{R_0}<2$) [26,67]. SH and UH amplitude of the $P_{sc}$ is plotted alongside each bifurcation diagram to highlight the effect of nonlinearities on the changes in the SH and UH strength.
\section{Results}
\subsection{Pressure dependent SH resonance frequency (PDSH)}
First we explored the bubble expansion ratio ($\frac{R}{R_0}$) as a function of peak excitation pressure for a range of frequencies between 1.4$f_r$- 2$f_r$, where $f_r$ is the linear resonance frequency. It has been hypothesized that a local maximum in the expansion ratio would be observed at $2f_r$, which would represent the subharmonic response of the bubble.  However, the maximum response shifted to lower frequencies as the excitation pressure was increased (Fig. 1). Theoretical studies have reported that the resonance frequency, which equates to $f_r$ only at very low excitation pressures (i.e. $< 50 kPa$) for a microbubble is inversely related with excitation pressure.  We postulate this  can explain the shift in the subharmonic response to lower frequencies relative to $2f_r$. Figure 1 shows the SH resonance frequency of a $R_0$=2$\mu m$ bubble sonicated with different pressure amplitudes. The linear SH resonance frequency is generated at ~60 kPa and f=$2f_r$. As the pressure increases, similar to the case of pressure dependent resonance [26], SH resonance frequency decreases. For example the SH resonance frequency is $\approxeq 2f_r$ at 60 kPa and is 1.7$f_r$ at 200 kPa. We call this shifted SH resonance frequency “pressure dependent SH resonance frequency ($Pdf_{sh}$)”.  In the next section, we will show the mechanism of SH enhancement when bubbles of different sizes are sonicated with their $Pdf_{sh}$.
\subsection{Bifurcation structure of the micron size bubbles (SH enhancement region)}
Figures 2a and 2b show the bifurcation structure of the $R_0$=2$\mu m$ and $R_0$=1$\mu m$ bubbles as a function of pressure when f=$2f_r$. The radial oscillations of the bubbles undergo period doubling at the lowest pressure threshold (60 kPa for $R_0$=2 $\mu m$), which evolve in the form of a bowtie to P2 oscillations of higher amplitude as the acoustic pressure increases. The oscillations undergo further period doubling before the appearance of chaos. In this case full amplitude ($\frac{R}{R_0}$=2) P2 non-destructive oscillations do not develop.\\
Figures 2c and 2d show the bifurcation structure of the $R_0$=2 $\mu m$ and $R_0$=1 $\mu m$ bubbles as a function of pressure when $f=1.8f_r$. The radial oscillations undergo a saddle node bifurcation from P1 to P2 oscillations of higher amplitude. The P2 oscillations have one maximum (red curve), above a second pressure threshold the second maximum appears (oscillations become P2 with two maxima); the maxima are  exactly on top of one of the branches of the conventional method which implies the wall velocity is in phase with the driving signal once every two acoustic cycles. Compared to the case of $f=f_{sh}$, bubbles sonicated by their $Pdf_{sh}$ have a higher pressure threshold for P2 oscillations; however, the amplitude of the P2 oscillations are higher. The oscillations undergo further period doubling and chaos eventually occurs. Similar to $f=2f_r$, when $f=1.8f_r$ full amplitude ($\frac{R}{R_0}$=2) P2 non-destructive oscillations did not develop.\\
Figures 2c and 2d show the bifurcation structure of the $R_0$=2 $\mu m$ and $R_0$=1 $\mu m$ size bubbles as a function of pressure when $f=1.6f_r$. The radial oscillations undergo a saddle node bifurcation from P2 (with two maxima) oscillations to P2 oscillations of higher amplitude (with one maxima). Compared to the case of f=$f_{sh}$, and f=1.8$f_r$ bubbles sonicated by their $Pdf_{sh}$=1.6$f_r$ have a higher pressure threshold (Pt) for P2 oscillations (e.g. for $R_0$=2$\mu m$ Pt$\approxeq$ 230 kPa). The amplitude of the P2 oscillations are higher than the previous cases at the occurrence of SN bifurcation (e.g. for $R_0$=2$\mu m$ P$\approxeq$ 242 kPa the $\frac{R_{max}}{R_0}$=1.94).b Similar to the previous cases, as pressure increases a second maximum re-emerges in the blue curve with its value being the same as one of the branches of the red curve. P2 oscillations then grow by pressure increase and P2 oscillations reach a large amplitude that result in non-destructive oscillations (e.g. for $R_0$=2$\mu m$ R=1.99 $R_0$at $P_A=$309 kPa).\\ 
Figures 2g and 2h show the bifurcation structure of the $R_0$=2 $\mu m$ and $R_0$=1 $\mu m$ bubbles as a function of pressure when $f=1.5f_r$. PD initiation is at the highest pressure threshold (e.g. for $R_0$=2$\mu m$ Pd occurs at 245 kPa). Above a second pressure threshold (e.g. for $R_0$=2 $\mu m$ at $P_A=$295 kPa) P2 oscillations (with two maxima ) undergo a SN bifurcation to P2 oscillations (with two maxima) of higher amplitude (($\frac{R_{max}}{R_0}\approxeq2.13$). In this case occurrence of PD is concomitant with bubble destruction as $\frac{R}{R_0}>2$ for both bubbles.  

\begin{figure*}
	\begin{center}
		\scalebox{0.43}{\includegraphics{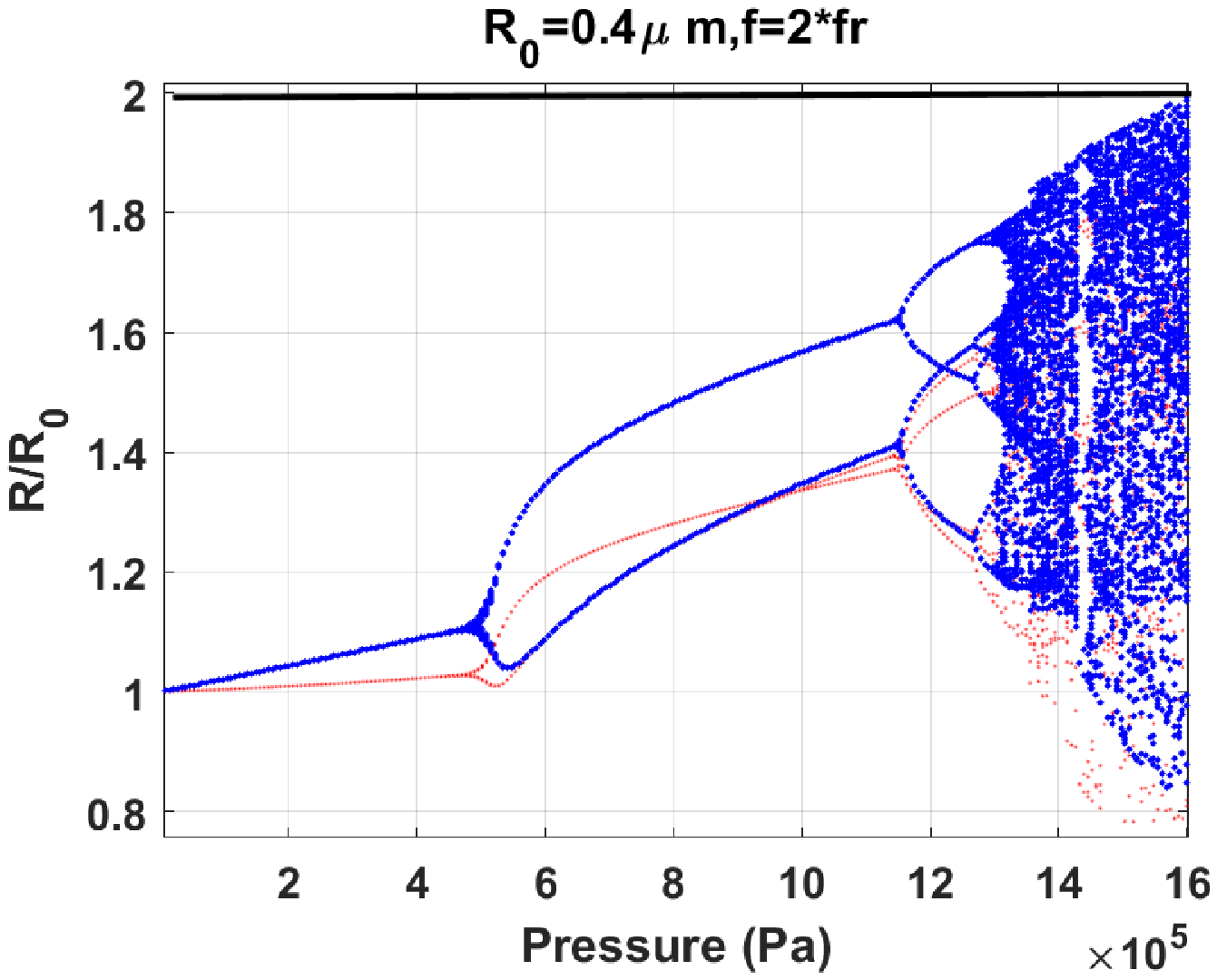}} \scalebox{0.43}{\includegraphics{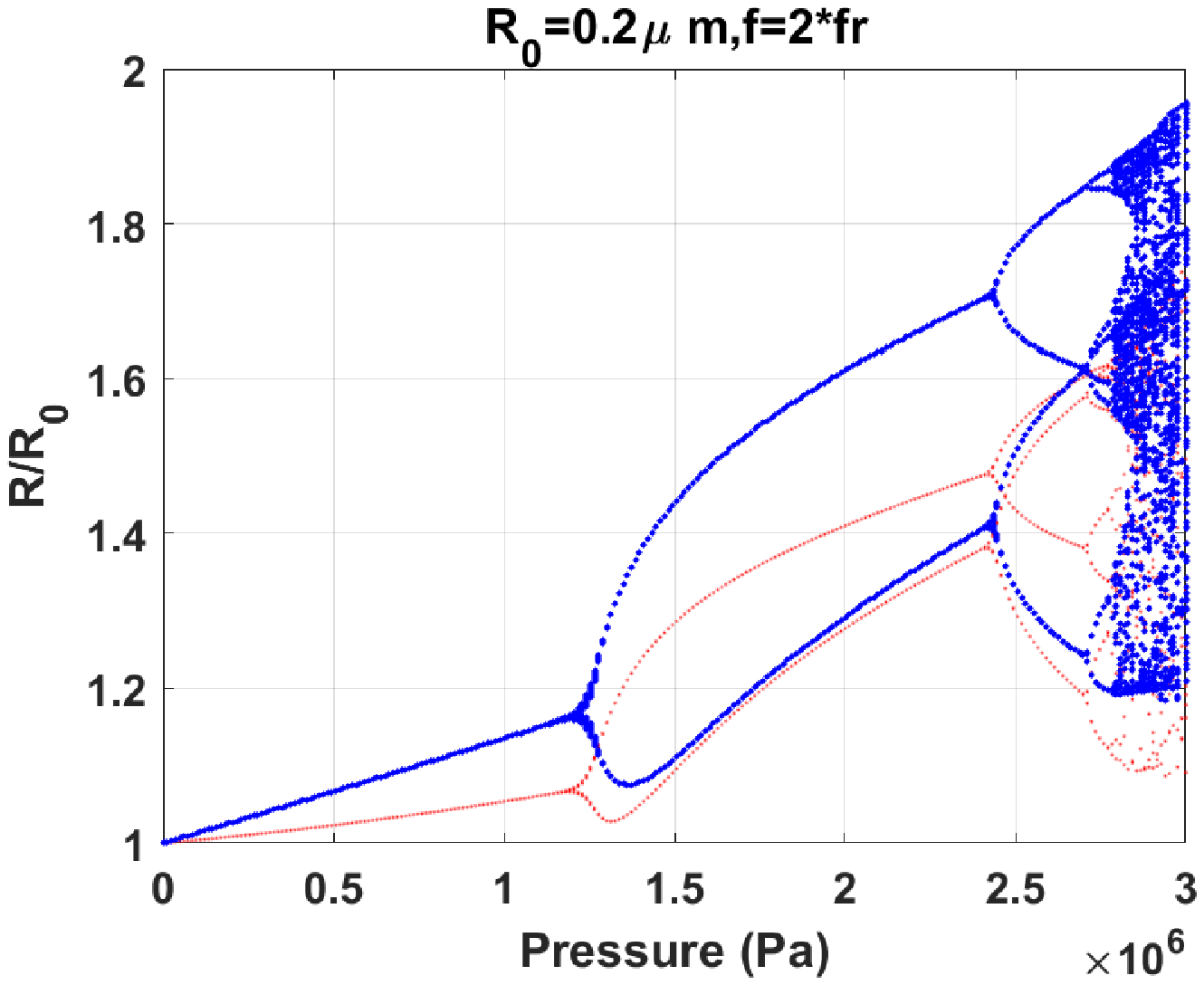}}\\
		\hspace{0.5cm} (a) \hspace{6cm} (b)\\
		\scalebox{0.43}{\includegraphics{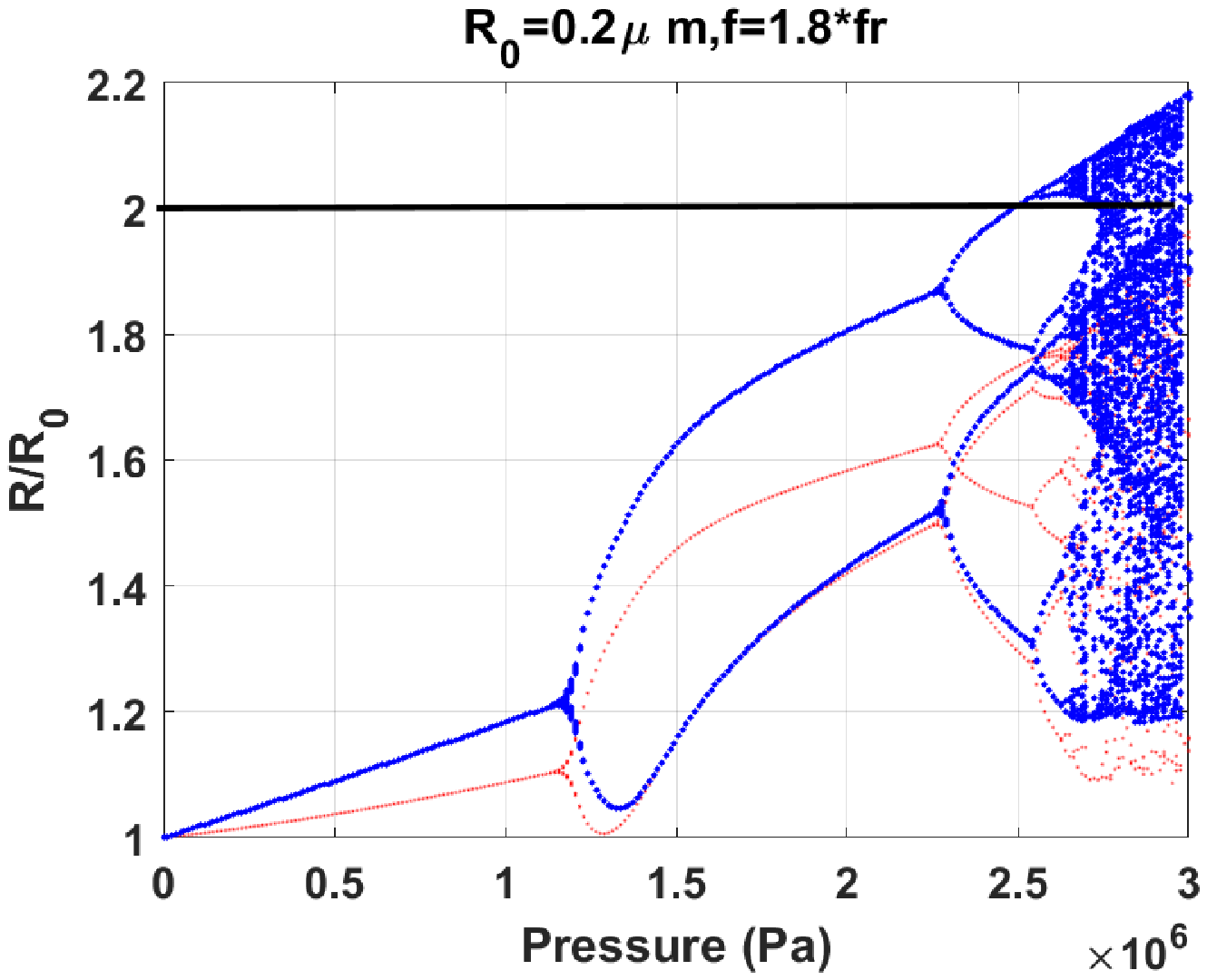}} \scalebox{0.43}{\includegraphics{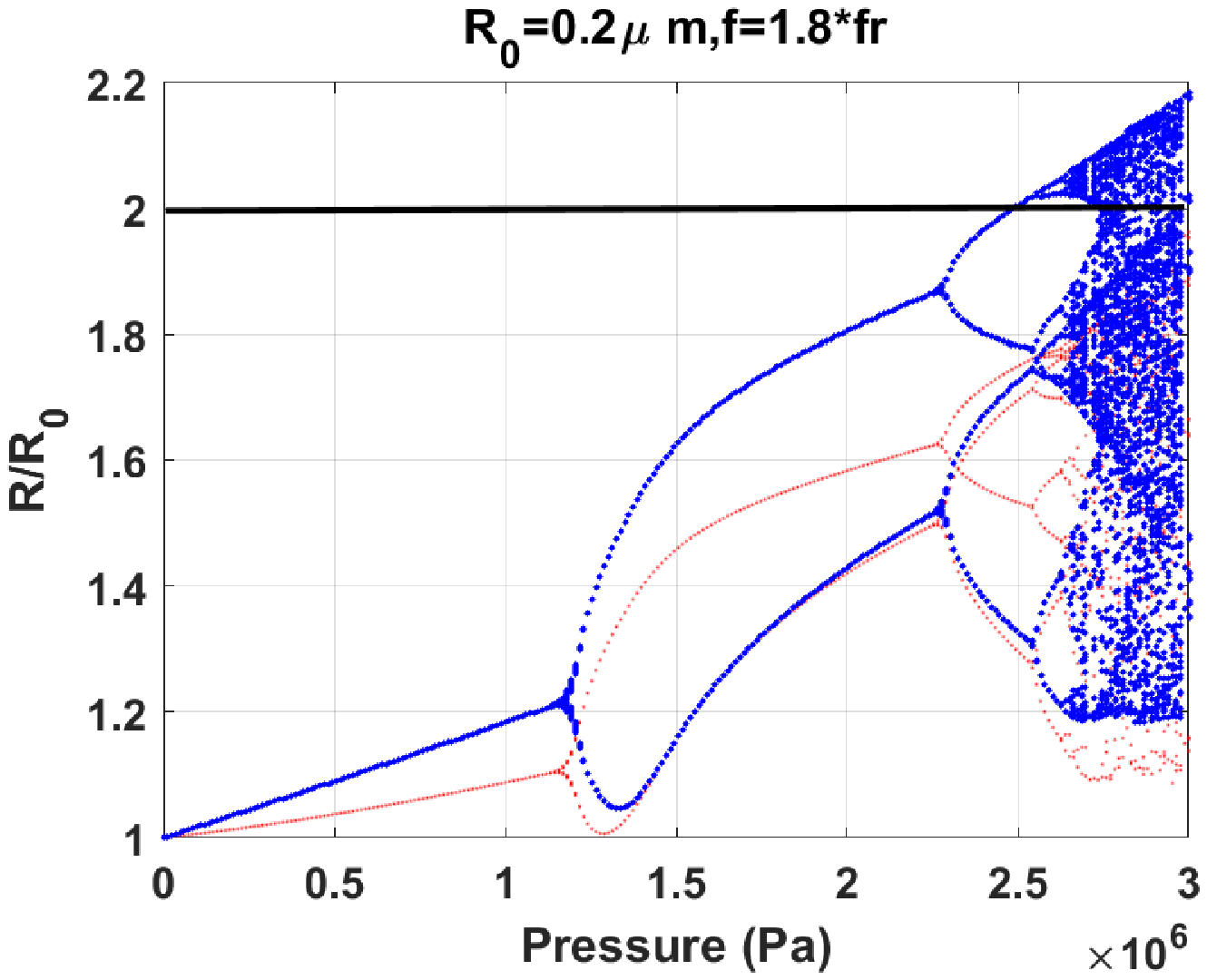}}\\
		\hspace{0.5cm} (c) \hspace{6cm} (d)\\
		\scalebox{0.43}{\includegraphics{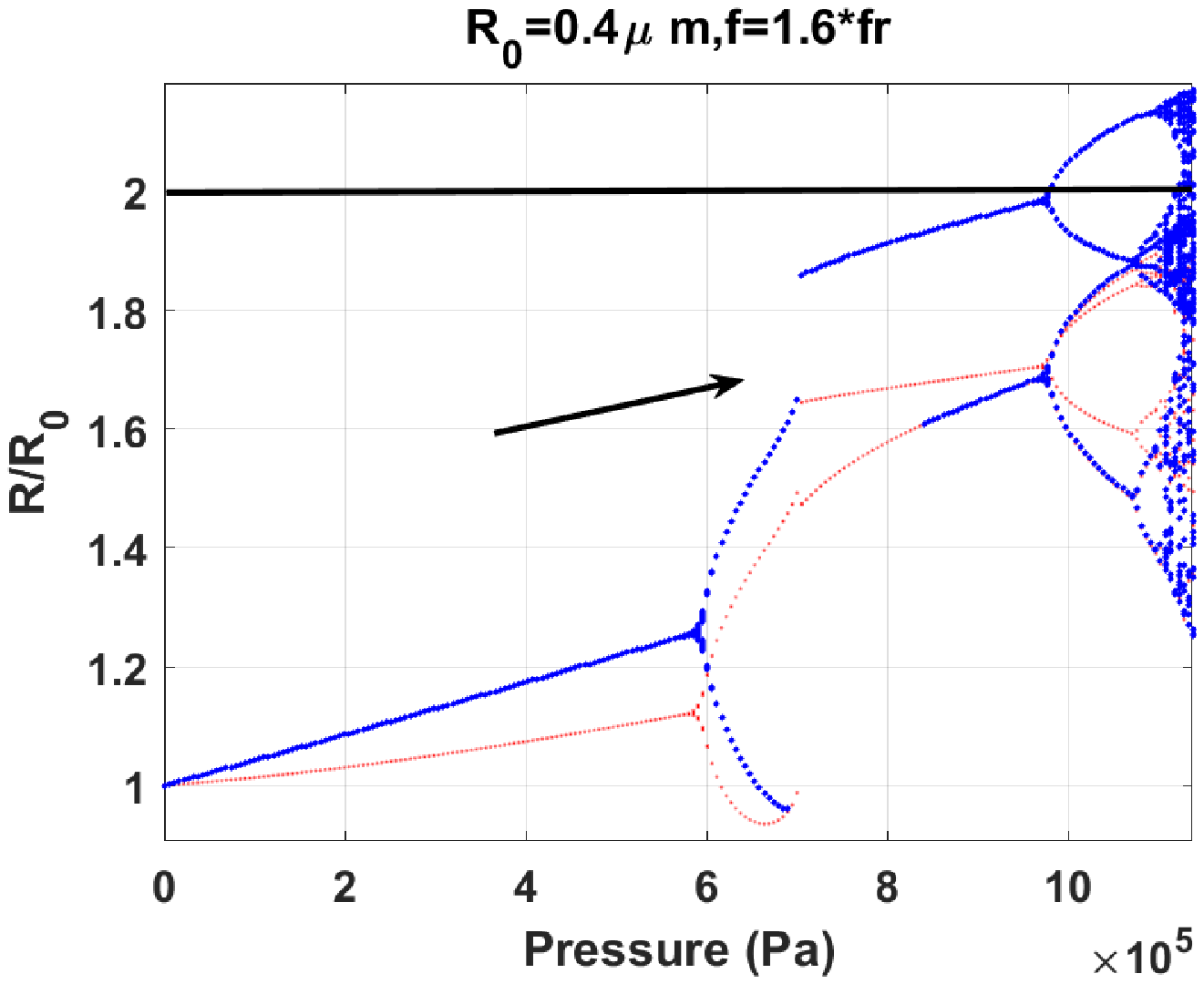}} \scalebox{0.43}{\includegraphics{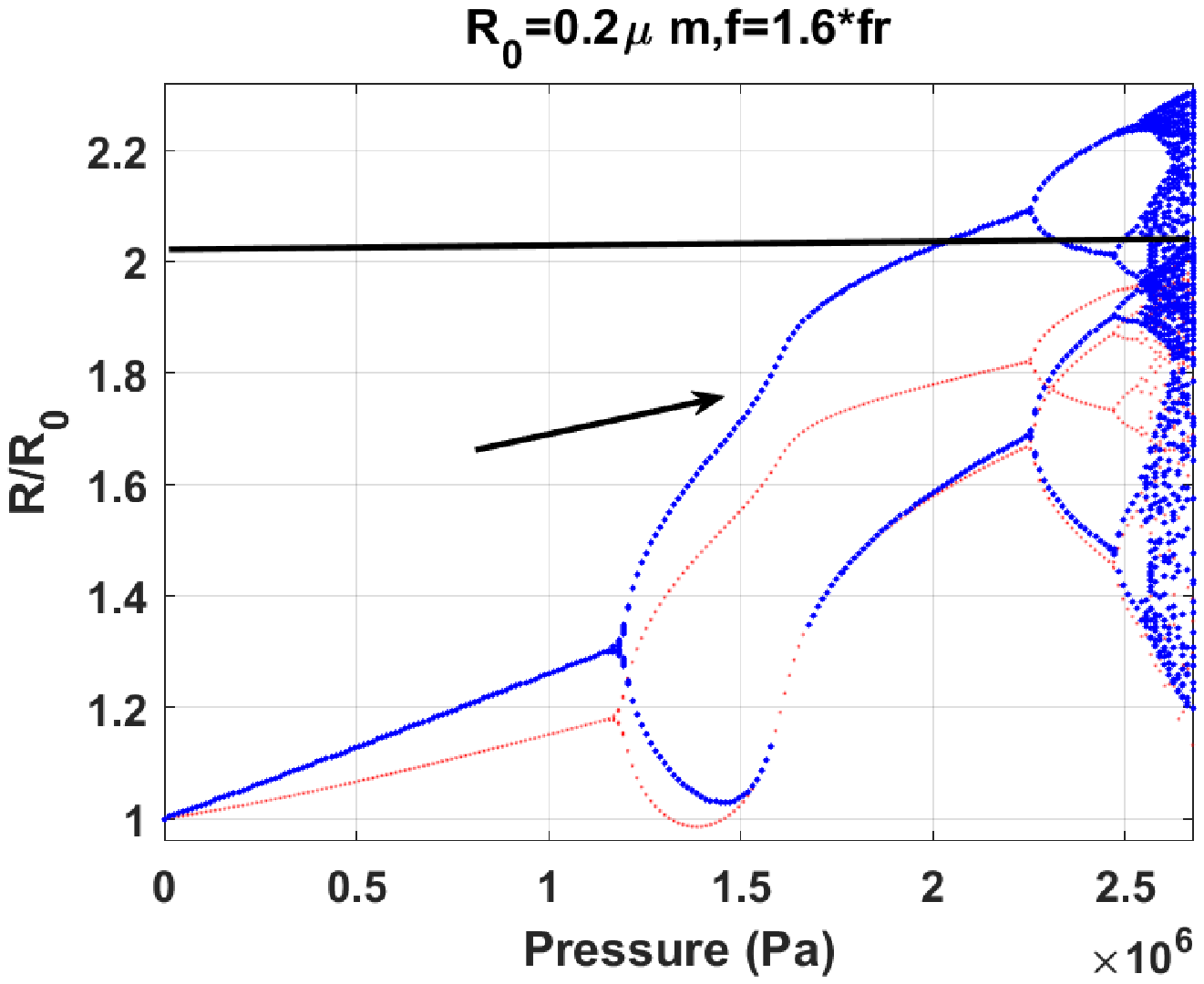}}\\
		\hspace{0.5cm} (e) \hspace{6cm} (f)\\
				\scalebox{0.43}{\includegraphics{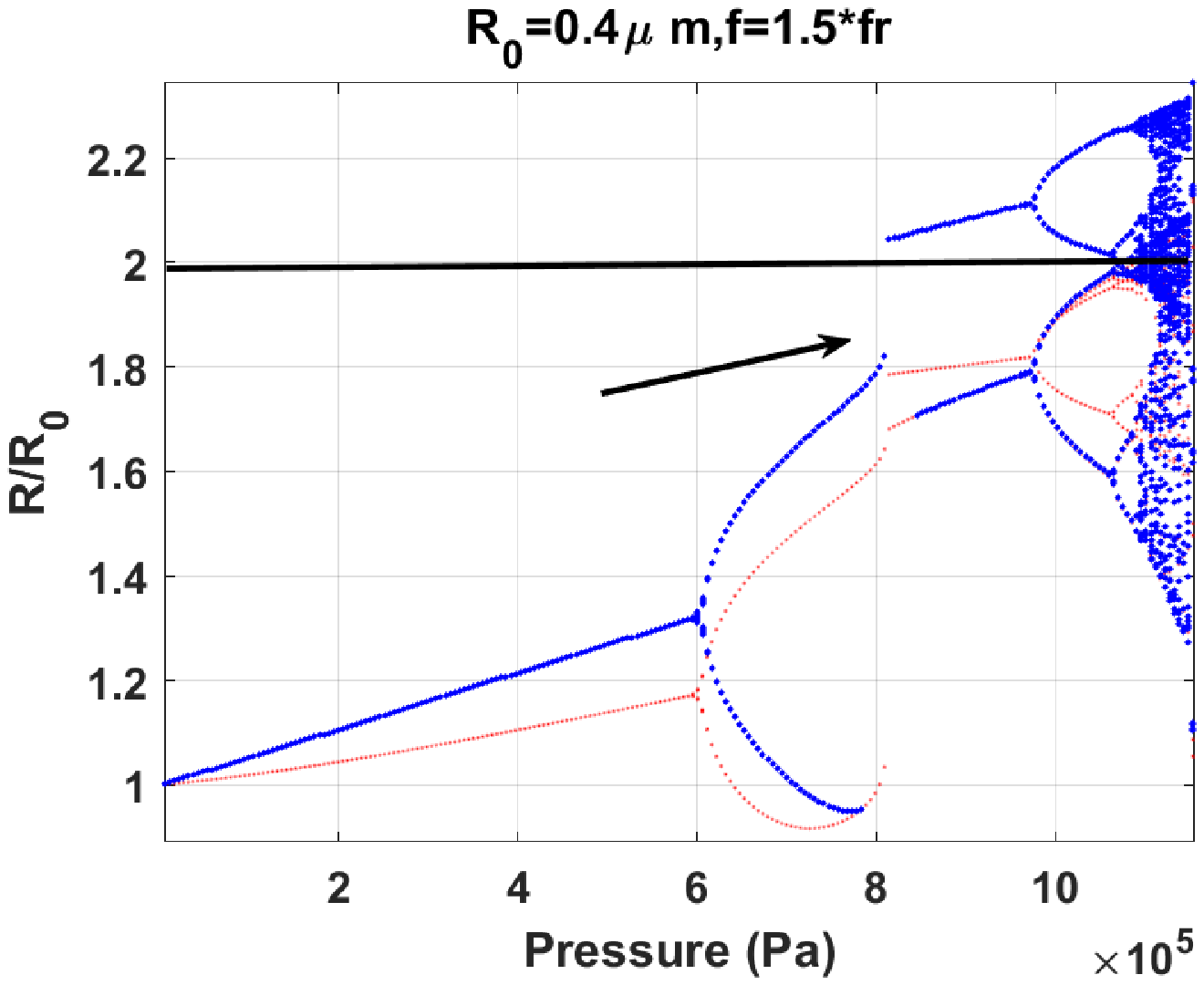}} \scalebox{0.43}{\includegraphics{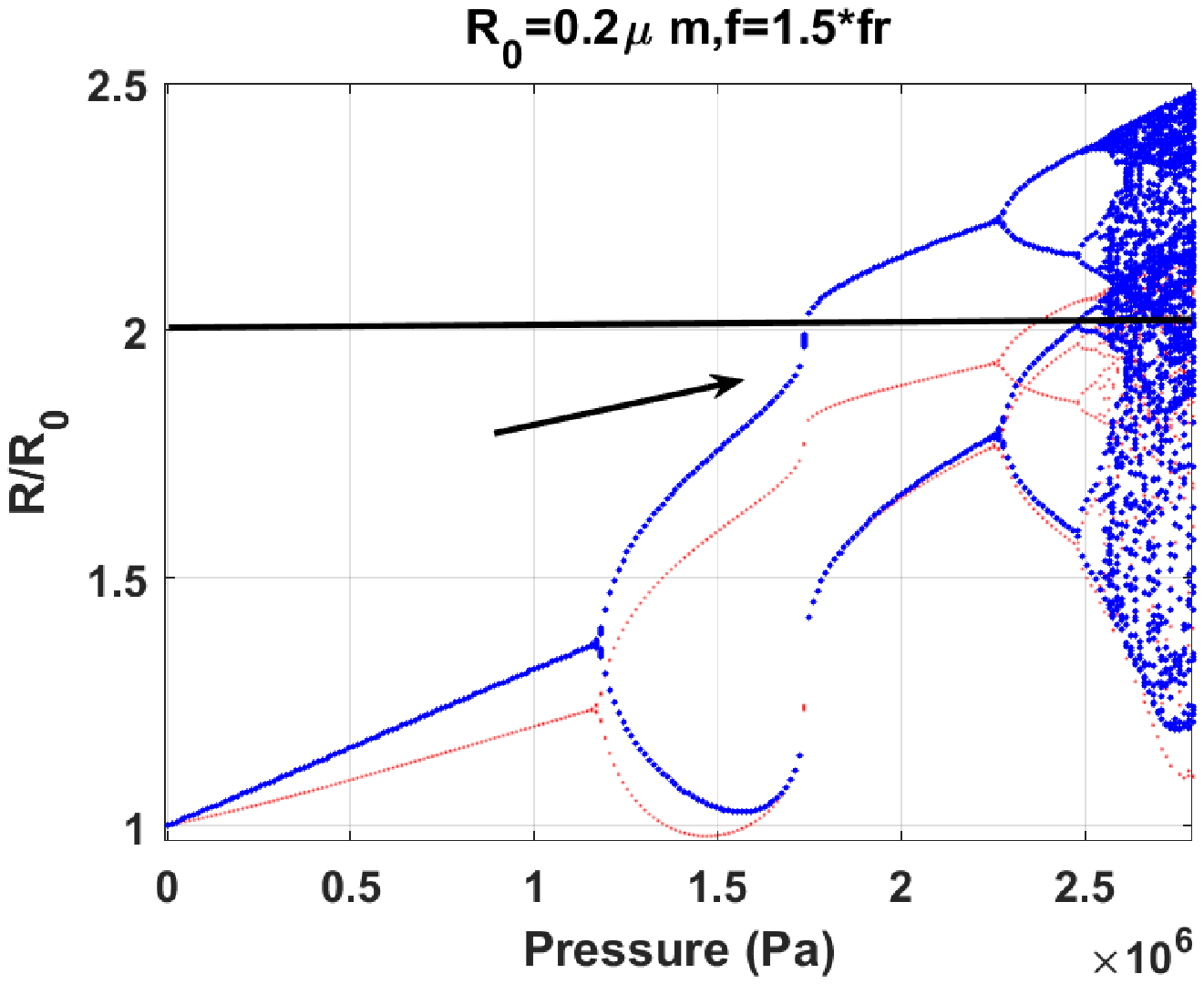}}\\
		\hspace{0.5cm} (e) \hspace{6cm} (f)\\
		\caption{Bifurcation structure (blue: method of peaks, red: conventional method) of the nano-size bubbles as a function of pressure when sonicated with $f_{sh}$ and $Pdf_{sh}$. Left column is for $R_0=0.4 \mu m$ and Right column is for $R_0=0.2 \mu m$ (arrow shows the pressure responsible for SN bifurcation)}
	\end{center}
\end{figure*}
\subsection{Bifurcation structure of the nano-bubbles (SH enhancement region)}
Figures 3a and 3b show the bifurcation structure of the bubbles with $R_=0= 400 nm$ $\&$ $R_0= 200 nm$ as a function of pressure when $f=2f_r$. The radial oscillations of the bubbles undergo period doubling at the lowest pressure threshold ($\approxeq$ 470 kPa for $R_0$=400 nm). P2 oscillations grow in amplitude as pressure increase and undergo further period doubling before the appearance of chaos. When $f=2f_r$ full amplitude ($\frac{R_{max}}{R0}$=2) non-destructive P2 oscillations do not develop.\\
Figures 3c and 3d show the bifurcation structure of $R_=0= 400 nm$ $\&$ $R_0= 200 nm$ bubbles as a function of pressure when $f=1.8f_r$. P2 oscillations undergo a sharp increase in amplitude (e.g. for $R_0$=400 nm at $P_A \approxeq$600 kPa). Compared to the case of $f=f_{sh}$, bubbles sonicated by their $Pdf_{sh}$  have a higher pressure threshold for P2 oscillations; however, the amplitude of the P2 oscillations are higher (similar for micron size bubbles in Fig 2c-d). The oscillations undergo further period doubling and chaos eventually occurs. Similar to $f=2f_r$, when $f=1.8f_r$, bubble doesn’t reach full amplitude ($\frac{R_{max}}{R0}$=2) P2 non-destructive oscillations (similar to the case of micron size bubbles). Unlike micron size bubbles, SN bifurcation is not observed; this is due to the stronger effect of liquid viscosity on smaller bubbles.\\
Figures 3e and 3f show the bifurcation structure of the $R_0$=400 nm and $R_0$=200 nm size bubbles as a function of pressure when $f=1.6f_r$. Radial oscillations undergo a saddle node bifurcation from P2 oscillations to P2 oscillations of higher amplitude. Compared to the case of $f=f_{sh}$, and $f=1.8f_r$, nano-bubbles sonicated by their $Pdf_{sh}$=1.6$f_r$ have a higher pressure threshold ($P_t$) for P2 oscillations (e.g. for $R_0$=400 nm $P_t\approxeq 570 kPa$). In this case the amplitude of the P2 oscillations are higher than the previous cases after the occurrence of SN bifurcation (e.g. for $R_0$=400nm $P_A\approxeq710 kPa$ and ($\frac{R_{max}}{R0}=1.86$). P2 oscillations then grow as pressure increases and P2 oscillations reach large amplitude of non-destructive oscillations (e.g. for $R_0=400nm$ $R_{max}=1.98 R_0$ at $P_A=$976 kPa).\\
\begin{figure*}
	\begin{center}
		\scalebox{0.43}{\includegraphics{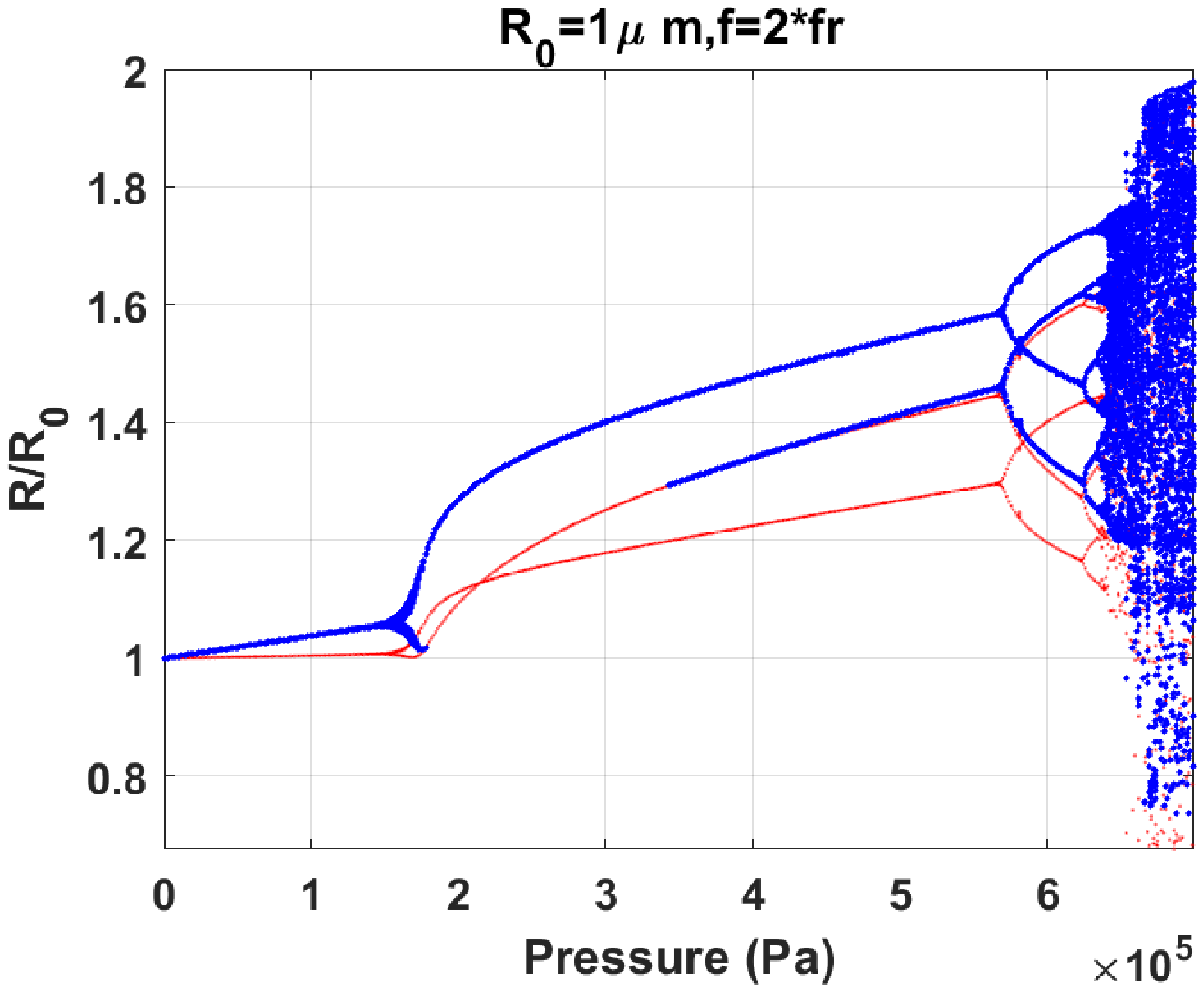}} \scalebox{0.43}{\includegraphics{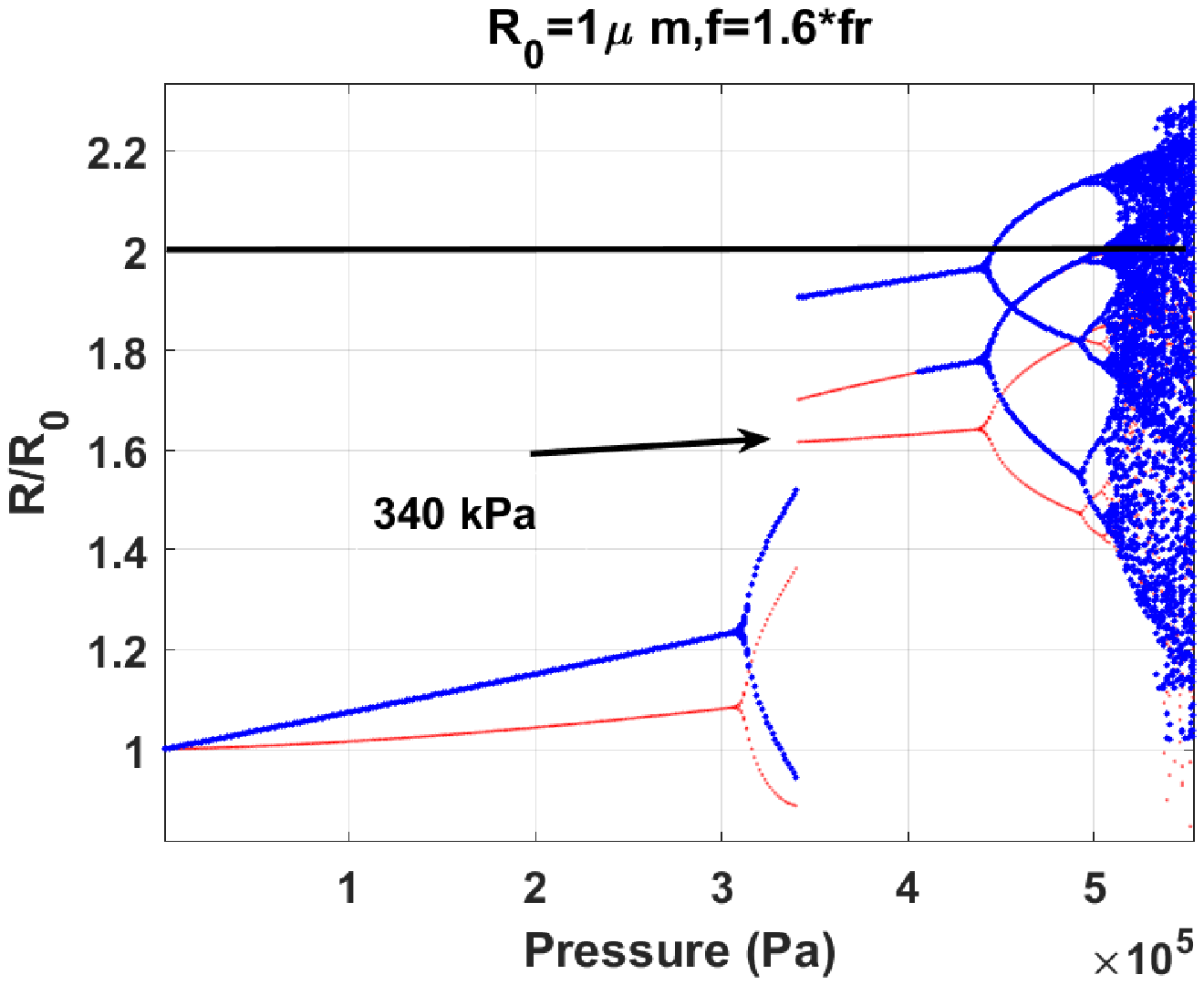}}\\
		\hspace{0.5cm} (a) \hspace{6cm} (b)\\
		\scalebox{0.43}{\includegraphics{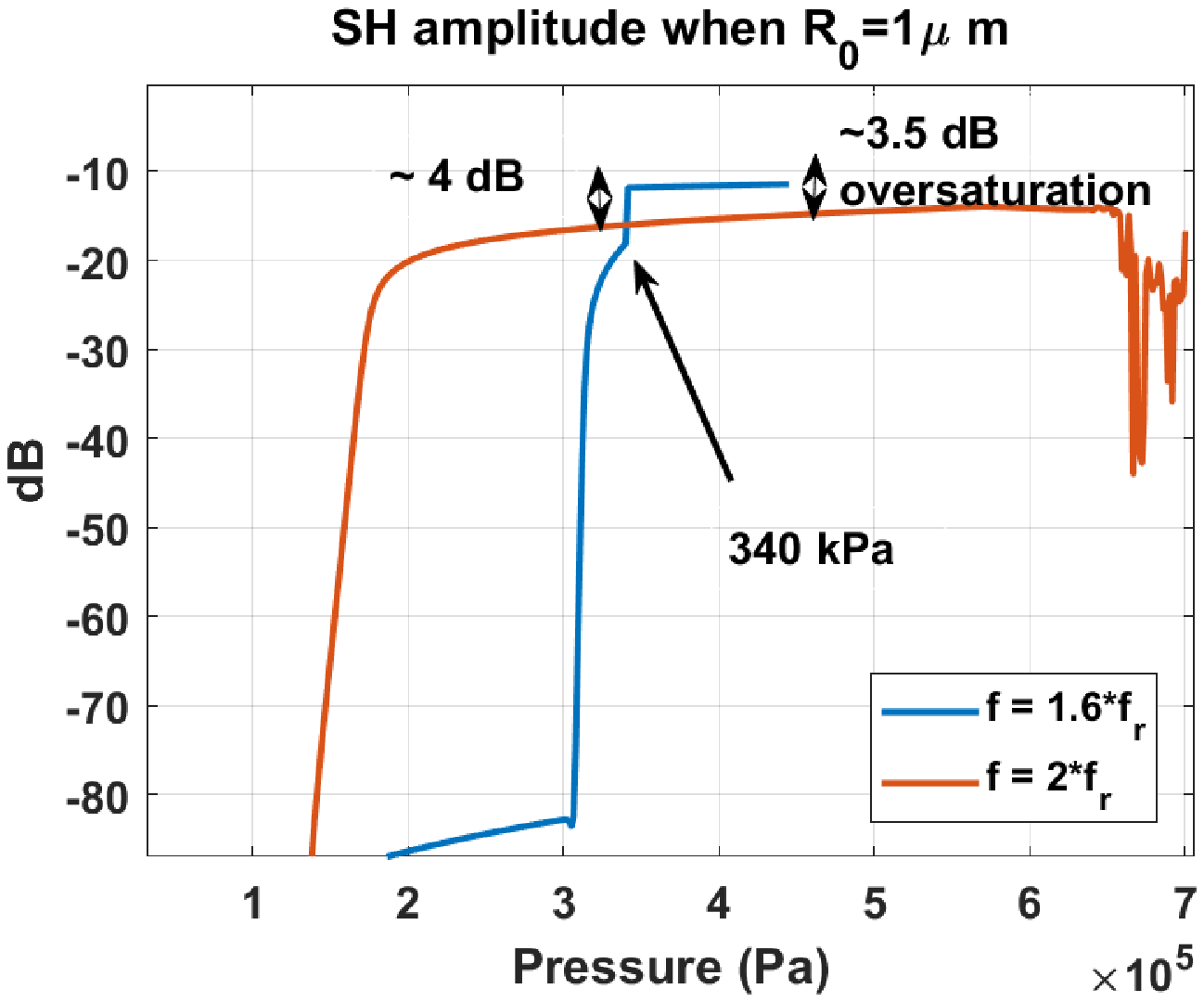}} \scalebox{0.43}{\includegraphics{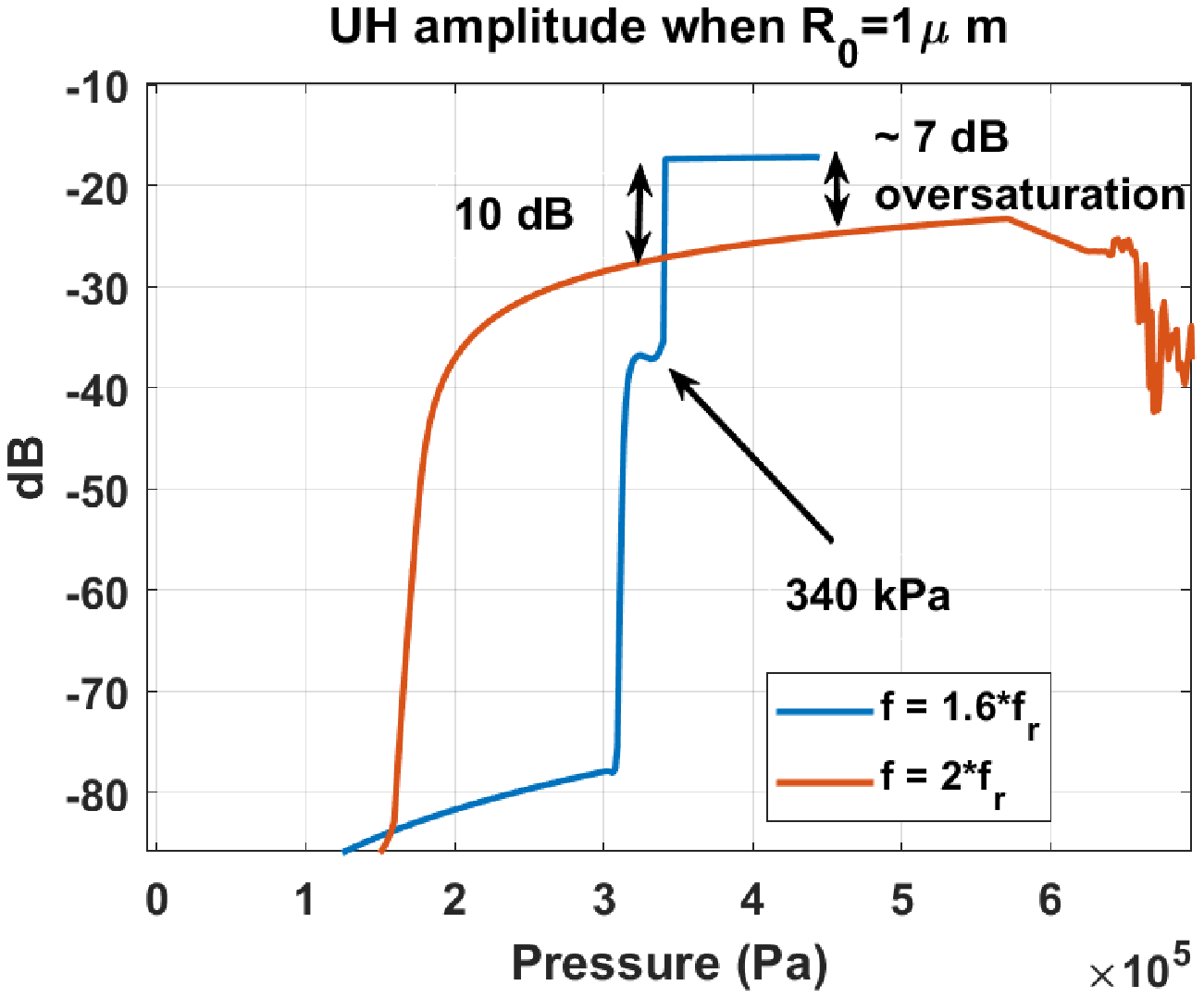}}\\
		\hspace{0.5cm} (c) \hspace{6cm} (d)\\
		\scalebox{0.43}{\includegraphics{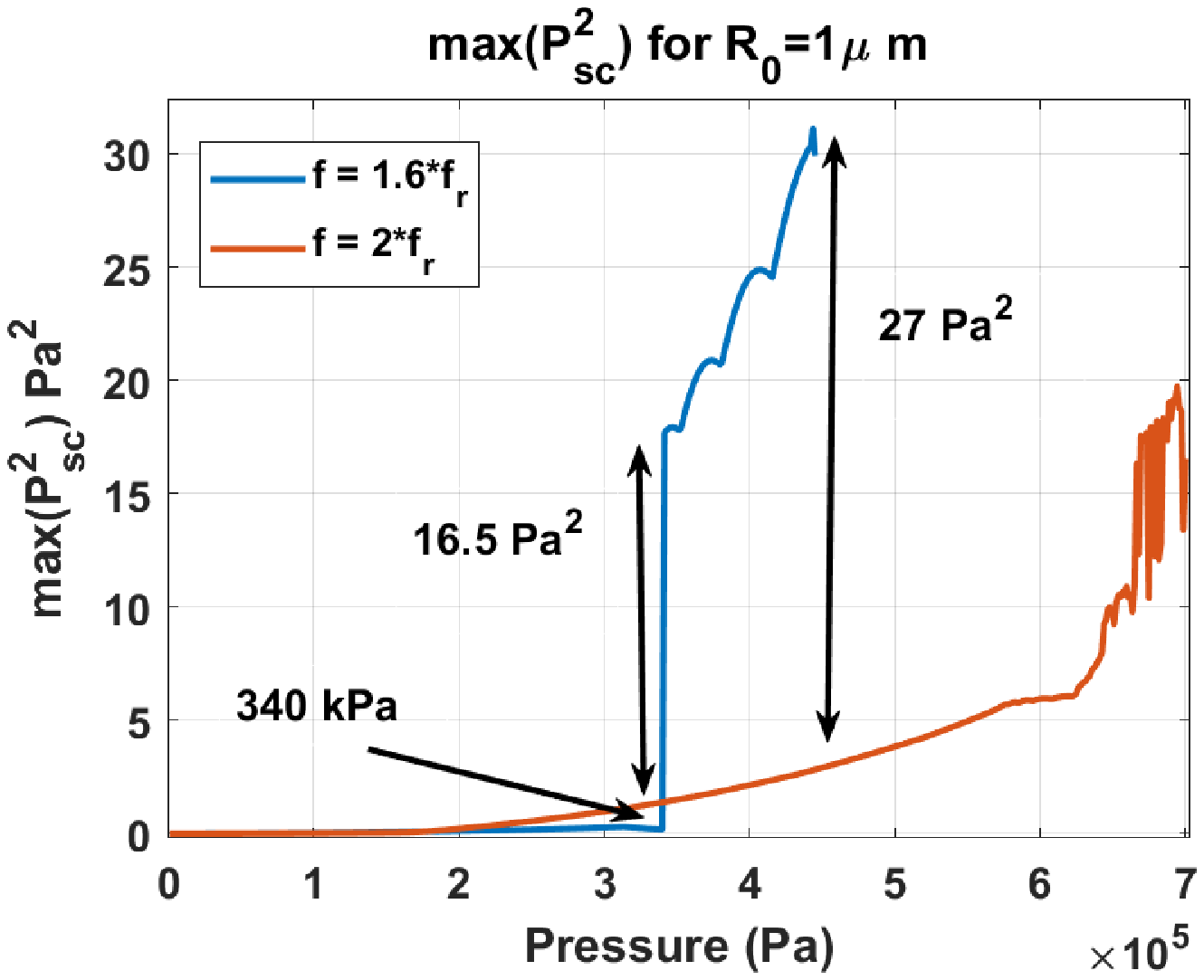}} \scalebox{0.43}{\includegraphics{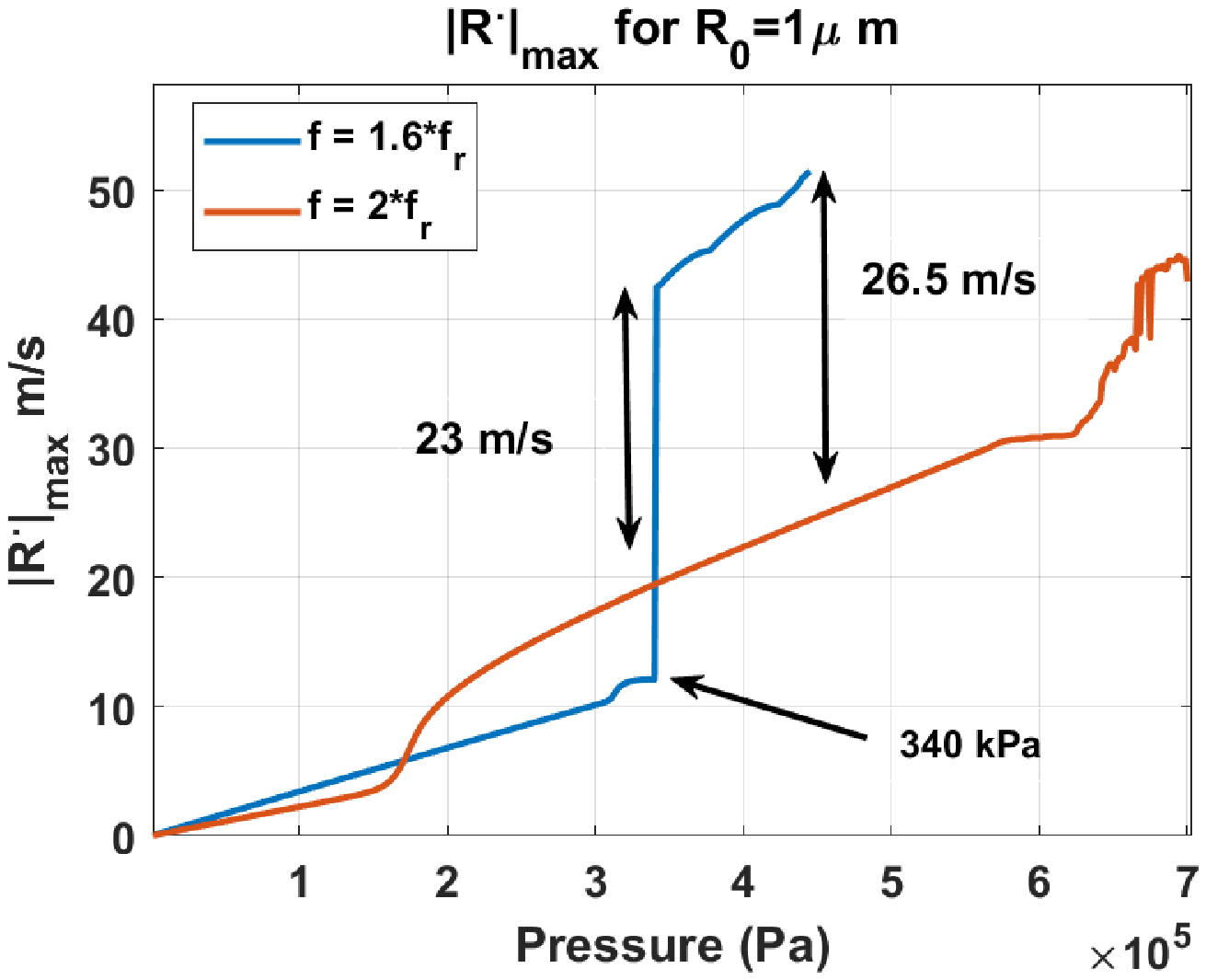}}\\
		\hspace{0.5cm} (e) \hspace{6cm} (f)\\
		\caption{Period doubling and the corresponding bifurcation structure: a) bifurcation structure of the bubble when $f=2f_r$,b) bifurcation structure of the bubble when $f=1.6 f_r$ and c) corresponding SH component of the $P_{sc}$, d) corresponding UH component of the $P_{sc}$, e) Maximum non-destructive $P_{sc}^2$ ($\frac{R_{max}}{R_0}<2$) and f) maximum absolute value of the wall velocity}
	\end{center}
\end{figure*}
Figures 3g and 3h show the bifurcation structure of the $R_0=400 nm$ and $R_0=200 nm$ size bubbles as a function of pressure when $f=1.5f_r$. PD initiation is at the highest pressure threshold (e.g. for $R_0$=400 nm Pd occurs at 597 kPa). Above a second pressure threshold (e.g. for $R_0=400nm$ at $P_A=$ 815 kPa) P2 oscillations undergo a SN bifurcation to P2 oscillations of higher amplitude ($\frac{R}{R_0}\approxeq 2.04$). In this case occurrence of PD is concomitant with bubble destruction as $\frac{R_{max}}{R0}>2$ for both bubbles.  
\subsection{Enhancement of the SH saturation level}
\begin{figure*}
	\begin{center}
		\scalebox{0.43}{\includegraphics{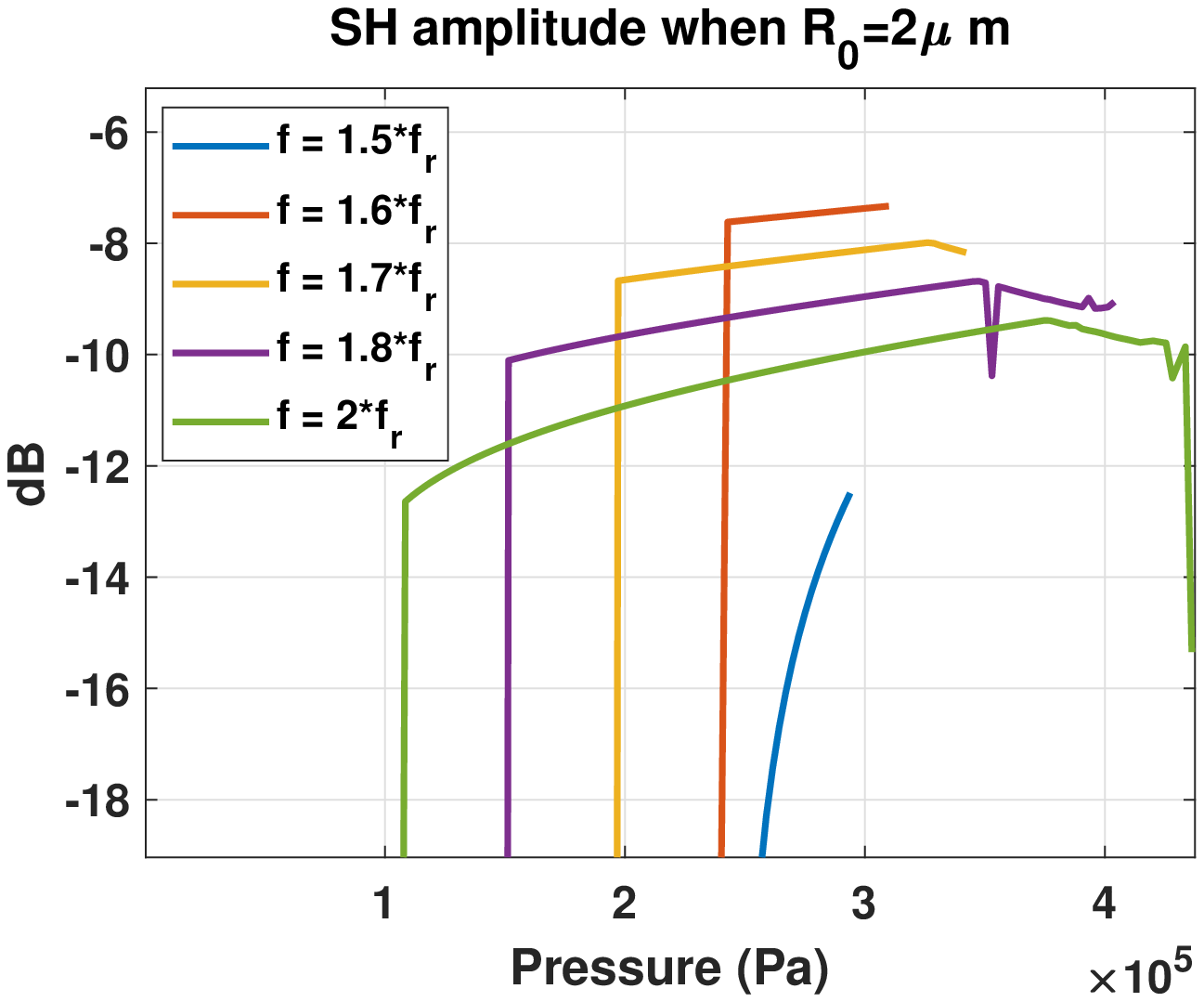}} \scalebox{0.43}{\includegraphics{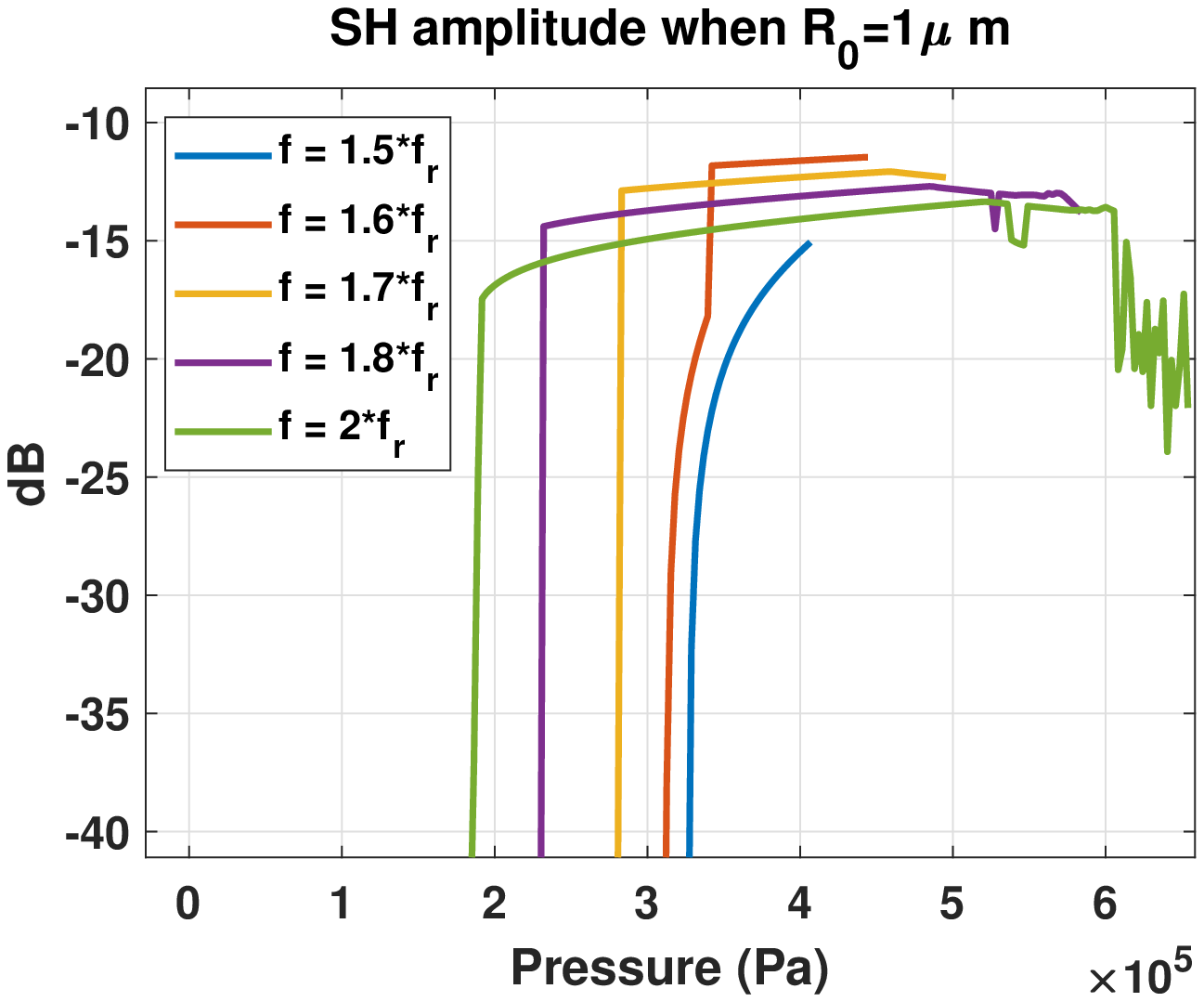}}\\
		\hspace{0.5cm} (a) \hspace{6cm} (b)\\
		\scalebox{0.43}{\includegraphics{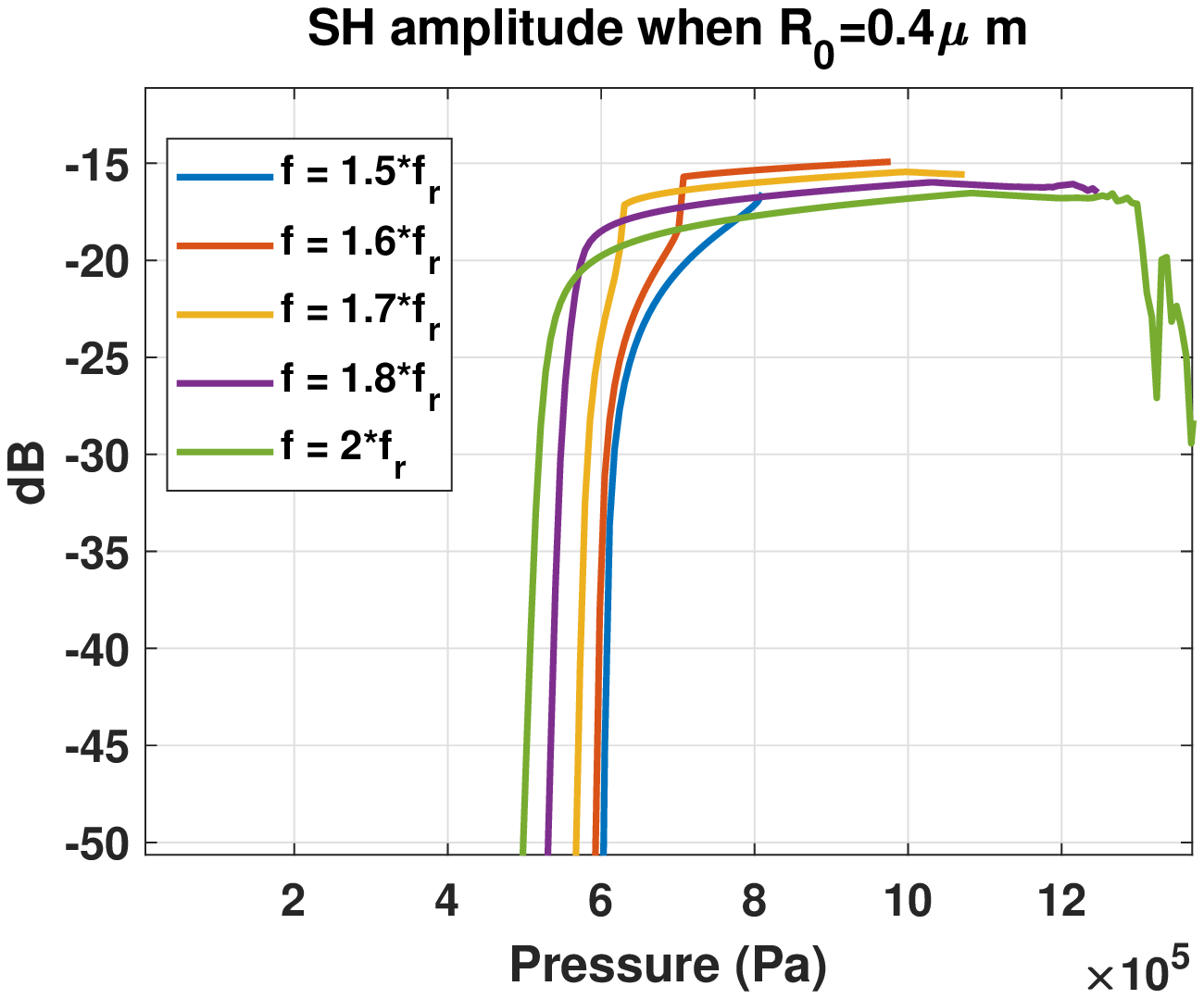}} \scalebox{0.43}{\includegraphics{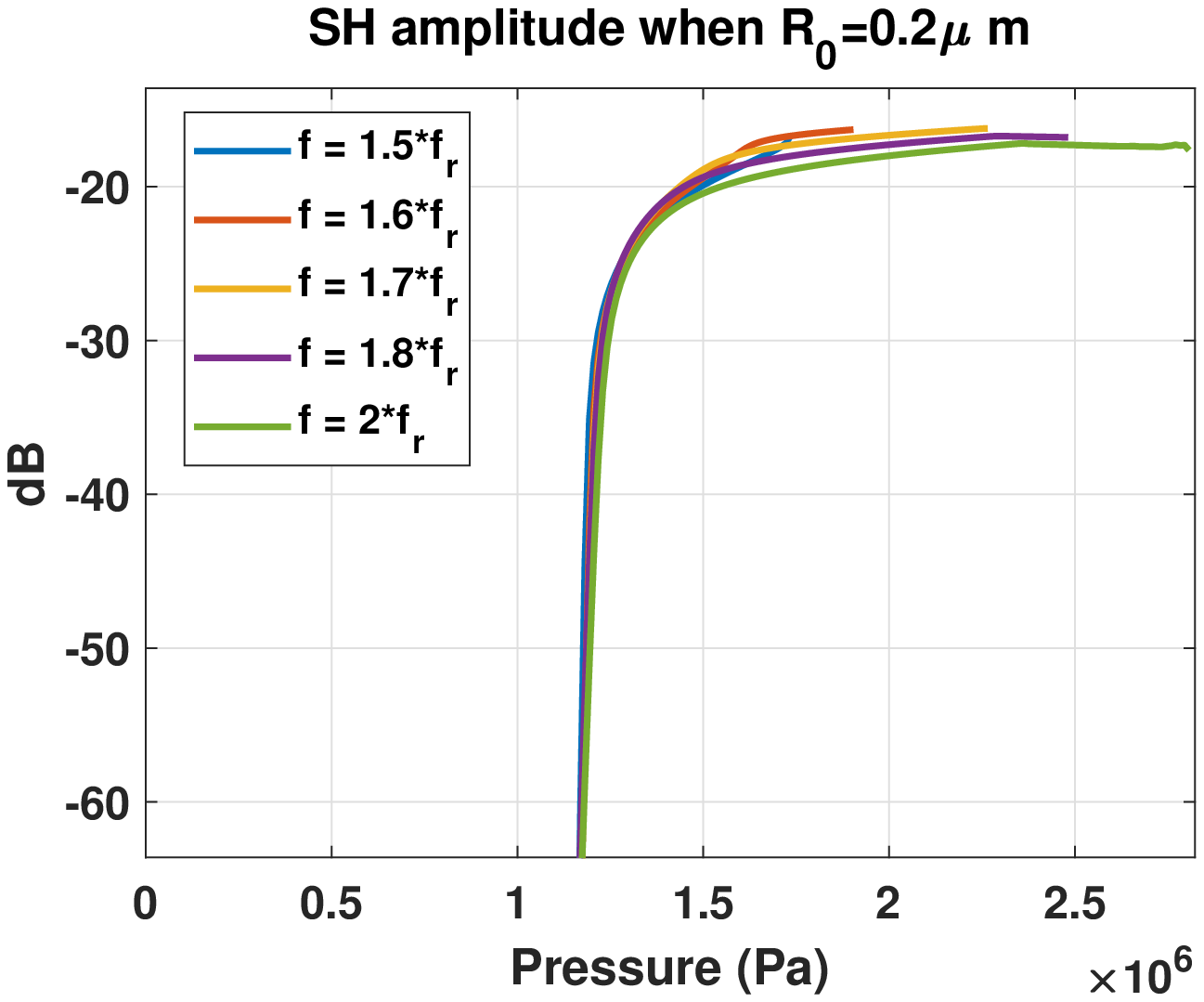}}\\
		\hspace{0.5cm} (c) \hspace{6cm} (d)\\
			\caption{Non-destructive ($\frac{R_{max}}{R_0}<2$) SH component of the $P_{sc}$ for $f_{sh}$ and $Pdf_{sh}$ of: a) $R_0=2 \mu m$, b) $R_0=1 \mu m$, c) $R_0=0.4 \mu m$ and d) $R_0=0.2 \mu m$.}
	\end{center}
\end{figure*}
In order to investigate the consequence of the occurrence of SN on the strength of the SH and UHs of the $P_{sc}$, figure 4 displays the bifurcation structure of the bubble with $R_0$=1 $\mu m$ when $f=2f_r$and $f=1.6f_r$ together with the SH and UH amplitude as well as the maximum value of both the $P_{sc}^2$ and $|\dot{R}|$. When $f=2f_r$, PD ($P_A\approxeq$135 kPa) is concomitant with SH and UH initiation. Consistent with experimental observations [61,62,63], SH and UH component of $P_{sc}$ grow quickly with pressure increase and are saturated. The amplitude of SHs and UHs decrease simultaneous with P4  oscillations ($P_A\approxeq$ 568 kPa) and chaos ($P_A\approxeq$ 651 kPa). Chaos result in increase in $P_{sc}^2$ and wall velocity; however, SHs and UHs amplitude of the $P_{sc}$ decrease. \\When $f=1.6f_r$, initiation of SH and UH oscillations are concomitant with P2 generation in the bifurcation diagram ($P_A\approxeq$ 310 kPa). When the SN bifurcation occur($P_A\approxeq$ 342kPa), SH and UH amplitude of $P_{sc}$ undergo a significant increase (4 dB and 10 dB larger than the case of sonication with $f=2f_r$). This results in oversaturation of the SH and UH amplitude.
When $f=1.6f_r$, the occurrence of a SN in the bifurcation diagram is concomitant with a significant increase in the maximum amplitude of $P_{sc}^2$ ($P_{sc}^2$ becomes 88 times larger than its value before the occurrence of SN) . At $P_A=$340 kPa (the pressure at the SN bifurcation) $P_{sc}^2$  and maximum wall velocity amplitude are respectively  16.5 $Pa^2$ and 23 m/s larger than the case of sonication with $f=2f_r$ (by a factor of $\approxeq$ 80 and 4 times respectively). Moreover, when $f=1.6f_r$, the maximum achievable non-destructive ($\frac{R}{R_0}<2$) SH and UH amplitude are respectively 3.5 and 7 dB larger than the case of $f=2f_r$. Thus, application of the $Pdf_{sh}$ (in this case $f=1.6f_r$) resulted in the oversaturation of the SH and UH amplitude. Maximum non-destructive P2 $P_{sc}^2$ and P2 non-destructive wall velocity amplitude are respectively 27 and 26.5 m/s higher than $f=2f_r$.\\ 
\begin{figure*}
	\begin{center}
		\scalebox{0.43}{\includegraphics{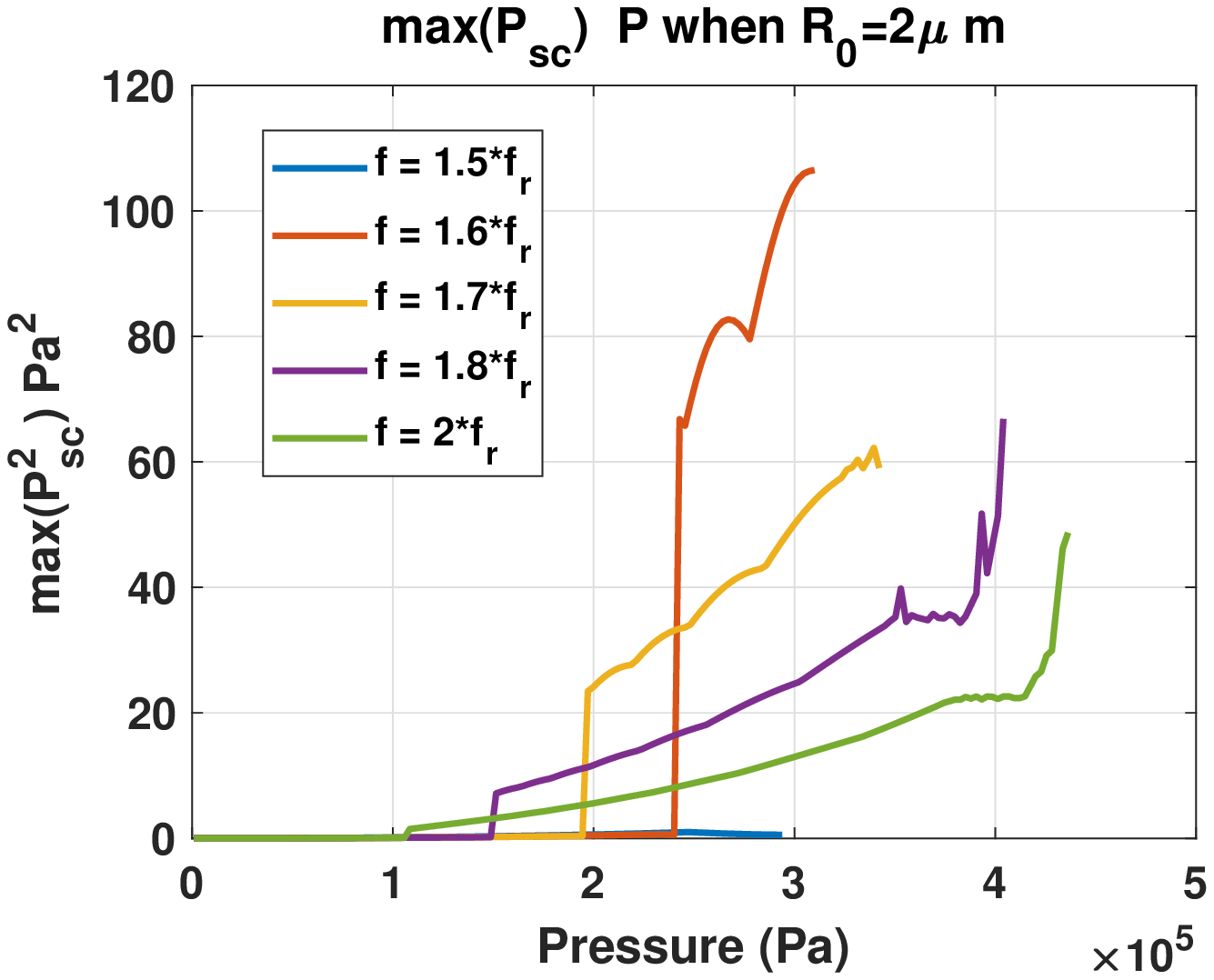}} \scalebox{0.43}{\includegraphics{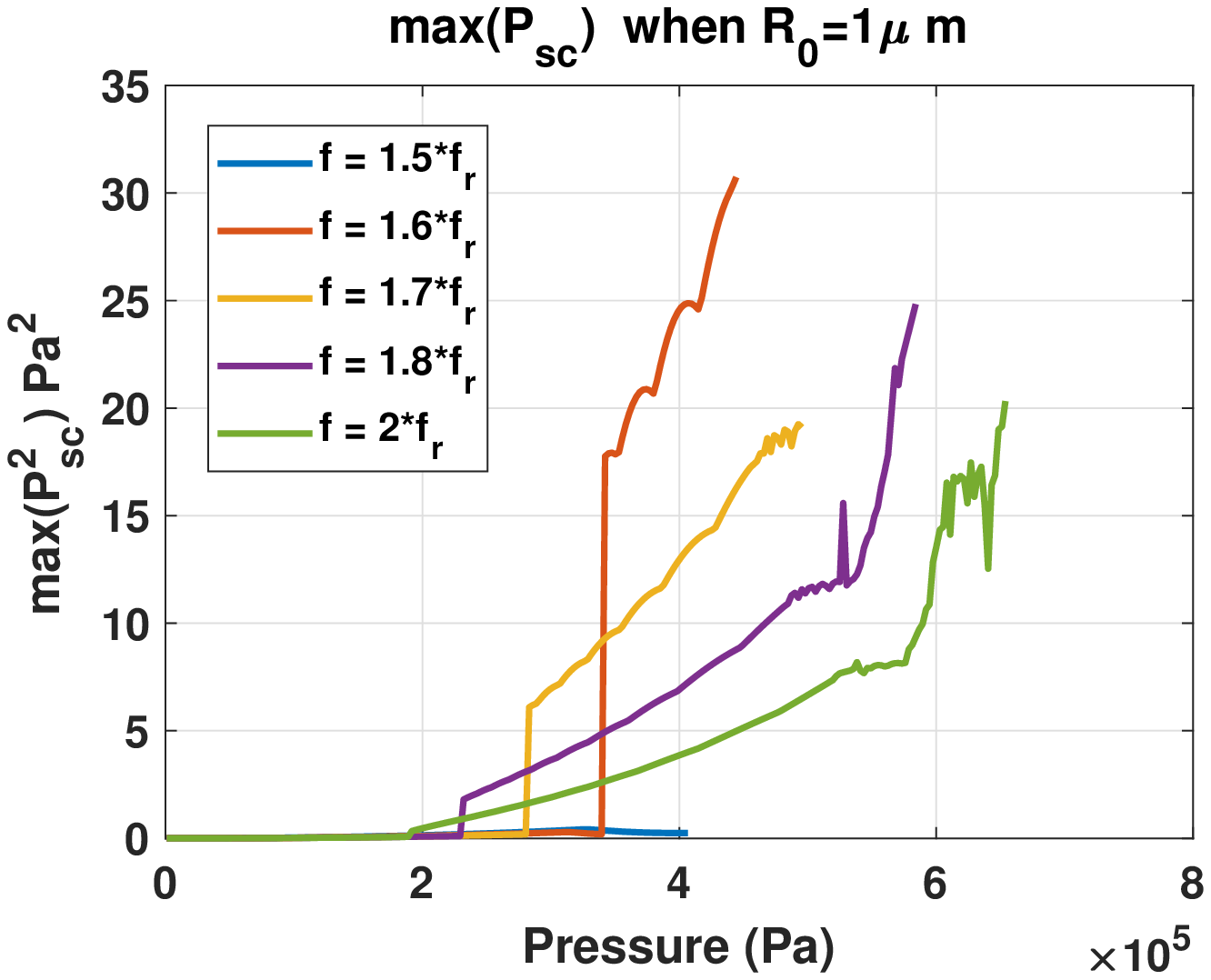}}\\
		\hspace{0.5cm} (a) \hspace{6cm} (b)\\
		\scalebox{0.43}{\includegraphics{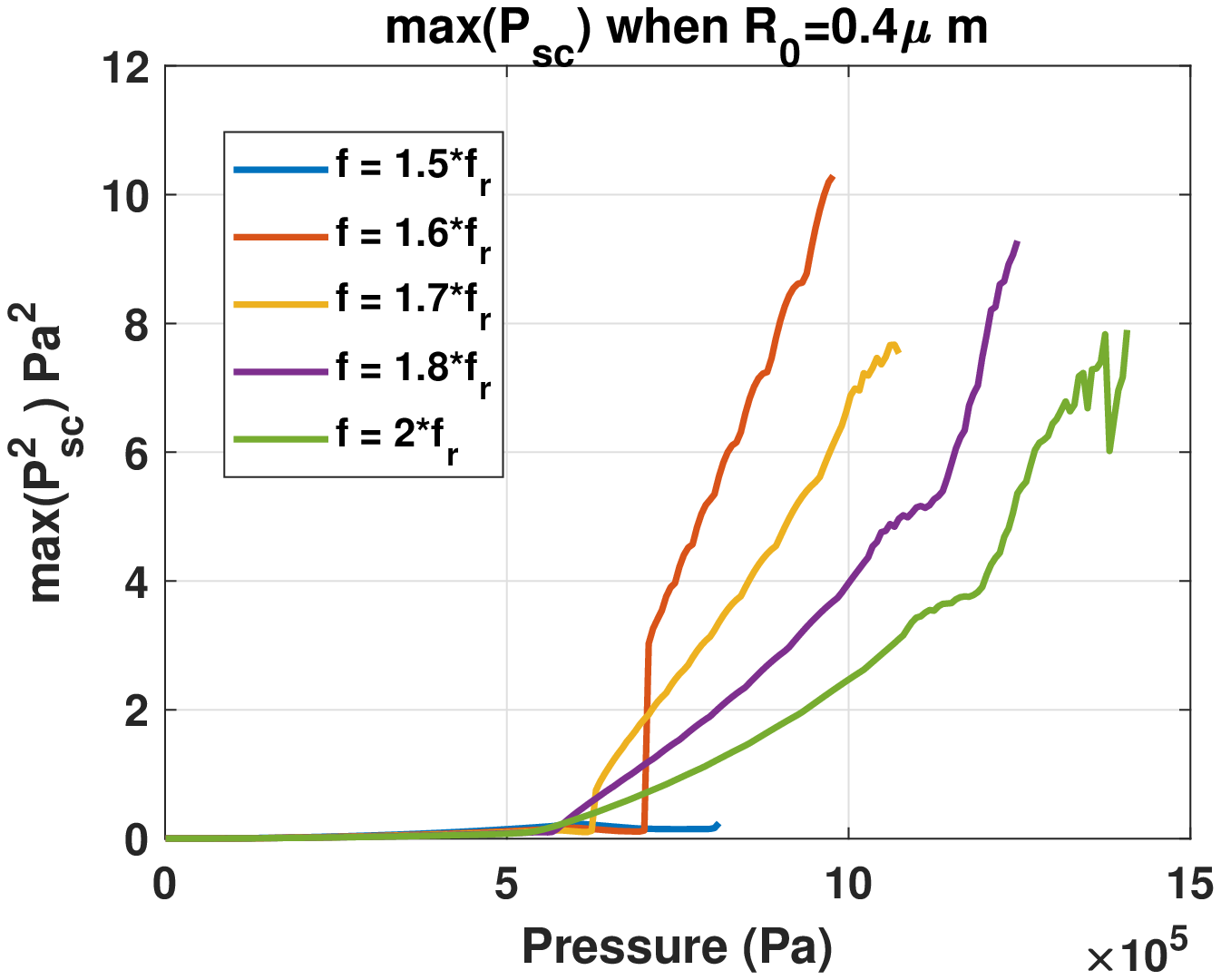}} \scalebox{0.43}{\includegraphics{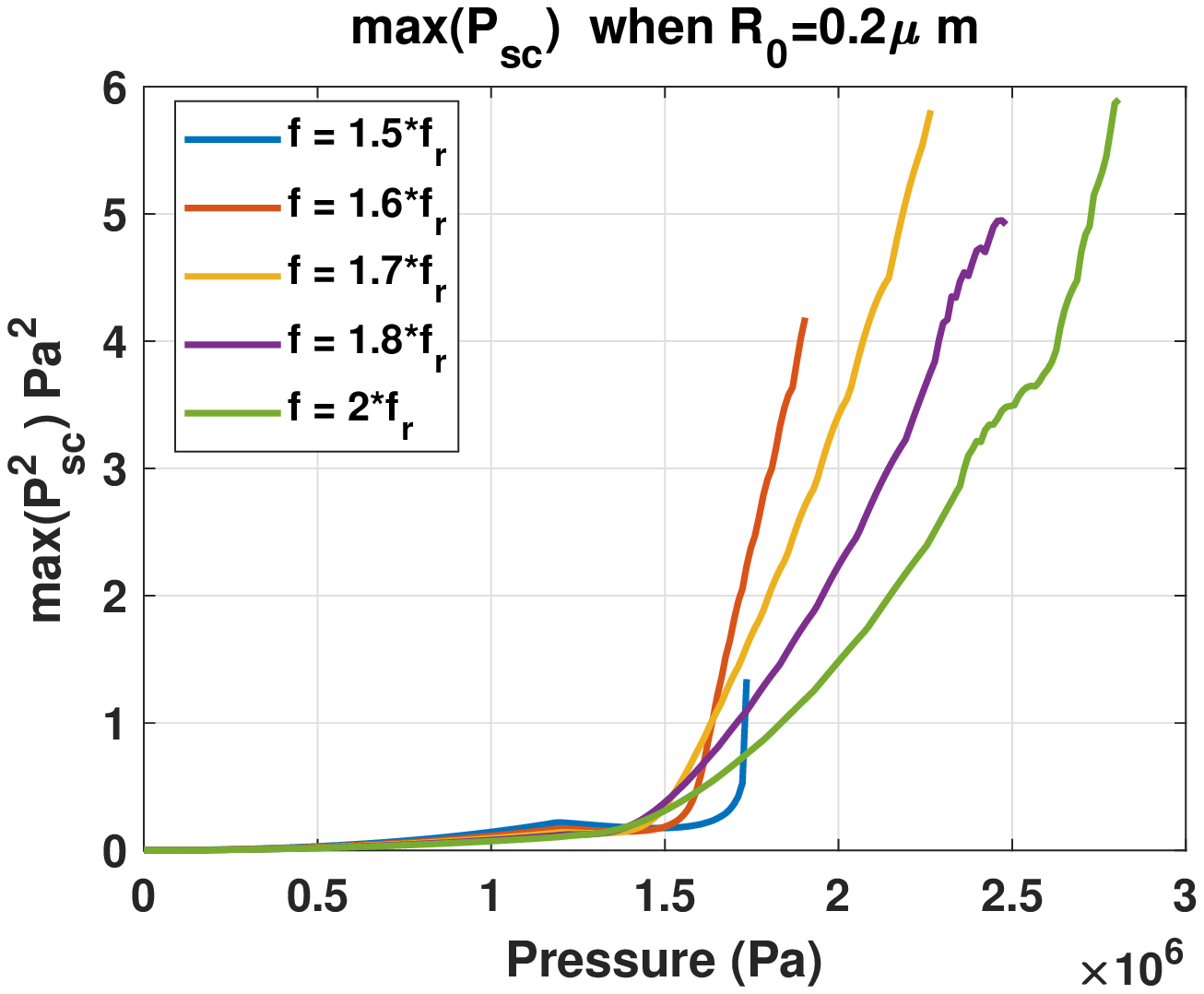}}\\
		\hspace{0.5cm} (c) \hspace{6cm} (d)\\
		\caption{Maximum value of Non-destructive $P_{sc}^2$ for $f_{sh}$ and $Pdf_{sh}$ of: a) $R_0=2 \mu m$, b) $R_0=1 \mu m$, c) $R_0=0.4 \mu m$ and d) $R_0=0.2 \mu m$.}
	\end{center}
\end{figure*}
Figure 5a-d illustrates the SH amplitude of the $P_{sc}$ as a function of acoustic pressure at different frequencies ($f=2f_r$, $1.8f_r$, $1.7f_r$, $1.6f_r$and $1.5f_r$) for $R_0$=2 $\mu m$ (5a), $R_0$=1 $\mu m$ (5b), $R_0$=400 nm (5c) and $R_0$=200 nm (5d). The SH amplitude of the $P_{sc}$ are only shown for non-destructive oscillation regimes where $\frac{R}{R_0}<2$. When bubbles are sonicated with f=$2f_r$, SHs are initiated at the lowest pressure, and the SHs amplitude grows with increasing pressure and then saturate. At higher pressures where P4 oscillations or chaos occurs, the SH amplitude decreases rapidly.  When bubbles are sonicated with their $Pdf_{sh}$, SHs are initiated at higher acoustic pressures. However, SHs grow rapidly after initiation (concomitant with SN bifurcation) and the SH amplitude becomes larger than the case of sonication with $f=2f_r$. For the frequencies shown in Fig. 5, the maximum SH amplitude occurs when $f=1.6f_r$ (red line). For all the bubble sizes studied here, when f=1.5$f_r$(blue line in Fig. 5), the SN bifurcation is concomitant with bubble destruction ($\frac{R}{R_0}>2$), therefore sonication with $f=1.5f_r$ does not result in any SH enhancement over the conventional method of sonication with $f=2f_r$. For frequencies less than 2$f_r$ the bubble undergoes destruction at a lower acoustic pressures.\\
Figure 6 displays the maximum non-destructive ($\frac{R}{R_0}<2$) $P_{sc}^2$ at different frequencies (f=$2f_r$, $1.8f_r$, $1.6f_r$ and $1.5f_r$) for $R_0$=2 $\mu m$ (6a), $R_0$=1 $\mu m$ (6b), $R_0$=400 nm (6c) and $R_0$=200 nm (6d). When f=$2f_r$, $P_{sc}^2$ is very small for pressures below PD; concomitant with generation of PD, $P_{sc}^2$ undergoes a rapid increase and then increases linearly with acoustic pressure. When P4 oscillations occur, $P_{sc}^2$ decreases for micron size bubbles (the growth rate decreases for nano-size bubbles); this is similar to the decrease of $P_{sc}$ concomitant with P2 oscillations when f=$f_r$ [26].  Further increase in the acoustic pressure results in chaotic oscillations which lead to significant enhancement of the $P_{sc}^2$; however, this enhancement in amplitude is associated with a rapid decrease in SH and UH amplitude (Fig 4 and 5).\\
\begin{figure*}
	\begin{center}
		\scalebox{0.43}{\includegraphics{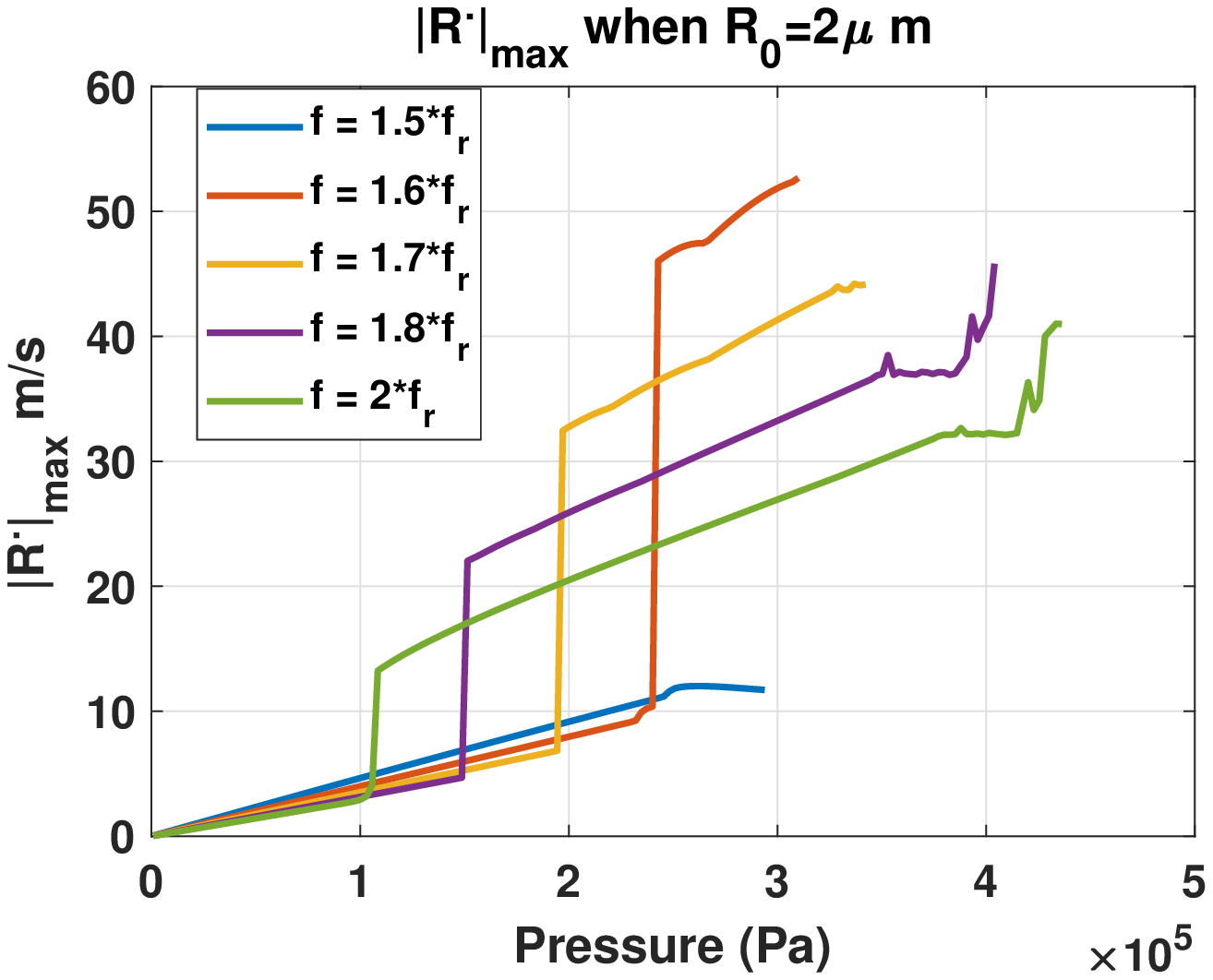}} \scalebox{0.43}{\includegraphics{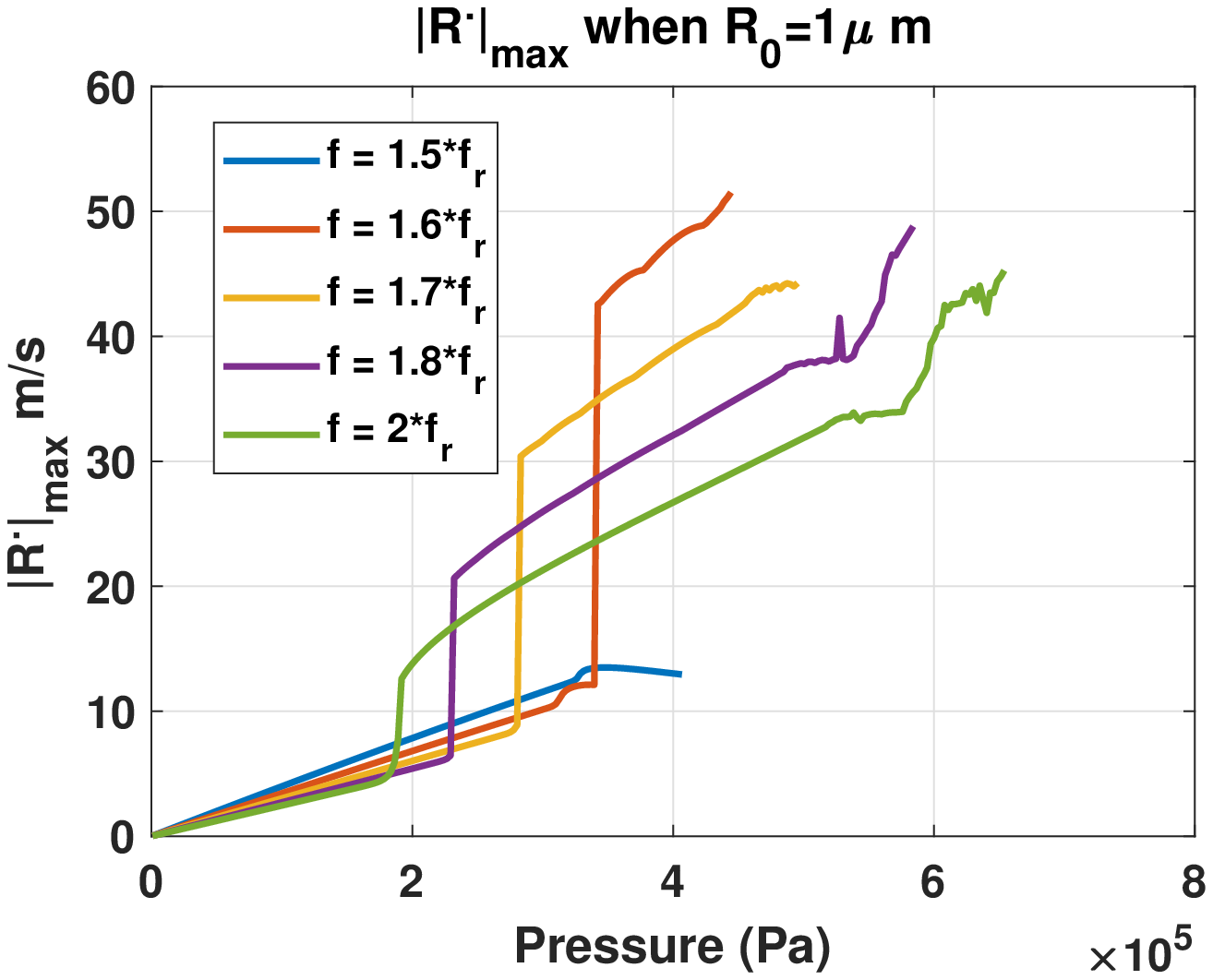}}\\
		\hspace{0.5cm} (a) \hspace{6cm} (b)\\
		\scalebox{0.43}{\includegraphics{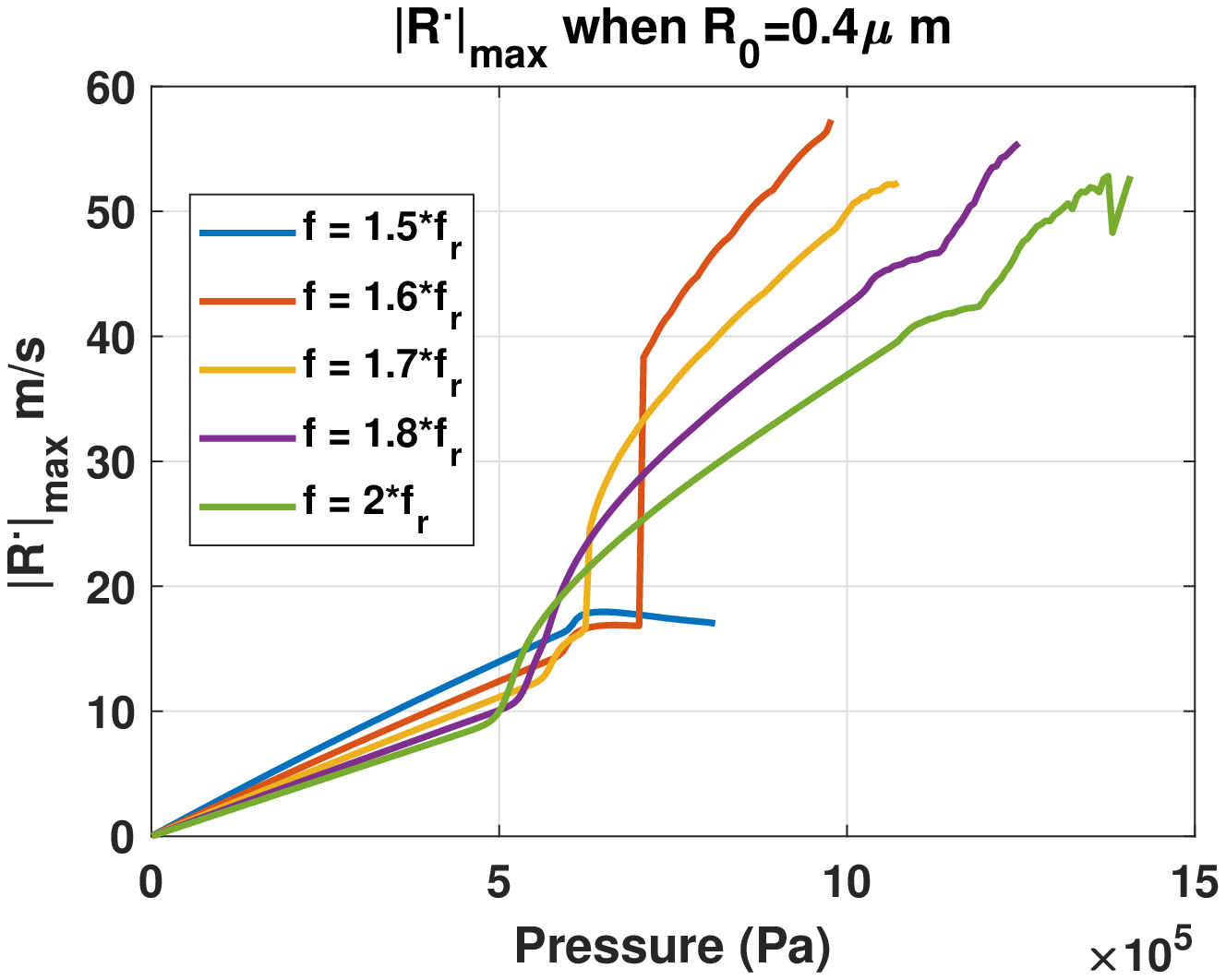}} \scalebox{0.43}{\includegraphics{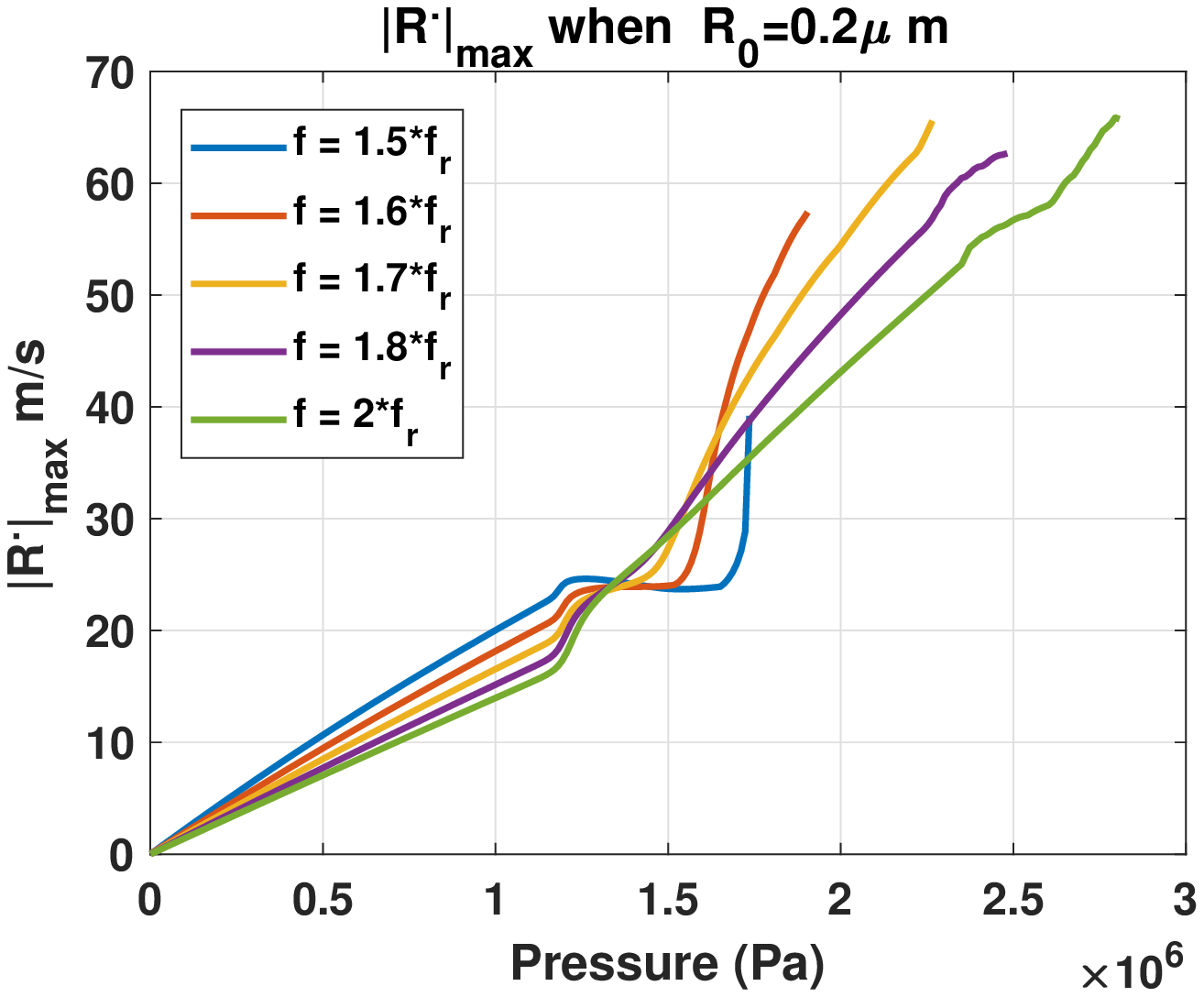}}\\
		\hspace{0.5cm} (c) \hspace{6cm} (d)\\
		\caption{Maximum value of Non-destructive absolute wall velocity for $f_{sh}$ and $Pdf_{sh}$ of: a) $R_0=2 \mu m$, b) $R_0=1 \mu m$, c) $R_0=0.4 \mu m$ and d) $R_0=0.2 \mu m$.}
	\end{center}
\end{figure*}
When $f=Pdf_{sh}$ ($f=1.8f_r$, $1.7f_r$ and $1.6f_r$), $P_{sc}^2$ is smaller than its counter part when $f=2f_r$. Above a pressure threshold, the SN bifurcation results in the generation of higher amplitude P2 oscillations. The SN bifurcation is coincident with a rapid increase in $P_{sc}^2$. $P_{sc}^2$ becomes significantly larger than the case of sonication with $f=2f_r$ (e.g. for the bubble with $R_0$=2 $\mu m$ and $f=1.6 f_r$, when SN occurs at $\approxeq$ 242 kPa, $P_{sc}^2$ becomes 8.4 times larger than its counterpart when $f=2f_r$). After the SN, $P_{sc}^2$ increase monotonically with pressure increase until the bubble is destroyed ($\frac{R_{max}}{R_0}>2$).\\  
Figure 7 displays the maximum non-destructive wall velocity at different frequencies ($f=2f_r$, $1.8f_r$, $1.6f_r$ and $1.5f_r$) for $R_0$=2 $\mu m$ (6a), $R_0$=1 $\mu m$ (6b), $R_0$=400 nm (6c) and $R_0$=200 nm (6d). Maximum wall velocity amplitude increases monotonically with pressure; however, as soon as PD occurs, the wall velocity undergoes a rapid increase and continues to increase monotonically after. The occurrence of P4 oscillations results in a decrease in the maximum wall velocities for micron size bubbles (or decrease in the growth rate of wall velocity for nano-size bubbles) which is similar to the decrease of wall velocity concomitant with the occurrence of P2 when $f=f_r$ [26]. At higher pressures, the generation of chaotic oscillations leads to a rapid increase in wall velocity; however, as seen before this is simultaneous with a decrease in SH and UH amplitudes. For the bubble with $R_0$=0.2 $\mu m$ the occurrence of SN bifurcation does not have a pronounced effect on the maximum wall velocity amplitude which is due to the dominant effect of liquid viscosity on smaller bubbles.
\subsection{Bifurcation structure of the micro-bubbles (for $fr<f<1.5f_r$)} 
\begin{figure*}
	\begin{center}
		\scalebox{0.43}{\includegraphics{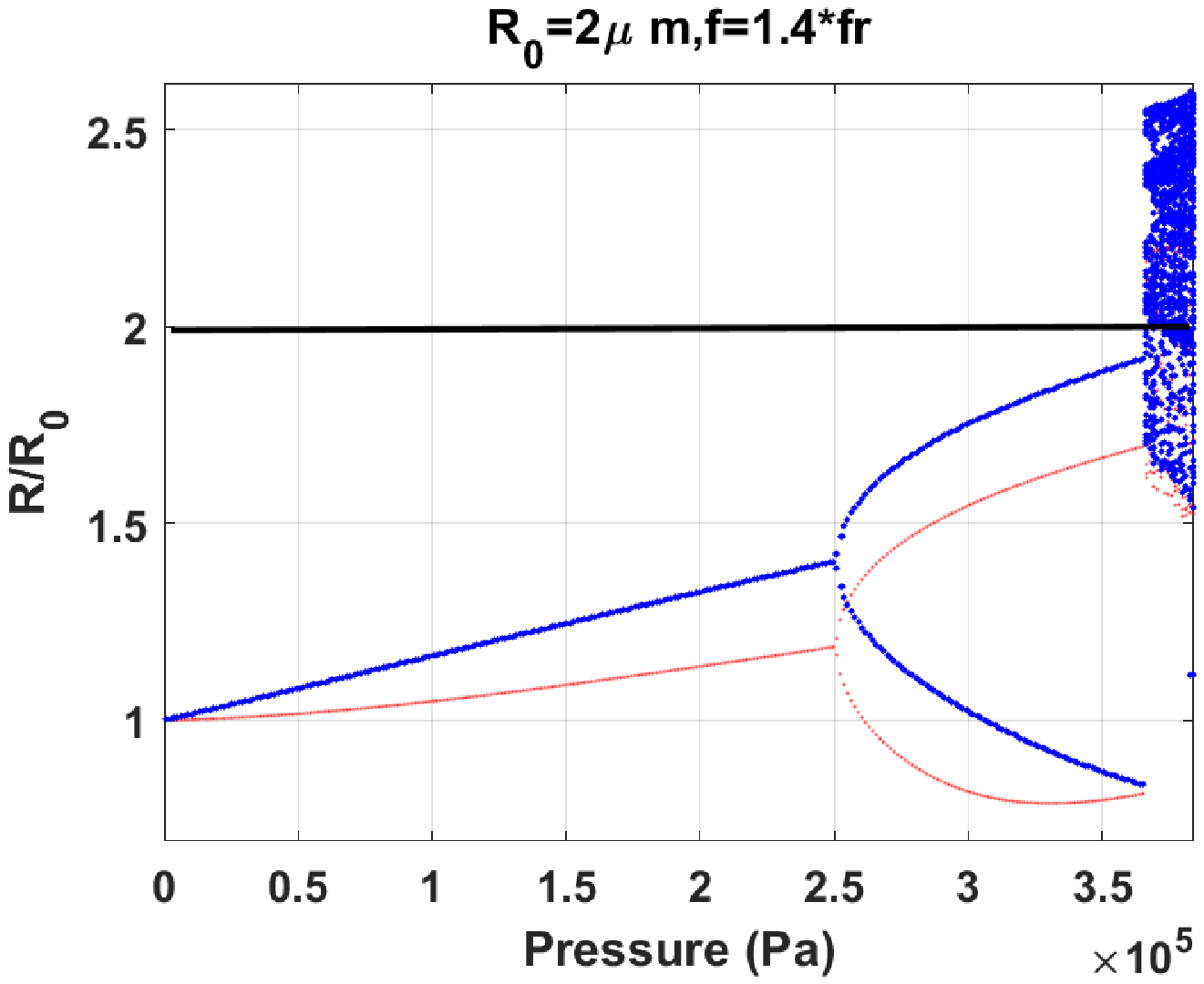}} \scalebox{0.43}{\includegraphics{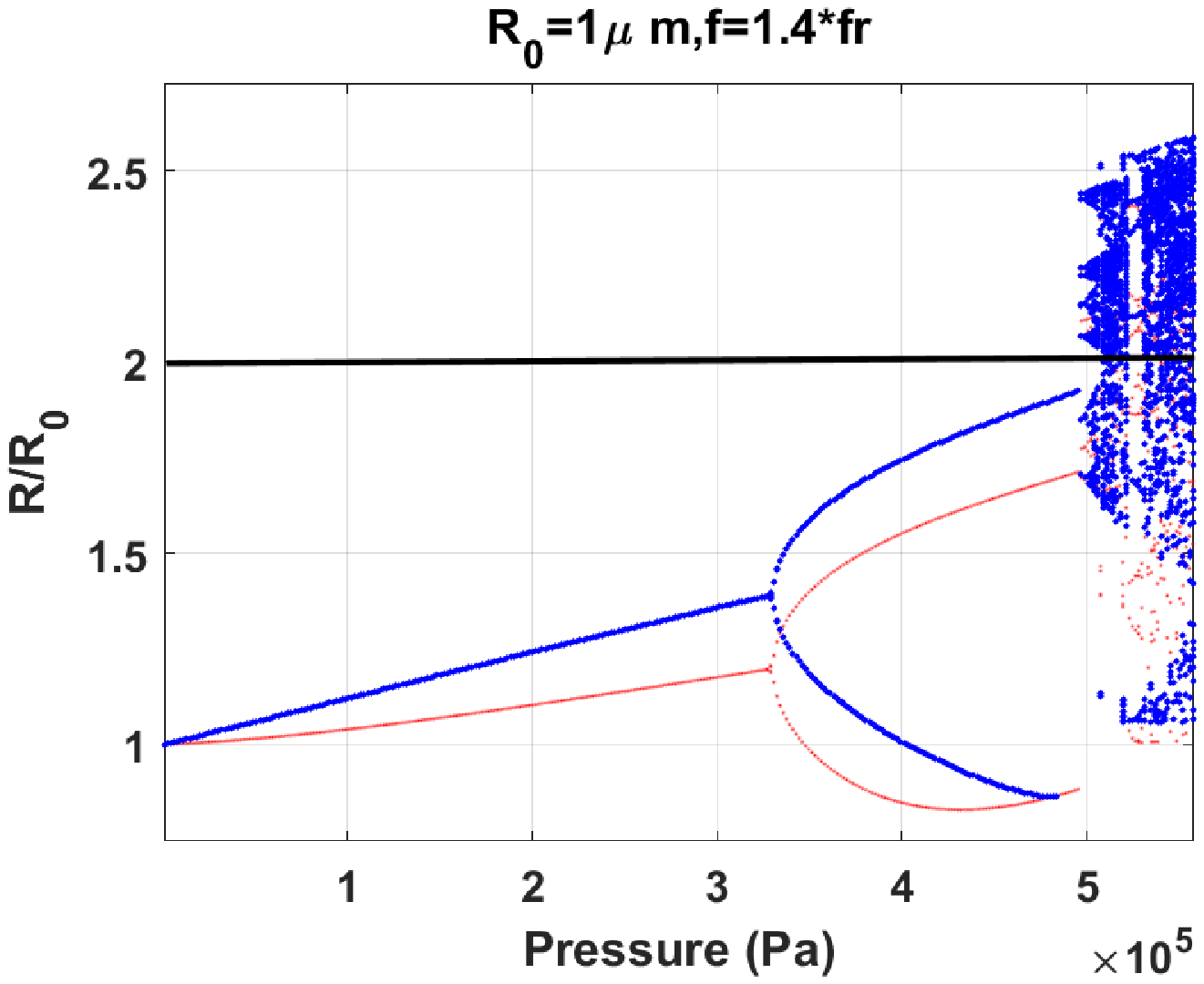}}\\
		\hspace{0.5cm} (a) \hspace{6cm} (b)\\
		\scalebox{0.43}{\includegraphics{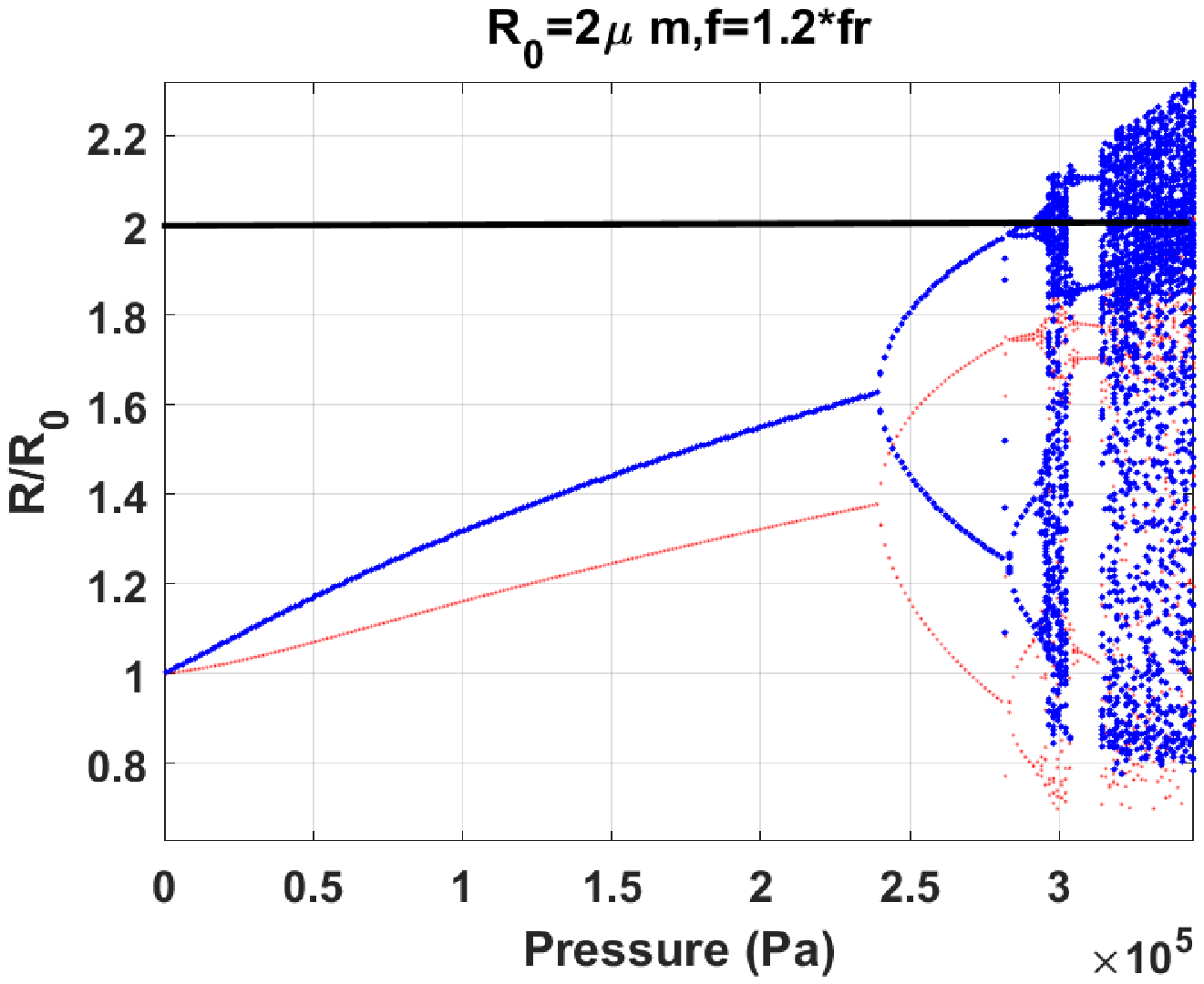}} \scalebox{0.43}{\includegraphics{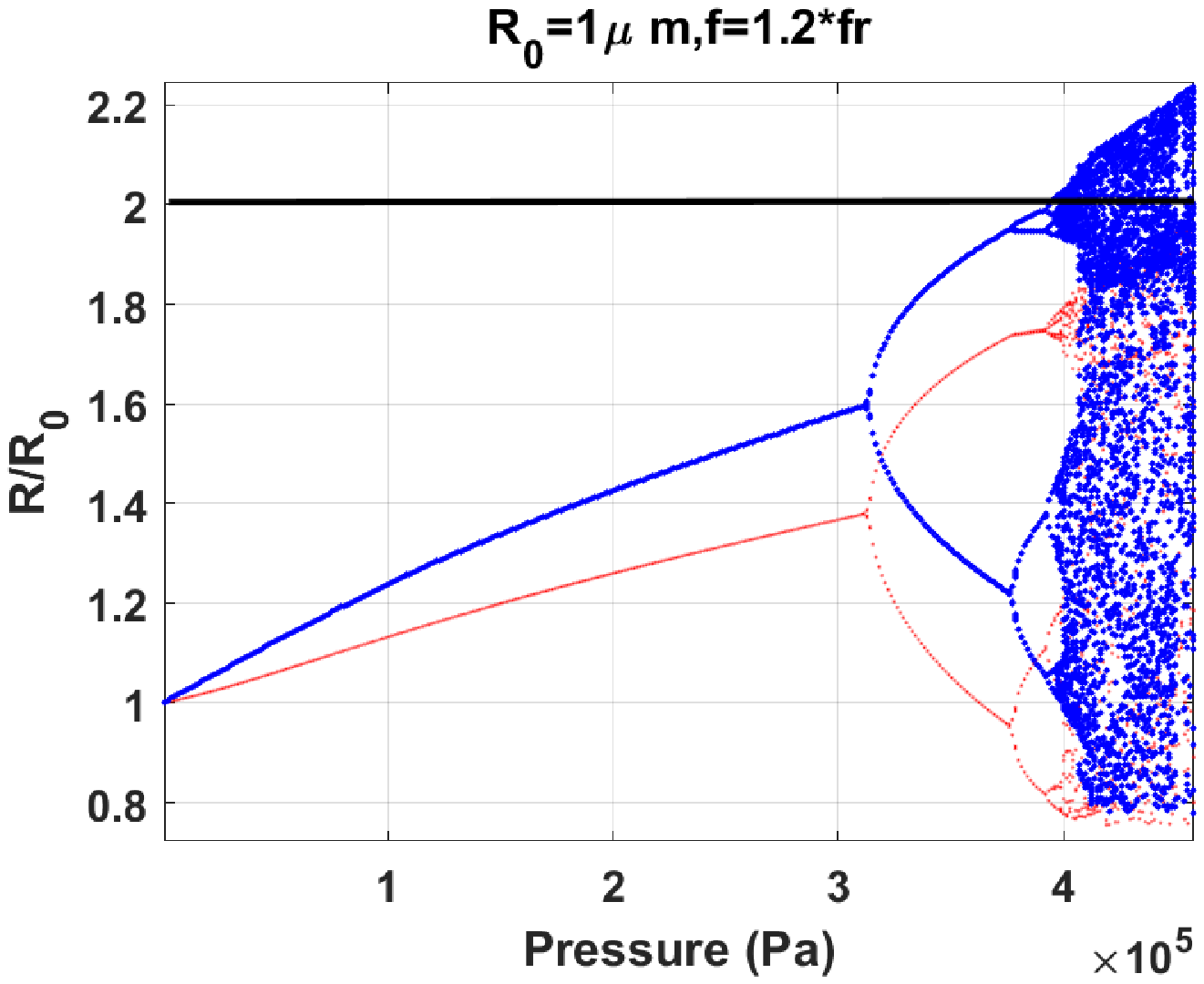}}\\
		\hspace{0.5cm} (c) \hspace{6cm} (d)\\
		\scalebox{0.43}{\includegraphics{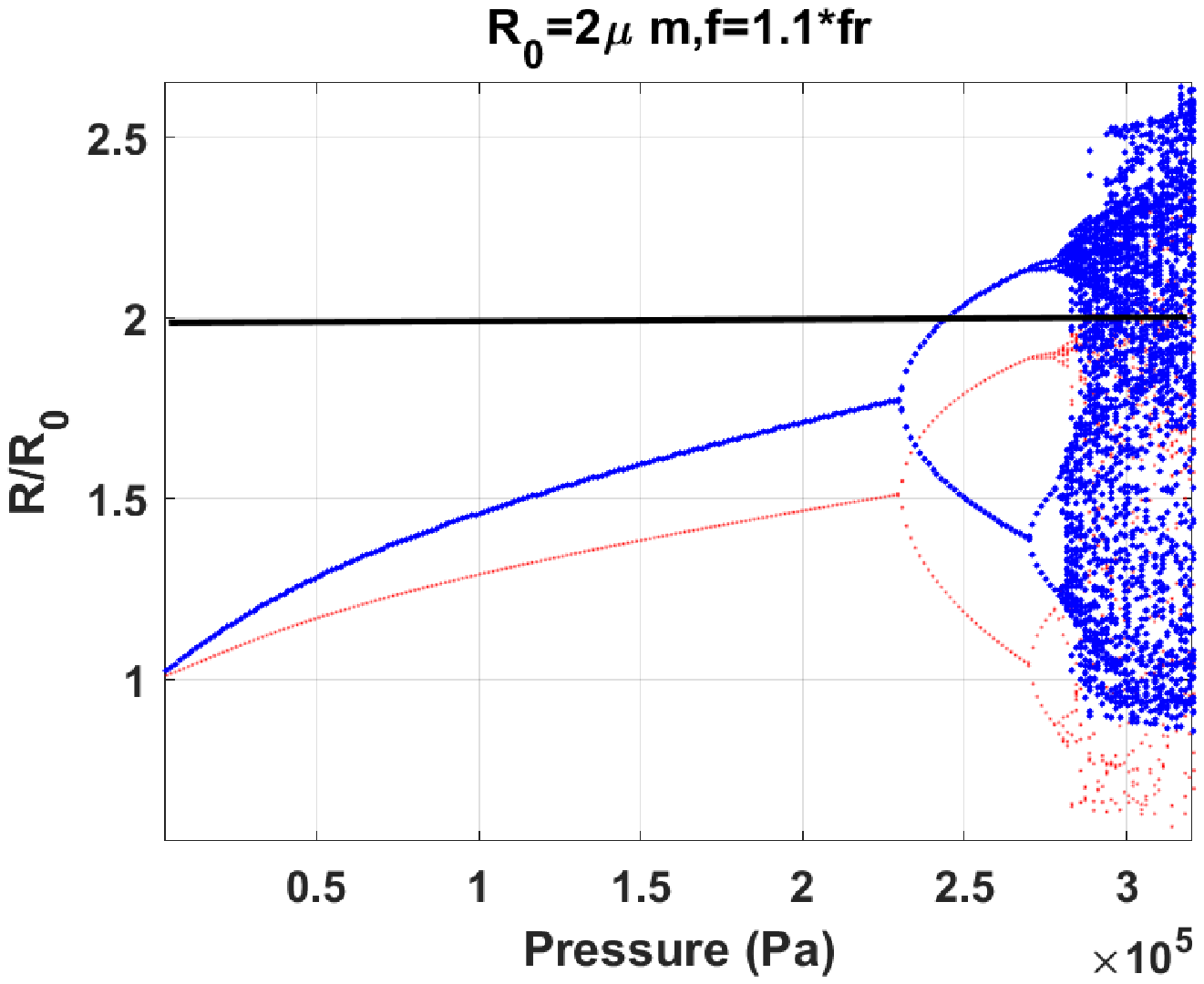}} \scalebox{0.43}{\includegraphics{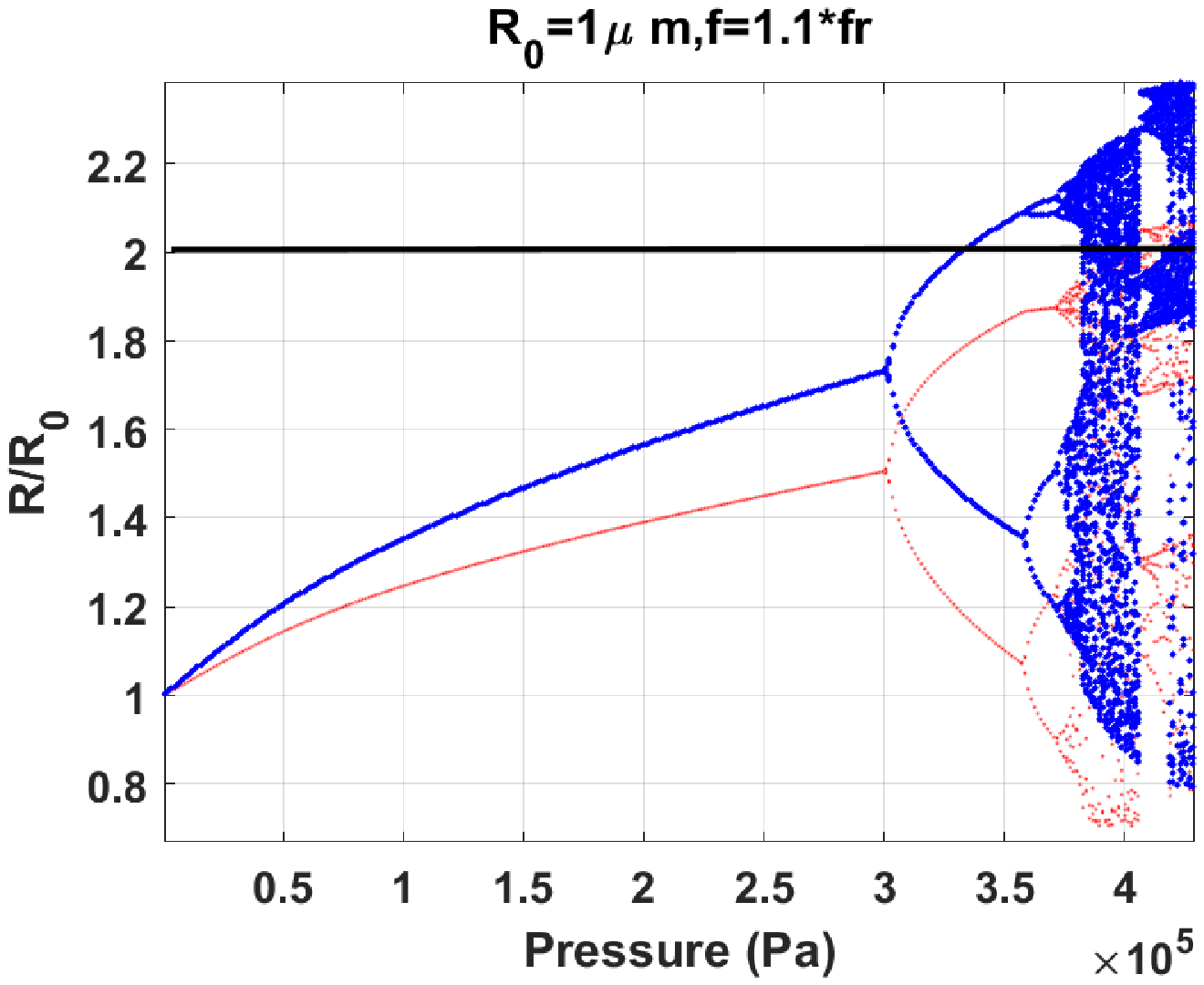}}\\
		\hspace{0.5cm} (e) \hspace{6cm} (f)\\
		\caption{Bifurcation structure (blue: method of peaks, red: conventional method) of the micron-size bubbles as a function of pressure when sonicated with $f=1.4f_r, 1.2f_r \&1.1 f_r$. Left column is for $R_0=2 \mu m$ and Right column is for $R_0=1 \mu m$ (arrow shows the pressure responsible for SN bifurcation)}
	\end{center}
\end{figure*}
In the previous sections we saw that when $f\approxeq1.5f_r$, SN bifurcation results in bubble destruction ($\frac{R}{R_0}>2$) and sonication with $1.5fr<f<2f_r$ results in an enhancement of the SH amplitude. In this section, we closely examine the bifurcation structure of the micro-bubbles when sonicated with  $f_r<f<1.5f_r$.  Figure 8 shows the bifurcation structure of the micron-size bubbles with $R_0$=2$\mu m$ and $R_0$=1$\mu m$ as a function of pressure. The frequencies are $1.4f_r$, 1.2$f_r$ and $1.1f_r$. Unlike the case of sonication with $1.5f_r<f<2f_r$; within the parameter ranges studies here (and for $\frac{R}{R_0}<2$), there is no SN bifurcation taking place in the diagrams. The evolution of the dynamics of the system is through a simple PD to P2 oscillations followed by a cascade of PDs to chaos. In case of $f=1.2f_r$and $1.1f_r$ full amplitude ($\frac{R_{max}}{R0}$=2) non-destructive P2 oscillations are developed; however, unlike $1.5f_r<f<2f_r$, PD is simultaneous with a decrease in wall velocity and scattered pressure. We have previously shown that when bubble is sonicated with $f=f_r$, the occurrence of PD is concomitant with a decrease in the scattered pressure and maximum wall velocity [26].  The red curve in figure 8 (constructed by conventional method) never meets the blue curve (constructed by the maxima method) suggesting that the wall velocity of the bubble never gets in phase with the acoustic driving force.
\subsection{Bifurcation structure of the nano-bubbles (for $f_r<f<1.5f_r$)}
\begin{figure*}
	\begin{center}
		\scalebox{0.43}{\includegraphics{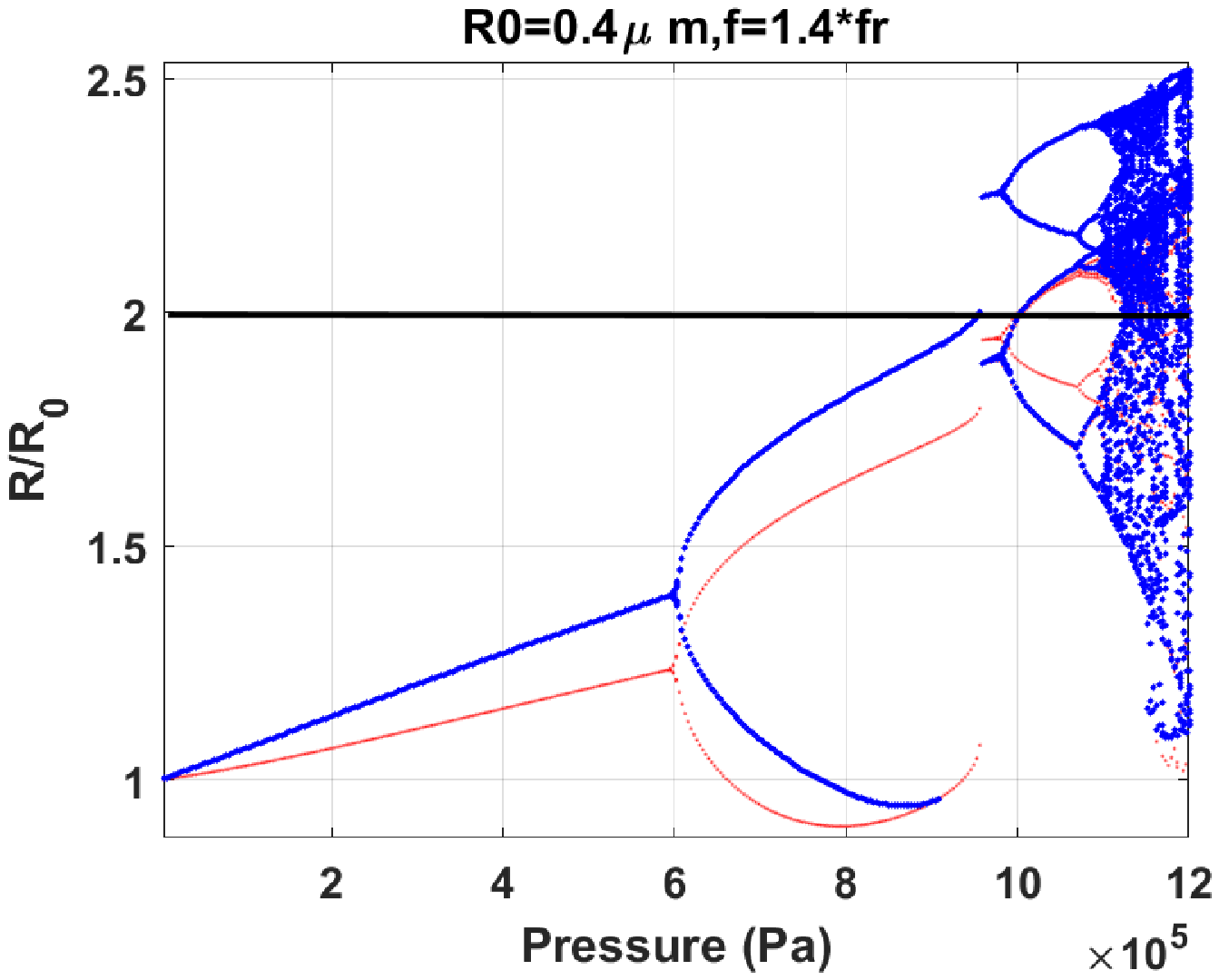}} \scalebox{0.43}{\includegraphics{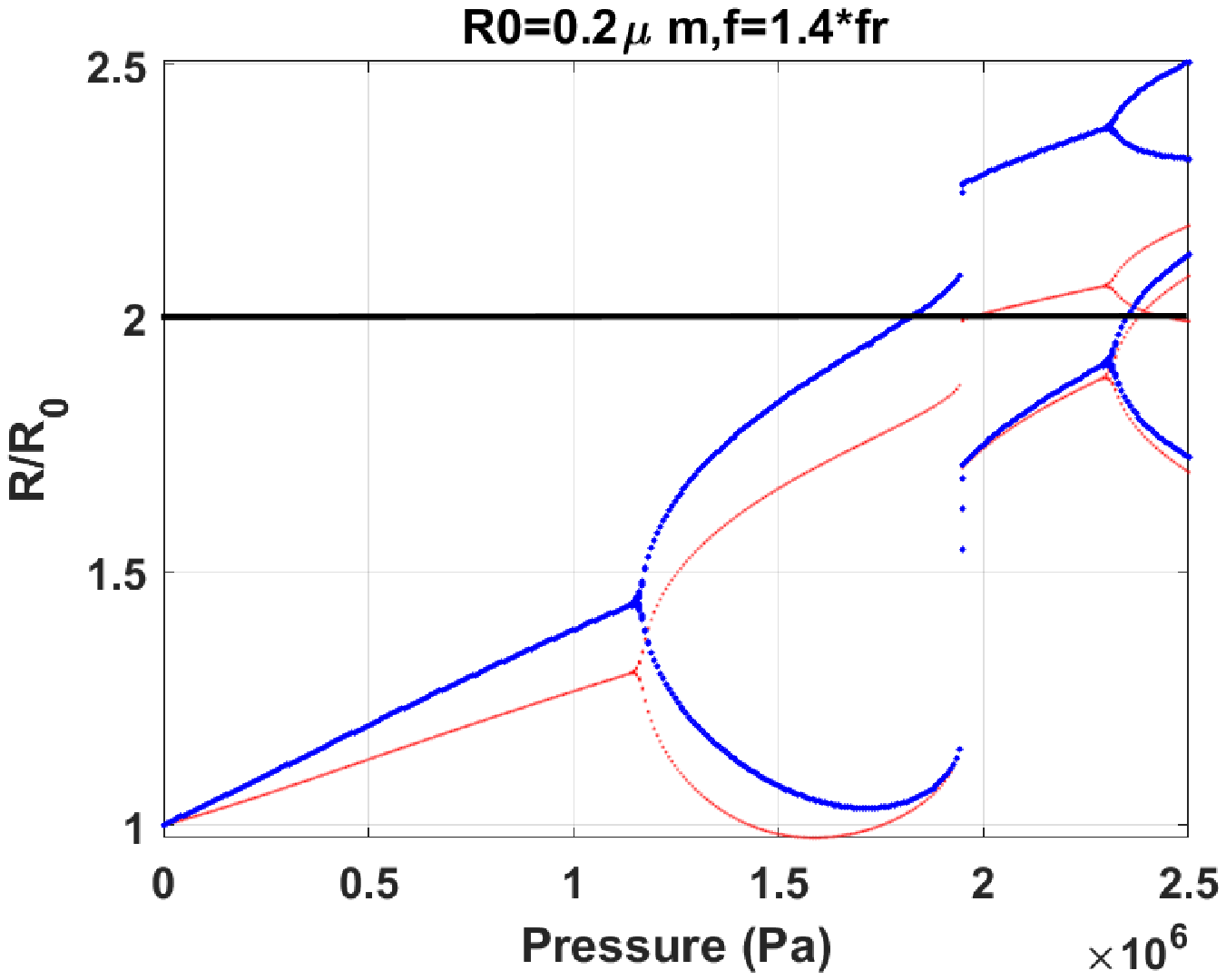}}\\
		\hspace{0.5cm} (a) \hspace{6cm} (b)\\
		\scalebox{0.43}{\includegraphics{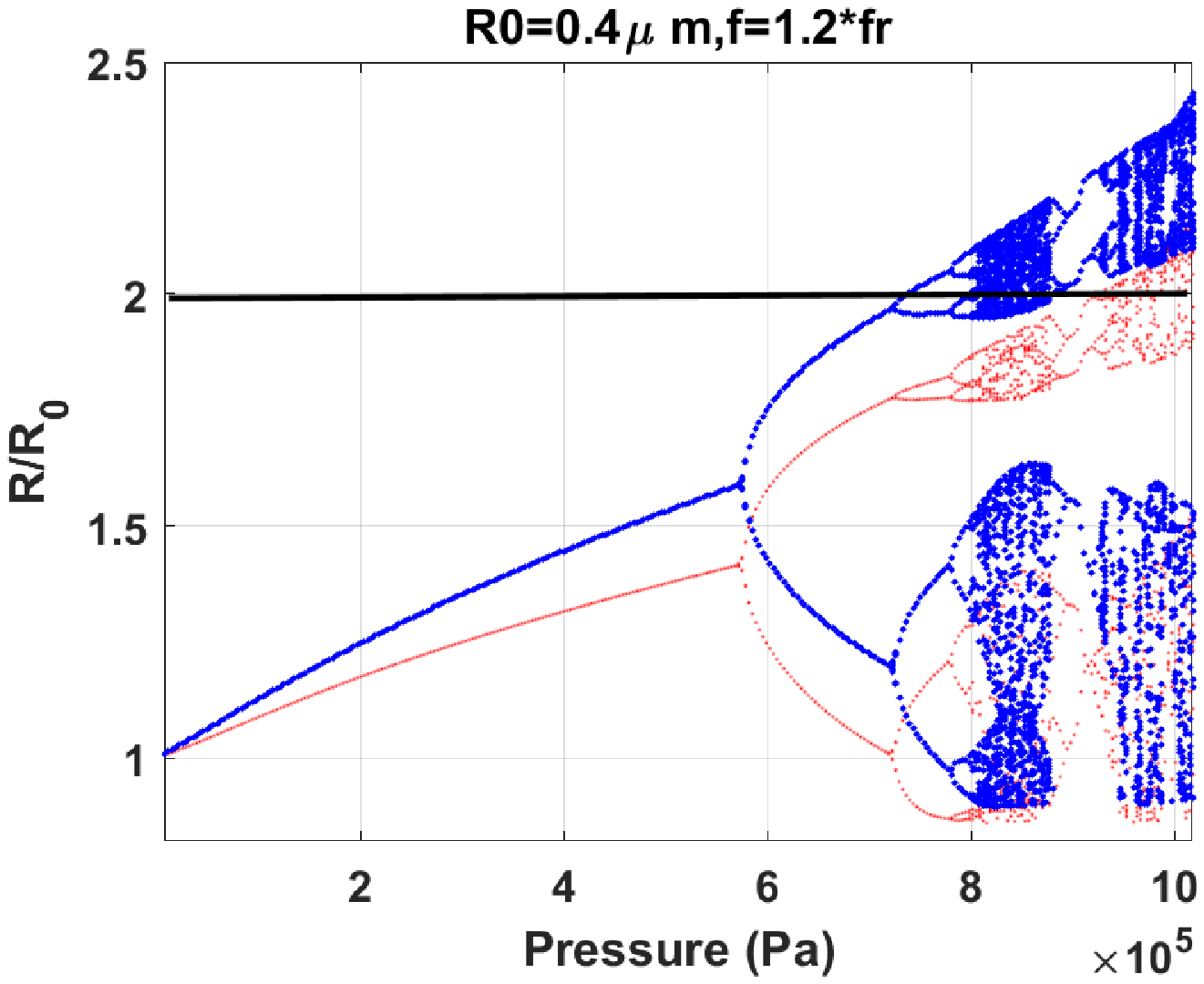}} \scalebox{0.43}{\includegraphics{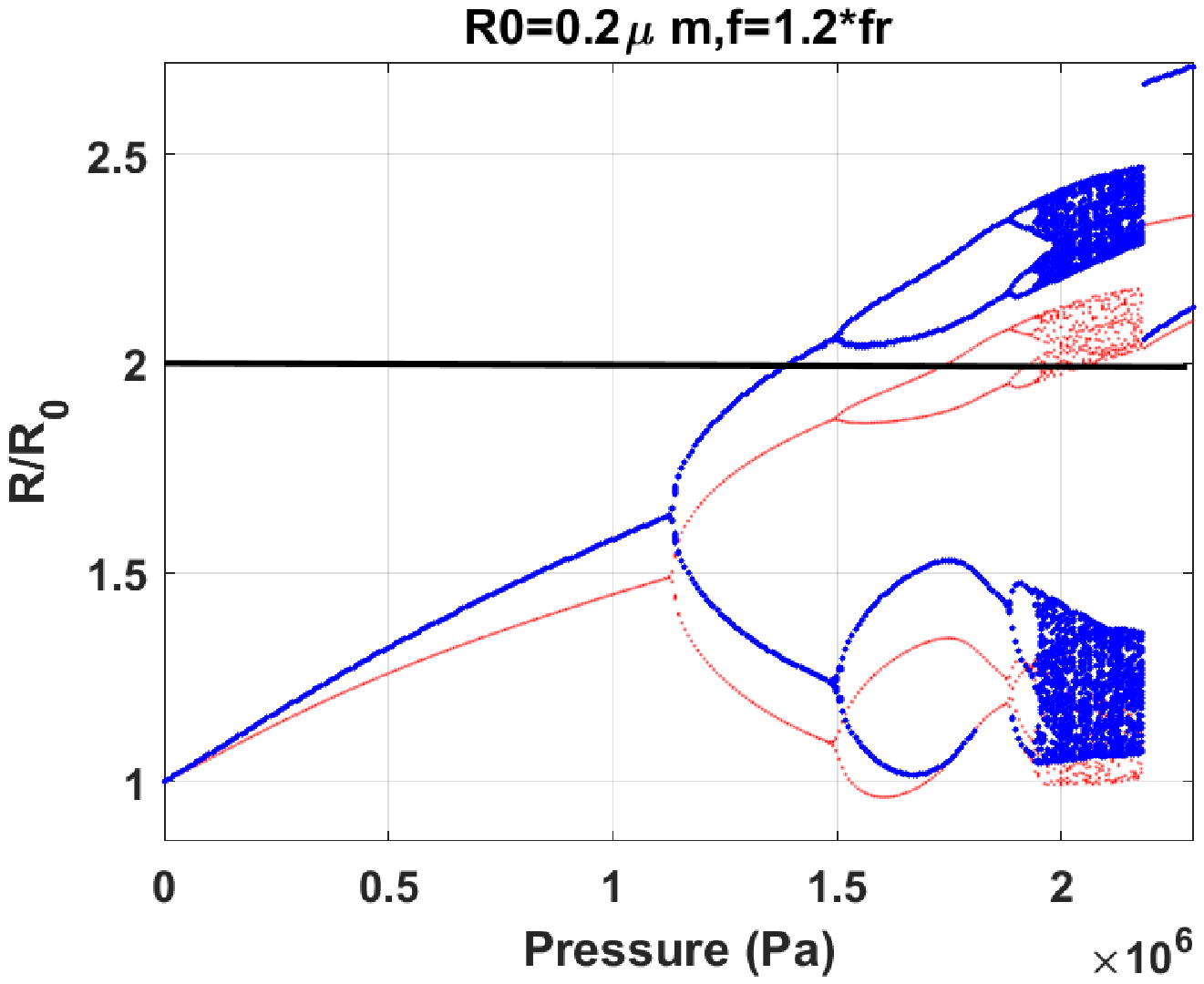}}\\
		\hspace{0.5cm} (c) \hspace{6cm} (d)\\
		\scalebox{0.43}{\includegraphics{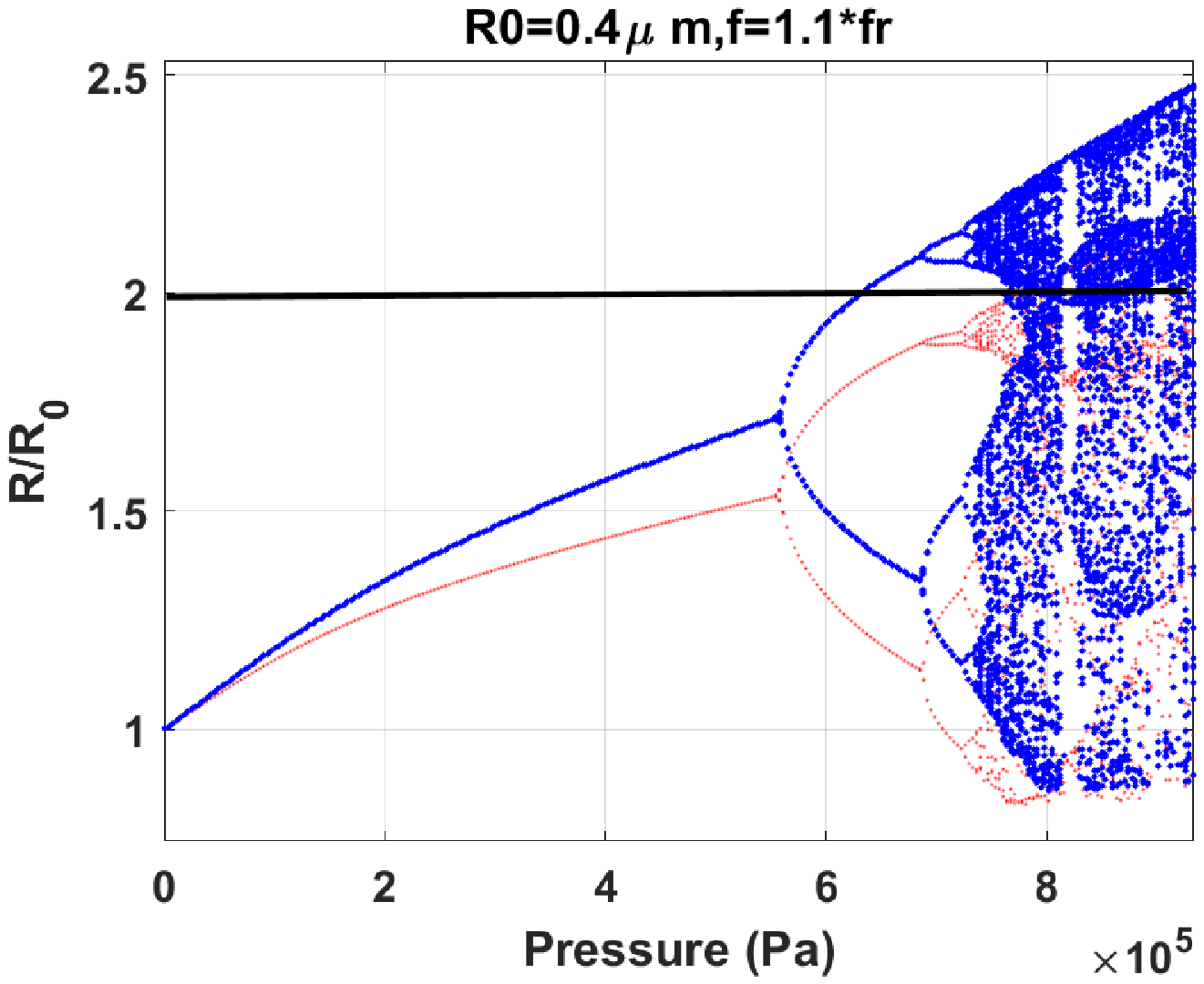}} \scalebox{0.43}{\includegraphics{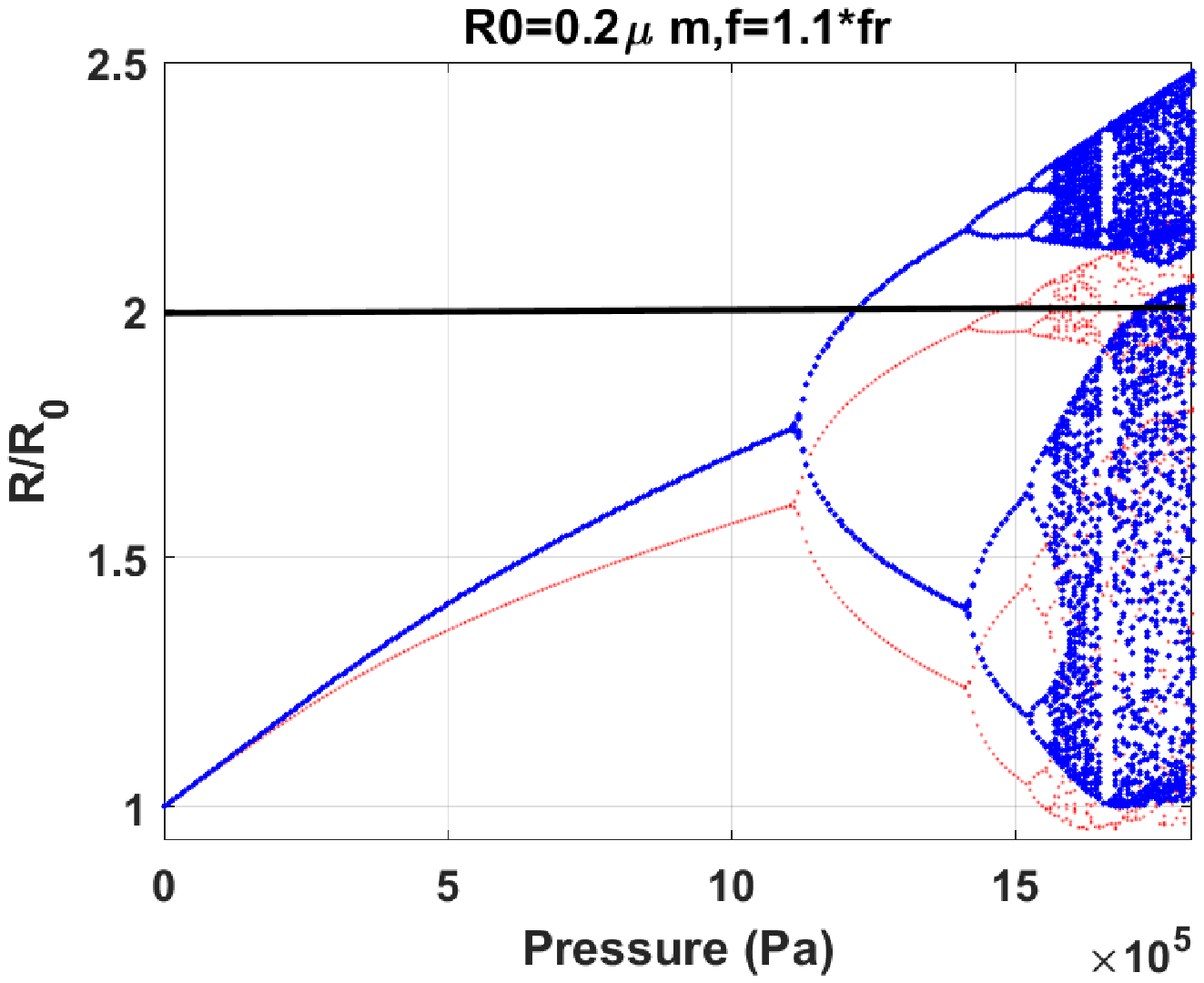}}\\
		\hspace{0.5cm} (e) \hspace{6cm} (f)\\
		\caption{Bifurcation structure (blue: method of peaks, red: conventional method) of the micron-size bubbles as a function of pressure when sonicated with $f=1.4f_r, 1.2f_r \&1.1 f_r$. Left column is for $R_0=0.4 \mu m$ and Right column is for $R_0=0.2 \mu m$ (arrow shows the pressure responsible for SN bifurcation)}
	\end{center}
\end{figure*}
Figure 9 shows the bifurcation structure of the nano-size bubbles as a function of pressure. Bubbles have initial radii of $R_0$= 400nm (left column) and $R_0$=200nm (right column). When $f=1.4f_r$, P1 bubble oscillations grow monotonically with increasing pressure and above a pressure threshold bubbles undergo a PD to P2 oscillations. P2 oscillations grow in amplitude and above a second pressure threshold; there is a SN bifurcation to P2 oscillations of higher amplitude. The SN bifurcation results in P2 oscillations that are resonant (one of the maxima in blue curve is on top of one of the branches of the red curve), however, similar to the case of the $f=1.5f_r$, the SN results in bubble destruction ($\frac{R_{max}}{R0}>2$). When $f=1.2f_r$ and $f=1.1f_r$, SN bifurcation does not take place and the bubble oscillations undergo a PD bifurcation to P2 oscillations followed by a cascade of PDs to chaos.
\subsection{Maximum achievable $P_{sc}^2$, wall velocity, SH and UH as a function of frequency}
\begin{figure*}
	\begin{center}
		\scalebox{0.43}{\includegraphics{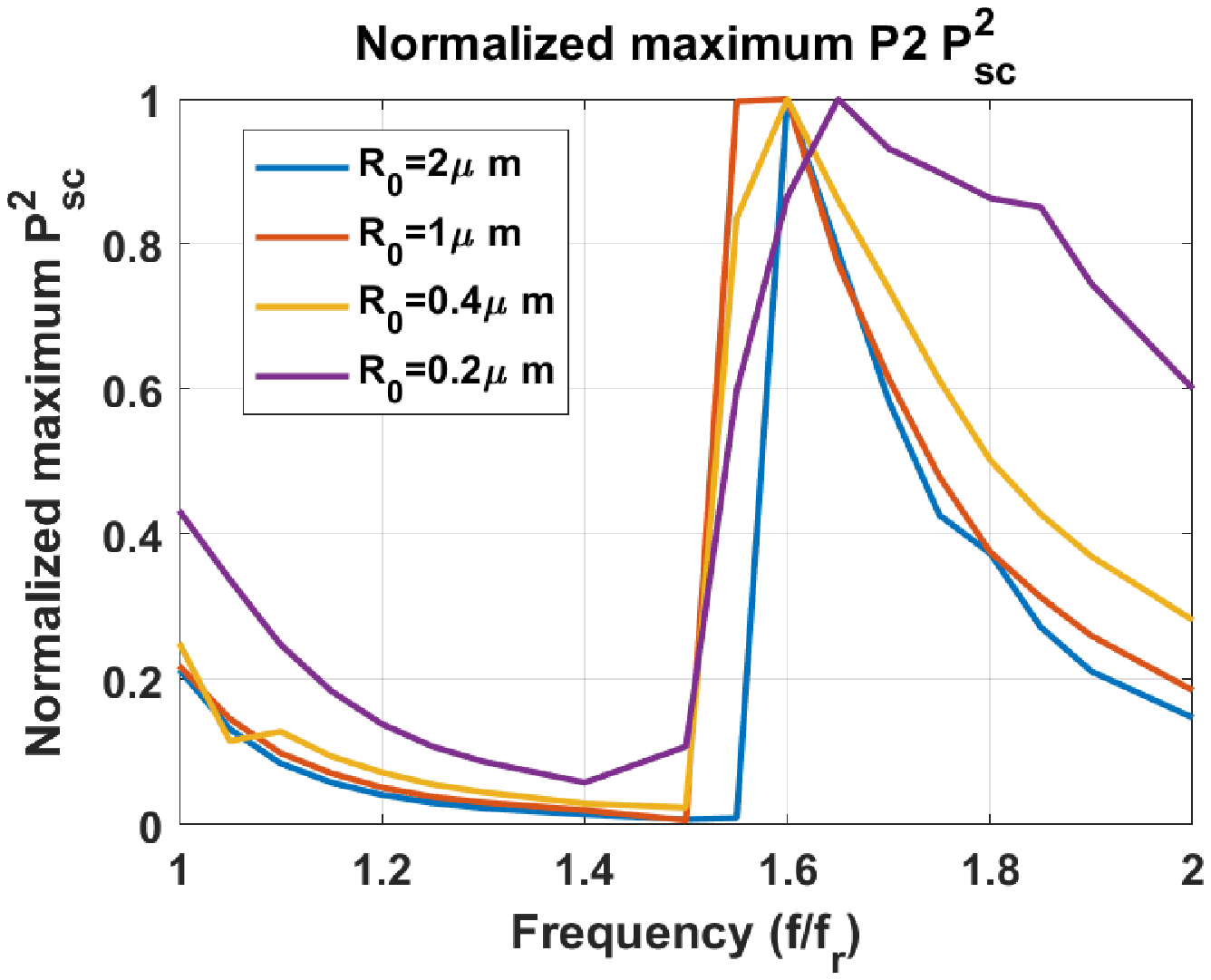}} \scalebox{0.43}{\includegraphics{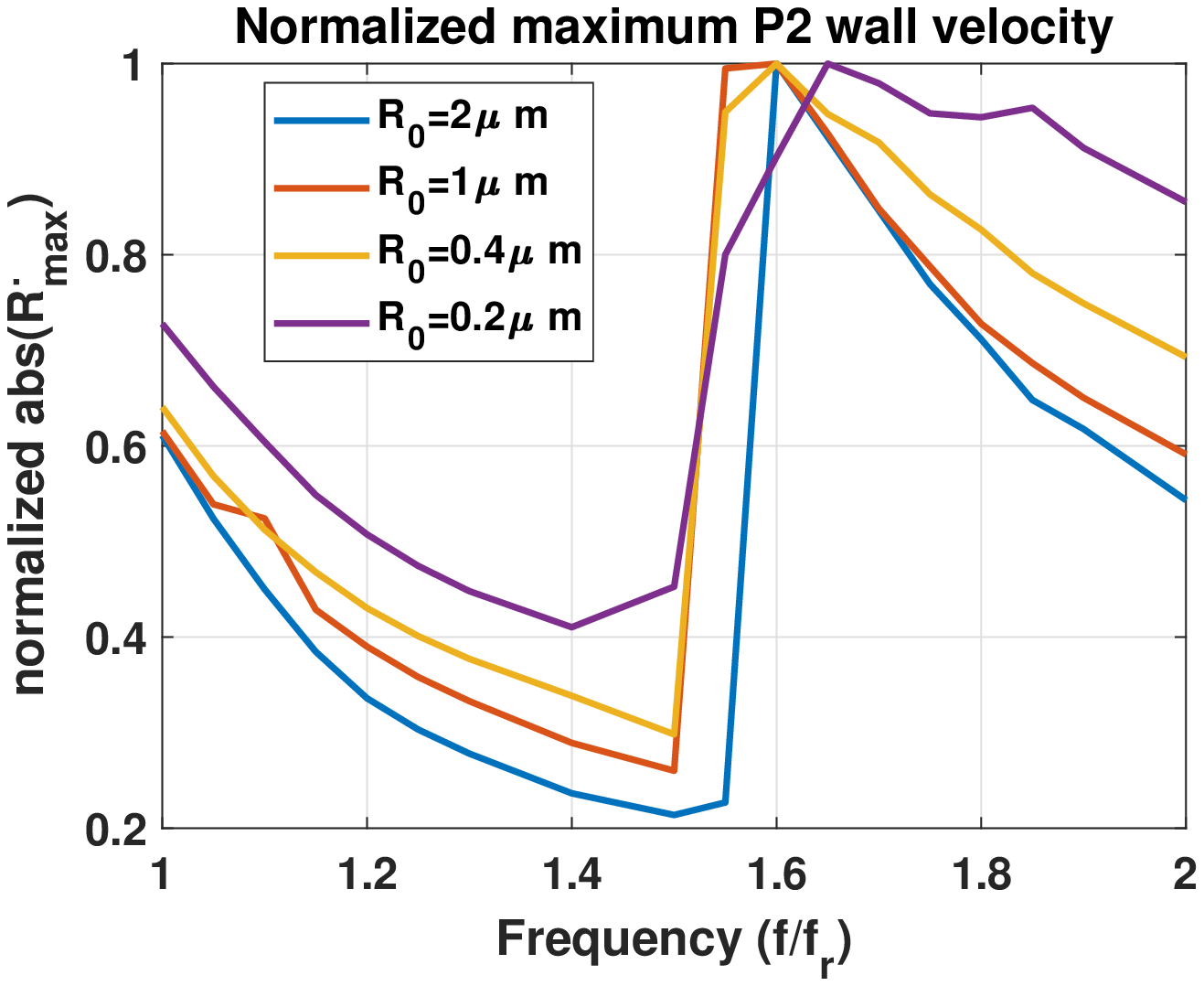}}\\
		\hspace{0.5cm} (a) \hspace{6cm} (b)\\
		\scalebox{0.43}{\includegraphics{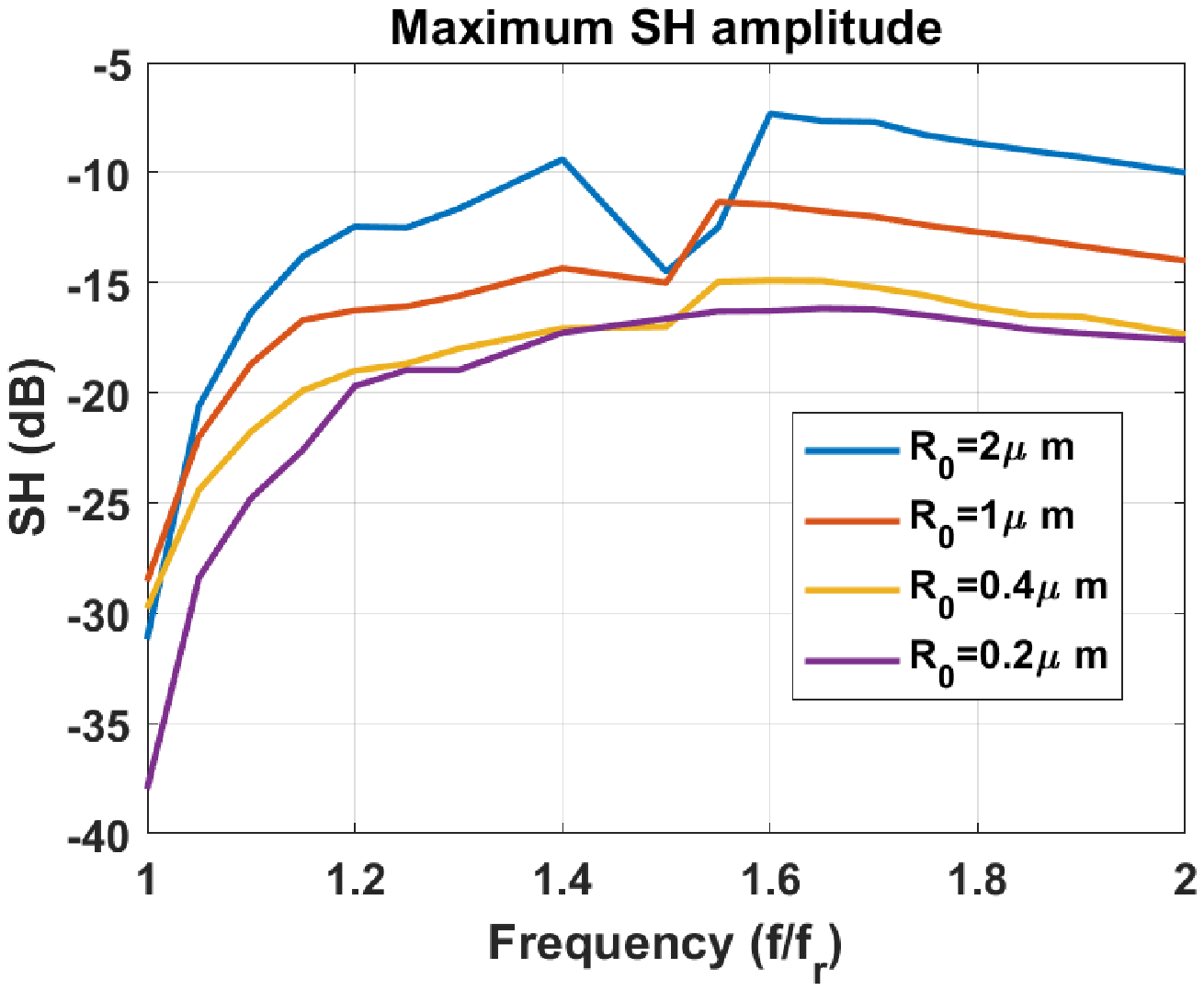}} \scalebox{0.43}{\includegraphics{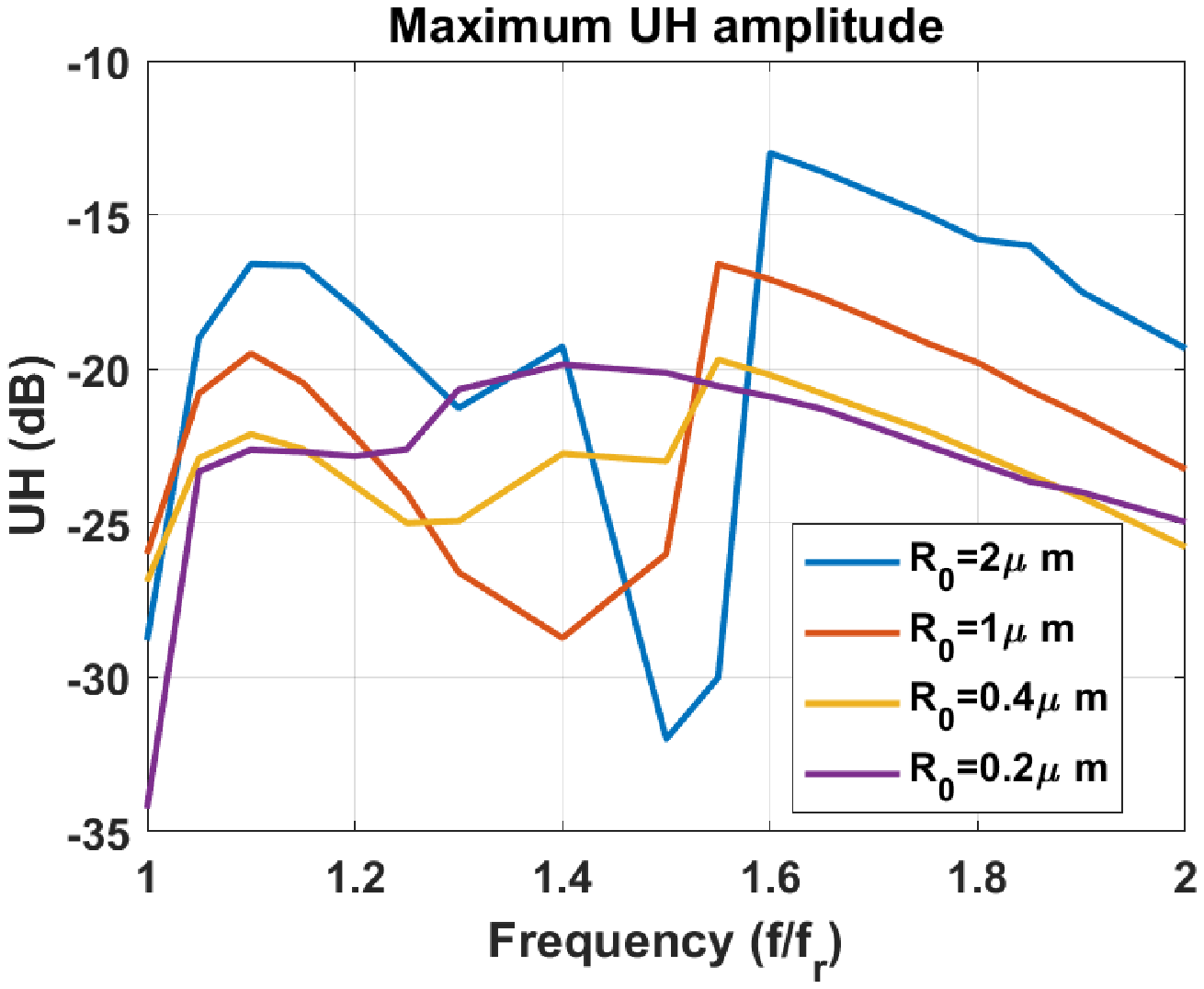}}\\
		\hspace{0.5cm} (c) \hspace{6cm} (d)\\
		\caption{a) Normalized maximum non-destructive P2 $P_{sc}^2$, b) Normalized maximum non-destructive P2 wall velocity, c) Maximum non-destructive SH amplitude, and d) Normalized maximum  UH amplitude.}
	\end{center}
\end{figure*}
In many applications, it is of interest to find exposure parameters that maximize the scattered pressure or enhances a specific frequency component in the scattered pressure.  Figure 10a-b show the normalized value of maximum non-destructive $P_{sc}^2$ ($\frac{R}{R_0}<2$) and wall velocity in the regime of P2 oscillations as a function of frequency. $P_{sc}^2$ and wall velocity reach their maximum at $f=1.6f_r$ for bubbles with $R_0$=0.4, 1 and 2$\mu m$ and it is maximum for $f=1.65f_r$ when $R_0$=0.2 $\mu m$. For all bubbles there is a universal minimum for $P_{sc}^2$ and wall velocity when $f=1.5-1.55fr$. We show that conventional practice of sonication with SH resonance frequency ($f_{sh}$=$2f_r$) does not lead to the generation of the maximum P2 scattered pressure or wall velocity. Instead there is a frequency range ($1.55<f<1.7$) that results in the maximum value of the two parameters. 
Figure 10c-d presents the maximum non-destructive ($\frac{R}{R_0}<2$) SH and UH amplitude as a function of frequency.  For all bubble sizes examined, the conventional sonication with the $f_{sh}$($2f_r$) does not result in the strongest SH or UH amplitude. The bubble with 
$R_0$=2$\mu m$ reaches the strongest SH and UH amplitude when $f=1.6f_r$; the bubbles with $R_0$=1$\mu m$ and $R_0$=0.4$\mu m$ reach the maximum at $f=1.55f_r$. For bubbles with $R_0$=2$\mu m$, 1$\mu m$ and 0.4$\mu m$ there exist a universal minimum for SH and UH at $1.5f_r$; this is because the SN bifurcation leads to bubble destruction at this frequency, thus non-destructive full amplitude P2 oscillations are not developed in this case. For $R_0$=200 nm, maximum SH and UH amplitudes occurs respectively at $f=1.6f_r$ and $1.4f_r$.
\section{Discussion and summary}
SH oscillation of bubbles are one of the most important nonlinear signatures of bubbles which are used in several medical and industrial applications [18-23,38-52]. Despite the importance of SH oscillations, most studies have only focused on investigating the minimum pressure threshold for SH oscillations [56-59,61-65]. Conditions to maximize the SH power remain uncertain. When the bubble is sonicated with twice its linear resonance frequency ($f_r$), SHs are developed at the lowest pressure threshold [55-59]. SHs grow quickly above this pressure threshold, however, they are saturated and any further increase in incident pressure may even lead to weakening of the SHs due to the occurrence of chaos or bubble destruction.  Knowledge of the conditions and exposure parameters that enhance the saturation level would allow to select exposure parameters that increase the contrast to tissue ratio (CTR) and signal to noise ratio (SNR) in applications. We have previously studied the two main routes of period doubling (PD)  in the bubble oscillator and showed that due to the very high oscillation amplitude ($\frac{R}{R_0}>2$ [67] and a for detailed review please see the discussion in [26]), non-destructive SH oscillations are less likely to be developed when bubbles are sonicated with $f_r$. However, sonication with $f_{sh}$=$2f_r$, results in the generation of SHs at very gentle oscillation regimes which increases the chance of the bubble survival during SH regime of oscillations. We have also previously shown that the scattered signal from bubbles can be enhanced if the bubbles are sonicated with its pressure dependent resonance frequency [26]. In this work the bifurcation structure of bubbles sonicated by its pressure dependent resonance frequencies ($Pdf_{sh}$) was investigated in detail.  
SH and UH amplitudes were examined between two pressure limits: The threshold for the onset of SH oscillations and the critical pressure at which the nonlinear response becomes chaotic (or results in bubble destruction). Knowledge of these limits is essential for the optimization of applications related to SHs as the SH amplitude drops rapidly when chaos occurs.
The findings of this study can be summarized as follows:\\
1-	When bubbles are sonicated with $f_{sh}$=$2f_r$, bubble oscillations undergo period doubling (PD) at the lowest pressure threshold. Period 2 (P2) oscillations result in the generation of SHs and UHs which then grow quickly with increasing pressure reaching a saturation value. Thus, there is an upper limit of achievable SH and UH strength under conventional exposure parameters for SH imaging.\\
2-	When sonicated with $f_{sh}$ or $Pdf_sh$, the occurrence of P4 or chaotic oscillations lead to a drop in the SH and UH amplitude.  Thus, in a clinical setting CTR decreases for these exposure conditions. Furthermore since P4 and chaos occur at higher acoustic pressures, higher backscatter from tissue will result in a decrease of bubble contrast enhancement compared to the tissue signal. Thus, the limit for the occurrence of P4 or chaos should be determined and avoided in practical situations where the goal is higher CTR and SNR.\\
3-	Pressure increase leads to a decrease in SH resonance frequency. This is similar to the decrease in resonance frequency with pressure [26].\\
4-	When the bubble is sonicated with $Pdf_sh$, PD initiation is at higher pressures compared to $f=f_{sh}$. Bubble oscillations undergo a SN bifurcation from P1 to P2 or from a P2 to P2 oscillations of higher amplitude. This is concomitant with a rapid growth of signal and oversaturation of the SH level.\\ 
5-	When $f=Pdf_sh$,  the SN bifurcation results in a sudden increase in the scattered pressure and wall velocity; this effect is more pronounced in bubbles $>800 nm$ as the higher viscous forces on smaller bubbles increases the pressure required for the onset of nonlinear oscillations.\\
6-	In this study for bubble sizes $>800nm$, the maximum non-destructive SH, UH, backscatter pressure and wall velocity are generated when bubbles are sonicated with f$\approxeq$1.55-1.6$f_r$. Conversely, there is a universal minimum for all these values at f$\approxeq$1.4-1.5$f_r$.\\
7-	When sonicated with $1.5f_r<f<2f_r$, the occurrence of PD is concomitant with an increase in the wall velocity and scattered pressure. This is in contrast to sonication with $f_r<f<1.5f_r$, where PD is simultaneous with a drop in scattered pressure and wall velocity. We have also previously shown that when bubble is sonicated with its pressure dependent resonance ($PDf_r$) and $f=f_r$, the maximum wall velocity drops when PD occurs [26]. This can be one of the reasons for the loss of echogenecity observed concomittant with Pd [72]\\
8-	When sonicated with $1.5f_r<f<2f_r$, the occurrence of P4 or chaotic oscillations lead to an increase in maximum scattered pressure and wall velocity; however, this increase is simultaneous with a drop in SH and UH strength; thus it is not an ideal situation for SH imaging applications.\\
9-	For bubble sizes $>400nm$ when sonicated with $f_r<f<1.5f_r$; oscillations undergo a simple PD from P1 to P2 oscillations which is followed by a PD cascade to chaos. \\
We have shown that exposure parameters ($f=2f_r$) that are used in conventional SH imaging do not result in the maximum SH or UH strength. We conclude that sonication with $f\approxeq 1.6f_r$ generates the highest achievable non-destructive ($\frac{R}{R_0}<2$) SH and UH amplitude (e.g.  depending on pressure, for the bubble with $R_0$=1$\mu m$ the enhancement in SH and UH were respectively 3.5-4 dB and 7-10 dB). \\
In this paper, we have derived the exposure parameters that maximize  the enhancement of the SHs, UHs, scattered pressure and wall velocities. The fundamental findings of this study can be used to optimize the outcome of ultrasound applications based on SH oscillations. Furthermore, in drug delivery applications, sonication parameters that lead to non-destructive oscillations with elevated wall velocities can be used to increase the long lasting shear stress on the nearby cells. \\  
SN bifurcation is concomitant with a fast increase in the scattered pressure. For example when the bubble with $R_0$=1$\mu m$ was sonicated with $f=1.6f_r$ scattered pressure underwent a 9.4 times increase as soon as SN occurred (Fig.  6b). This has several advantages for amplitude modulation imaging used in medical ultrasound [73-75]. Amplitude modulation (AM) is a method that takes advantage of the nonlinear response of the bubble to increase in acoustic pressure; in this method, two pulses are sent to the target with one having twice the amplitude of the other. The signals are scaled and subtracted upon return. Because of the linear response of tissue to pressure increase the signal from the tissue cancels and the residual signal from the bubble enhances the CTR. When bubble is sonicated with $Pdf_{sh}$, sonication with pressures below and above the pressure threshold for the SN can significantly enhance the residual signal; furthermore, because of the higher frequencies of PDsfh compared to conventional AM sonication at $f_r$, higher resolution is expected.\\
In this paper, we have analyzed the nonlinear dynamics of the free bubble in the absence of coating (shell). Coated bubbles and most importantly lipid shell bubbles [76] are used in medical applications of ultrasound from SH imaging [40] to blood brain barrier opening [46] and thrombolysis [77]. The nonlinear behavior of lipid coating (e.g. buckling and rupture) makes the dynamics of the bubble system more complex. In other words, the nonlinear shell dynamics are interwoven with the inherent nonlinear behavior of the bubble. This makes it very difficult to understand the behavior of the system and decouple effects due to the shell compared to nonlinear effects inherent in the forced bubble oscillator. It has been shown experimentally and numerically [40,52,78-84] that addition of the lipid shell reduces the pressure threshold for nonlinear oscillations including SH oscillations [59,64-65]. Followed by experimental observations of the pressure threshold of SH oscillations [78] and numerical results [64-65], Prosperetti [59] theoretically investigated the SH threshold of coated bubbles. He showed that consistent with experimental observations [78], the SH threshold can be considerably lowered with respect to that of an uncoated free bubble.  This happens when the mechanical response of the coating varies rapidly in the neighbourhood of certain specific values of the bubble radius (e.g. changes in shell parameters due to buckling of the shell [76]).\\
In this paper, we have neglected the shell effects. To better understand the dynamics of the more complex coated bubble, we first need to establish the nonlinear response of the less complex uncoated system. In this regard, understanding the nonlinear dynamics of the bubble system in the absence of the shell is the first step in developing a comprehensive framework for the understanding of the complex nonlinear behavior of bubbles. Future studies will include the effect of coating and since we know the behavior of the free bubble system, it is much easier to understand the shell effects on the bubble system.\\
We have neglected the effects of thermal damping [85-88].  Thermal damping especially in bigger bubbles can potentially have a strong effect on the dynamics of the system and changes the resonant behavior of the system. At higher frequencies (above resonance), the effect of thermal damping is weaker and neglecting the thermal effects in this paper may not change the general conclusions presented here. A more complete understanding of the thermal damping, however, is necessary for accurate prediction of the bubble behavior. Another important factor that should be considered is the interaction between bubbles [89-93]. In applications bubbles exist in poly-disperse clusters and their oscillations affect each other. We have recently shown that SH behavior of a polydisperse interacting cluster of bubbles is dictated by the SH response of the bigger bubbles in the cluster. Conclusions of this study can be useful in optimizing the SH strength of a poly-disperse cluster by optimizing the exposure parameters required to enhance the SH response of the clusters bigger bubbles.

\end{document}